%% file: HIG-23-013_temp.tex
\pdfoutput=1
\documentclass[11pt,twoside,a4paper,cmspaper,final,collab]{cms-tdr}
\def\svnVersion{d1aa487}\def\svnDate{2026/04/23}\def\cmsCernNoTag{CERN-EP-2025-065}\def\cmsCernDate{\today}\def\cmsMessage{Published in the Journal of High Energy Physics as \href{http://dx.doi.org/10.1007/JHEP04(2026)093}{\doi{10.1007/JHEP04(2026)093}.}}
\begin{document}\cmsNoteHeader{HIG-23-013}

\newcommand{\hgamgam}{\ensuremath{\PH \to \PGg \PGg}\xspace}
\newcommand{\hzz}{\ensuremath{\PH \to \PZ \PZ^{(*)} \to 4\Pell}\xspace}
\newcommand{\hww}{\ensuremath{\PH \to \PW \PW^{(*)} \to \Pe^{\pm} \PGm^{\mp} \PGn_{\Pell} \PAGn_{\Pell}}\xspace}
\newcommand{\htt}{\ensuremath{\PH \to \PGtp \PGtm}\xspace}
\newcommand{\higgs}{\ensuremath{\PH}\xspace}

\newcommand{\pth}{\ensuremath{\pt^{\PH}}\xspace}
\newcommand{\njets}{\ensuremath{N_{\mathrm{jets}}}\xspace}
\newcommand{\yh}{\ensuremath{\abs{y_{\PH}}}\xspace}
\newcommand{\ptjz}{\ensuremath{\pt^{\mathrm{j}_1}}\xspace}
\newcommand{\mjj}{\ensuremath{m_{\mathrm{jj}}}\xspace}
\newcommand{\deta}{\ensuremath{\abs{\Delta\eta_{\mathrm{jj}}}}\xspace}
\newcommand{\taujc}{\ensuremath{\tau^{\mathrm{j}}_{\mathrm{C}}}\xspace}
\newcommand{\dpjj}{\ensuremath{\Delta\phi_{\mathrm{jj}}}\xspace}
\newcommand{\dpjjabs}{\ensuremath{\abs{\Delta\phi_{\mathrm{jj}}}}\xspace}

\newcommand{\kappab}{\ensuremath{\kappa_{\PQb}}\xspace}
\newcommand{\kappac}{\ensuremath{\kappa_{\PQc}}\xspace}
\newcommand{\kappat}{\ensuremath{\kappa_{\PQt}}\xspace}
\newcommand{\cg}{\ensuremath{c_{\mathrm{g}}}\xspace}

\newcommand{\chg}{\ensuremath{c_{\mathrm{HG}}}\xspace}
\newcommand{\chb}{\ensuremath{c_{\mathrm{HB}}}\xspace}
\newcommand{\chw}{\ensuremath{c_{\mathrm{HW}}}\xspace}
\newcommand{\chwb}{\ensuremath{c_{\mathrm{HWB}}}\xspace}
\newcommand{\chgt}{\ensuremath{\tilde{c}_{\mathrm{HG}}}\xspace}
\newcommand{\chbt}{\ensuremath{\tilde{c}_{\mathrm{HB}}}\xspace}
\newcommand{\chwt}{\ensuremath{\tilde{c}_{\mathrm{HW}}}\xspace}
\newcommand{\chwtb}{\ensuremath{\tilde{c}_{\mathrm{HWB}}}\xspace}

\newcommand{\ttH}{\ensuremath{\PQt\PAQt\PH}\xspace}
\newcommand{\ggH}{\ensuremath{\Pg\Pg\PH}\xspace}
\newcommand{\VH}{\ensuremath{\PV\PH}\xspace}

\newcommand{\com}{\ensuremath{\sqrt{s} = 13\TeV}\xspace}
\newcommand{\intlumi}{\ensuremath{138\fbinv}\xspace}

\interfootnotelinepenalty=10000

\cmsNoteHeader{HIG-23-013}
\title{Combination and interpretation of differential Higgs boson production cross sections in proton-proton collisions at \texorpdfstring{$\sqrt{s}=13\TeV$}{sqrt(s)=13 TeV}}

\author*[cern]{The CMS Collaboration}

\date{\today}

\abstract{
   Precision measurements of Higgs boson differential production cross sections are a key tool to probe the properties of the Higgs boson and test the standard model. New physics can affect both Higgs boson production and decay, leading to deviations from the distributions that are expected in the standard model. In this paper, combined measurements of differential spectra in a fiducial region matching the experimental selections are performed, based on analyses of four Higgs boson decay channels ($\PGg\PGg$, $\PZ\PZ^{(*)}$, $\PW\PW^{(*)}$, and $\PGt\PGt$) using proton-proton collision data recorded with the CMS detector at $\sqrt{s}=13\TeV$, corresponding to an integrated luminosity of 138\fbinv. The differential measurements are extrapolated to the full phase space and combined to provide the differential spectra. A measurement of the total Higgs boson production cross section is also performed using the $\PGg\PGg$ and $\PZ\PZ$ decay channels, with a result of $53.4^{+2.9}_{-2.9}\stat^{+1.9}_{-1.8}\syst\unit{pb}$, consistent with the standard model prediction of $55.6\pm 2.5$\unit{pb}. The fiducial measurements are used to compute limits on Higgs boson couplings using the $\kappa$-framework and the SM effective field theory.
}

\hypersetup{
pdfauthor={CMS Collaboration},
pdftitle={Combination and interpretation of differential Higgs boson production cross sections in proton-proton collisions at sqrt(s) =13 TeV},
pdfsubject={CMS},
pdfkeywords={CMS, differential cross sections, combination, Higgs boson coupling modifiers}}

\maketitle

\section{Introduction}
\label{sec:introduction}

Since the discovery of the Higgs boson (\higgs) by the CMS and ATLAS Collaborations in 2012~\cite{cms_higgs_discovery,CMS:2013btf,atlas_higgs_discovery}, extensive efforts have been devoted to investigating its properties and couplings, as well as to searching for possible deviations from the standard model (SM) predictions. The Higgs boson is produced in proton-proton (pp) collisions through four main mechanisms: gluon-gluon fusion (ggH), vector boson fusion (qqH), Higgs-strahlung (\VH), and Higgs boson production associated with two top quarks (\ttH).
Among these, the dominant production mechanism is ggH, which has a production cross section roughly one order of magnitude larger than the other mechanisms combined.

Differential fiducial measurements represent the most model-independent way to measure Higgs boson production cross sections.
The term \textit{fiducial} refers to the fact that measurements are performed in a specific region of the phase space, close to the detector acceptance.
Whenever we mention differential cross sections in the rest of the paper, we are specifically referring to differential fiducial cross sections.
The CMS Collaboration has measured differential Higgs boson production cross sections using data recorded during the years 2016--2018 at \com (corresponding to an integrated luminosity of about \intlumi~\cite{CMS-LUM-17-003}) in a number of decay channels: \hgamgam~\cite{h_gamgam}, \hzz~\cite{h_zz}, \hww~\cite{h_ww}, \htt~\cite{h_tt}, and boosted \htt~\cite{htt_boosted}.
In this paper we present combined measurements of differential cross sections, using the results above as input.
The differential spectra are reported for the following observables: the Higgs boson transverse momentum \pth, the number of hadronic jets \njets, the absolute value of the Higgs boson rapidity \yh, the transverse momentum of the leading hadronic jet \ptjz, the invariant mass of the dijet system containing the two leading-\pt jets \mjj, the absolute value of the difference in pseudorapidity between these two jets \deta, and the variable \taujc, defined in Ref.~\cite{Gangal:2014qda}, which is the jet \pt weighted by a function of its rapidity.

Differential distributions provide information on the Higgs boson couplings.
When couplings to quarks, leptons and other bosons differ with respect to their SM values, distortions appear in the predicted differential cross section spectra, particularly noticeable in the \pth distribution.

Information is extracted by fitting parametrized spectra to a combination of differential cross sections in different decay channels.
Several frameworks for parametrizing differential spectra exist. Two of these are the $\kappa$-framework \cite{higgs_handbook,LHCHiggsCrossSectionWorkingGroup:2012nn} and the standard model effective field theory (SMEFT)~\cite{Grzadkowski_2010}: in the former, couplings are modified at the vertex level by scaling factors $\kappa$, while in the latter, the SM is extended by adding higher-dimensional operators to the Lagrangian. These operators may induce Higgs boson couplings with a different Lorentz structure than that present in the SM, and hence modify the kinematical properties of the Higgs boson. Therefore, the SMEFT language is considered more general than the $\kappa$-framework.

In the case of the $\kappa$-framework, we follow the same procedure as in Ref.~\cite{tk_paper}, which provides interpretations using data collected in 2016.
The measured \pth spectra are parametrized in terms of Higgs boson couplings, with different sets studied simultaneously:
\begin{itemize}
\item the modifier of the Higgs boson coupling to the charm quark \kappac and the bottom quark \kappab; 
\item the modifier of the Higgs boson coupling to the top quark \kappat and the coefficient \cg of the anomalous direct coupling to the gluon field in the heavy top mass limit;
\item \kappat and \kappab.
\end{itemize}
In the context of the SMEFT, CP-even and CP-odd pairs of coefficients are extracted from the \pth and \dpjj spectra, the latter being defined as the signed difference in azimuthal angle (in radians) between the two highest \pt jets in the event.

In the main interpretation obtained in the SMEFT framework, given the complexity of the fit and the impossibility of constraining all the parameters of the effective theory, a principal component analysis (PCA) is performed to identify sensitive directions of the likelihood function in the parameter space.
The linear combinations of the coefficients that have the largest eigenvalues are fitted, and constraints on these linear combinations are reported.

The results presented in this paper, tabulated and provided in the HEPData record for this analysis~\cite{hepdata}, constitute a step forward in the characterization of the properties of the Higgs boson and in the search for beyond-the-SM (BSM) physics: the combined spectra provide measurements of Higgs boson observables at the highest level of precision presently achievable; the interpretation in the $\kappa$-framework extends and improves the results obtained in Ref.~\cite{tk_paper}. 
Interpretations of Higgs boson differential distributions in the SMEFT have been performed for the first time by the CMS Collaboration, providing complementary information to the results already published in Refs.~\cite{hgg_atlas,ATLAS:2024lyh}.

This paper is organized as follows:
Section~\ref{sec:cms_detector} describes the CMS detector, with the measurements used as input to the combination presented in Section~\ref{sec:inputs}. The statistical procedure used to combine the measurements and extract the results is described in 
Section~\ref{sec:statistical_analysis}, and the systematic uncertainties are discussed in Section~\ref{sec:systematics}.
The combined differential spectra and the total Higgs boson production cross section measurement are presented in Section~\ref{sec:combination}, with the interpretations of the results in the $\kappa$ and SMEFT frameworks presented in Sections~\ref{sec:kappa} and~\ref{sec:smeft}, respectively.
Section~\ref{sec:conclusions} summarizes the results and presents the conclusions.

\section{The CMS detector}
\label{sec:cms_detector}

The central feature of the CMS apparatus is a superconducting solenoid of 6\unit{m} internal diameter, providing a magnetic field of 3.8\unit{T}.
Within the solenoid volume are a silicon pixel and strip tracker, a lead tungstate crystal electromagnetic calorimeter, and a brass and scintillator hadron
calorimeter, each composed of a barrel and two endcap sections.
Forward calorimeters extend the pseudorapidity ($\eta$) coverage provided by the barrel and endcap detectors.
Muons are detected in gas-ionization chambers embedded in the steel flux-return yoke outside the solenoid.
A more detailed description of the CMS detector, together with a definition of the coordinate system used and the relevant kinematic variables, can be found in~\cite{CMS:2023gfb}.

\section{Inputs to the combined analysis}
\label{sec:inputs}

The differential cross section measurements used as input to the combination, mentioned in the introduction, are \hgamgam, \hzz, \hww, \htt, and Lorentz-boosted \htt.
The \ggH production dominates in all the measurements considered.
The inclusion of the \htt analyses provides better sensitivity at high \pth, where the three other channels are less sensitive.

A larger number of observables are measured compared with Ref.~\cite{tk_paper}: \deta, \mjj, and \taujc are measured in addition to \pth, \njets, \ptjz, and \yh.
Tables \ref{tab:ptbins}--\ref{tab:taujc} show the bin boundaries used for each observable in each analysis.
Table~\ref{tab:deltaphijj} reports the bin boundaries used for \dpjjabs (\hgamgam, defined as in Ref.~\cite{h_gamgam}) and \dpjj (\hzz, defined as in Ref.~\cite{h_zz}), which are used only to set limits on Wilson coefficients in the SMEFT interpretation.

Each analysis is performed in a different fiducial phase space and applies a different event categorization.
The fiducial phase spaces are defined by the selection criteria applied to the particles at generator level (\ie, before detector simulation).

In the \hgamgam measurement~\cite{h_gamgam}, the leading (subleading) photon transverse momentum divided by the diphoton mass must be greater than 1/3 (1/4).
The total hadronic energy in a cone of radius $\Delta R = 0.3$ around the photon candidate is required to be less than 10\GeV, with the angular distance between two particles $i$ and $j$ defined as $\Delta R(i, j)=\sqrt{\smash[b]{(\Delta \eta_{i, j})^2+(\Delta \phi_{i, j})^2}}$, and only photons within $\abs{\eta^{\PGg}} < 2.5$ are accepted.

In the case of \hzz~\cite{h_zz}, leptons at the fiducial level are considered as \textit{dressed}, \ie, final-state radiation (FSR) photons are collected within $\Delta R=0.3$ from the lepton, and added to the lepton momentum.
In the \hzz analysis, FSR photons are identified by selecting all stable photons and then verifying, in simulation, whether the lepton to which the FSR recovery procedure is applied is a parent of those photons.
Electrons (muons) are required to have $\pt>7 \ (5) \GeV$ and $\abs{\eta}<2.5\, (2.4)$.
Pairs of same-flavor, opposite-charge leptons are used to form \PZ boson candidates, which are retained if the leading (subleading) dressed lepton has $\pt>20 \ (10) \GeV$.
To ensure the leptons are isolated, the scalar \pt sum of all stable particles within a cone of $\Delta R=0.3$, with the exception of FSR photons and other leptons, must be less than $0.35$ times the lepton \pt.
Events passing these requirements are retained if they have at least two lepton pairs.
The lepton pair with invariant mass closest to the true \PZ boson mass ($\PZ_{1}$) must have $40<m_{\PZ_{1}}<120\GeV$.
The second $\PZ$ boson candidate ($\PZ_{2}$) must have $12<m_{\PZ_{2}}<120\GeV$.
Each lepton pair $\Pell_i, \Pell_j$ must be separated by $\Delta R(\Pell_i, \Pell_j) > 0.02$, while any opposite-charge lepton pair must have  $m_{\ell^{+}\ell'^{-}} > 4\GeV$.

In the case of \hww~\cite{h_ww}, the leptons are dressed and must be either electrons or muons with opposite charge and $\abs{\eta} < 2.5\,(2.4)$ for electrons (muons).
The leading (subleading) lepton is required to have $\pt^{\Pell_1} > 25\GeV$ ($\pt^{\Pell_2} > 13\GeV$). The dilepton system is required to have $m^{\Pell\Pell} > 12\GeV$ and $\pt^{\Pell\Pell} >30\GeV$. Furthermore, the transverse mass of the subleading lepton must be $m_{\text{T}}^{l_2} > 30\GeV$ and the Higgs boson transverse mass must be greater than 60\GeV.
Both these quantities are defined in Ref.~\cite{h_ww}.

Electrons and muons are also dressed in the \htt analysis~\cite{h_tt, htt_boosted}. In the $\Pe\tauh$ ($\PGm\tauh$) final state, where $\tauh$ denotes a hadronically decaying \PGt lepton, the electron (muon) is required to have $\pt>25\,(20)\GeV$ and $\abs{\eta}<2.1$. The $\tauh$ candidate must have a \textit{visible} $\pt > 30$\GeV and visible $\abs{\eta}<2.3$.
Here, the term visible refers to the kinematic variables constructed from the momenta of the visible decay products of the \PGt leptons, thus excluding neutrinos. 
In addition, the transverse mass $\mT(\Pe/\PGm,\ptvecmiss)$, with \ptvecmiss computed by summing the \pt of the neutrinos, must be less than 50\GeV.

In the $\tauh\tauh$ final state, the visible \pt\ of both $\tauh$ candidates must exceed 40\GeV, while their visible $\abs{\eta}$ must be less than 2.1, and there must be at least one jet with $\pt>30\GeV$.
In the $\Pe\PGm$ final state, the leading (subleading) lepton must have
$\pt>24$ (15)\GeV, both leptons must have $\abs{\eta}<2.4$, and the $\mT$ of the dilepton system and $\ptvecmiss$ must be below 60\GeV.

SM predictions for the four main Higgs boson production modes (ggH, qqH, VH, \ttH) are generated following the procedure described in Ref.~\cite{h_gamgam}, using \MGvATNLO (version 2.6.5) at next-to-leading order (NLO) accuracy of the strong coupling constant $\alpS$ in perturbative quantum chromodynamics~(QCD).

\begin{table}[!ht]
    \centering
    \topcaption{The $\pt^{\PH}$ bin boundaries used in the analyses that are input to the combination.}
    \resizebox{\columnwidth}{!}{%
    \begin{tabular}{lccccc}
        Channel & \hgamgam & \hzz & \hww & \htt & \htt\ boosted \\
        \hline
        \multirow{24}{*}{\pth bin boundaries (\GeVns{})} & 0--5 & \multirow{2}{*}{0--10} & \multirow{6}{*}{0--30} & \multirow{8}{*}{0--45} & \multirow{18}{*}{} \\
        & 5--10 & & & & \\
        & 10--15 & \multirow{2}{*}{10--20} & & & \\
        & 15--20 & & & & \\
        & 20--25 & \multirow{2}{*}{20--30} & & & \\
        & 25--30 & & & & \\
        & 30--35 & \multirow{2}{*}{30--45} & \multirow{2}{*}{30--45} & & \\
        & 35--45 & & & & \\
        & 45--60 & 45--60 & \multirow{2}{*}{45--80} & \multirow{2}{*}{45--80} & \\
        & 60--80 & 60--80 & & & \\
        & 80--100 & \multirow{2}{*}{80--120} & \multirow{2}{*}{80--120} & \multirow{2}{*}{80--120} & \\
        & 100--120 & & & & \\
        & 120--140 & \multirow{3}{*}{120--200} & \multirow{3}{*}{120--200} & 120--140 & \\
        & 140--170 & & & 140--170 & \\
        & 170--200 & & & 170--200 & \\
        & 200--250 & \multirow{5}{*}{200--$\infty$} & \multirow{5}{*}{200--$\infty$} & \multirow{2}{*}{200--350} & \\
        & 250--350 & & & & \\
        & 350--450 & & & 350--450 & \\
        & \multirow{2}{*}{450--$\infty$} & & & \multirow{2}{*}{450--$\infty$} & 450--600 \\
        & & & & & 600--$\infty$ \\
        \\
        \\
    \end{tabular}
    }
    \label{tab:ptbins}
\end{table}

\begin{table}[!ht]
    \topcaption{The $N_{\text{jets}}$ bins used in the analyses that are input to the combination.}
    \centering
    \begin{tabular}{lccccc}
        Channel & \multicolumn{5}{l}{$N_{\text{jets}}$ bins} \\
        \hline
        \hgamgam & 0 & 1 & 2 & 3 & $\geq 4$ \\
        \hzz & 0 & 1 & 2 & 3 & $\geq 4$ \\
        \hww & 0 & 1 & 2 & 3 & $\geq 4$ \\
        \htt & 0 & 1 & 2 & 3 & $\geq 4$ \\
        \\
        \\
    \end{tabular}
    \label{tab:njetsbins}
\end{table}

\begin{table}[!ht]
    \topcaption{The \ptjz\ bin boundaries used in the analyses that are input to the combination.}
    \centering
    \resizebox{\columnwidth}{!}{%
    \begin{tabular}{lcccccccccc}
        Channel & \multicolumn{9}{c}{\ptjz\ bin boundaries (\GeVns{})} \\
        \hline
        \hgamgam & 30--40 & 40--55 & 55--75 & 75--95 & 95--120 & 120--150 & 150--200 & \multicolumn{2}{c }{200--$\infty$} \\
        \hzz & \multicolumn{2}{c}{30--55} & \multicolumn{2}{c}{55--95} & \multicolumn{3}{c}{95--200} & \multicolumn{2}{c}{200--$\infty$} \\
        \htt\ boosted & \multicolumn{7}{c}{} & 450--600 & 600--$\infty$ \\
        \\
        \\
    \end{tabular}%
    }
    \label{tab:ptjz}
\end{table}

\begin{table}[!ht]
    \topcaption{The \yh\ bin boundaries used in the analyses that are input to the combination.}
    \centering
    \resizebox{\columnwidth}{!}{%
    \begin{tabular}{lcccccccccc}
        Channel & \multicolumn{10}{c}{\yh\ bin boundaries (\GeVns{})} \\
        \hline
        \hgamgam & 0.0--0.15 & 0.15--0.3 & 0.3--0.45 & 0.45--0.6 & 0.6--0.75 & 0.75--0.9 & 0.9--1.2 & 1.2--1.6 & 1.6--2.0 & 2.0--2.5 \\
        \hzz & 0.0--0.15 & 0.15--0.3 & 0.3--0.45 & 0.45--0.6 & 0.6--0.75 & 0.75--0.9 & 0.9--1.2 & 1.2--1.6 & \multicolumn{2}{c}{1.6--2.5} \\
        \\
        \\
    \end{tabular}
    }
    \label{tab:yhbins}
\end{table}

\begin{table}[!ht]
    \topcaption{The \deta bin boundaries used in the analyses that are input to the combination.}
    \centering
    \begin{tabular}{lccccc}
        Channel & \multicolumn{5}{c}{\deta bin boundaries} \\
        \hline
        \hgamgam & 0.0--0.7 & 0.7--1.6 & 1.6--3.0 & 3.0--5.0 & 5.0--$\infty$ \\
        \hzz & \multicolumn{2}{c}{0.0--1.6} & 1.6--3.0 & \multicolumn{2}{c}{3.0--$\infty$} \\
    \\
    \\
    \end{tabular}
    \label{tab:detabins}
\end{table}

\begin{table}[!ht]
    \topcaption{The \mjj bin boundaries used in the analyses that are input to the combination.}
    \centering
    \begin{tabular}{lccccccc}
        Channel & \multicolumn{7}{c}{\mjj bin boundaries (\GeVns{})} \\
        \hline
        \hgamgam & 0--75 & 75--120 & 120--180 & 180--300 & 300--500 & 500--1000 & 1000--$\infty$ \\
        \hzz & \multicolumn{2}{c}{0--120} & \multicolumn{2}{c}{120--300} & \multicolumn{3}{c}{300--$\infty$} \\
    \\
    \\
    \end{tabular}
    \label{tab:mjj}
\end{table}

\begin{table}[!ht]
    \centering
    \topcaption{The \taujc bin boundaries used in the analyses that are input to the combination.}
    \begin{tabular}{lccccc}
        Channel & \multicolumn{5}{c}{\taujc bin boundaries} \\
        \hline
        \hgamgam & 15--20 & 20--30 & 30--50 & 50--80 & 80--$\infty$ \\
        \hzz & 15--20 & 20--30 & 30--50 & 50--80 & 80--$\infty$ \\
        \\
        \\
    \end{tabular}
    \label{tab:taujc}
\end{table}

\begin{table}[!ht]
    \centering
    \topcaption{The $|\Delta\phi_{jj}|$ (\hgamgam, Ref.~\cite{h_gamgam}) and $\Delta\phi_{jj}$ (\hzz, Ref.~\cite{h_zz}) bin boundaries, used to set constraints on Wilson coefficients.}
    \begin{tabular}{lcccccccccccc}
        Channel & \multicolumn{12}{c}{$|\Delta\phi_{jj}|$, $\Delta\phi_{jj}$ bin boundaries} \\
        \hline
        \hgamgam & \multicolumn{6}{c}{} & 0--0.5 & 0.5--0.9 & 0.9--1.3 & 1.3--1.7 & 1.7--2.5 & 2.5--$\pi$ \\
        \hzz & \multicolumn{3}{c}{(-$\pi$)--(-$\pi$/2)} & \multicolumn{3}{c}{(-$\pi$/2)--0} & \multicolumn{3}{c}{0--($\pi$/2)} & \multicolumn{3}{c}{($\pi$/2)--$\pi$} \\
        \\
        \\
    \end{tabular}
    \label{tab:deltaphijj}
\end{table}

\section{Statistical analysis}
\label{sec:statistical_analysis}

The parameters of interest are estimated through a simultaneous extended maximum likelihood fit in all the analysis categories of the following distributions:
\begin{itemize}
    \item diphoton invariant mass in \hgamgam;
    \item four-lepton invariant mass in \hzz;
    \item two-dimensional dilepton invariant mass and transverse mass in \hww;
    \item di-$\PGt$\ invariant mass in \htt;
    \item output of the neural network used to distinguish signal from backgrounds in boosted \htt.
\end{itemize}

The numbers of expected events $n^{\mathrm{exp}}$ in a given reconstructed kinematic bin $i$, analysis category $k$, and decay channel $m$ is written as:

\begin{equation}
    n_i^{\mathrm{exp, } k m}(\vec{\mu} \mid \vec{\nu}, \vec{\nu_{\mathrm{th}}}) = n_i^{\mathrm{sig, } k m}(\vec{\mu} \mid \vec{\nu}, \vec{\nu_{\mathrm{th}}}) + n_i^{\mathrm{bkg, } k m}(\vec{\nu}, \vec{\nu_{\mathrm{th}}}),
    \label{eq:nexp}
\end{equation}
where $n_i^{\text{sig, } k m}$ and $n_i^{\text{bkg, } k m}$ are the expected number of signal and background events, respectively.

The signal component $n_i^{\text{sig, } k m}$ is explicitly written as:

\begin{equation}
    n_i^{\mathrm{sig, } k m}(\vec{\mu} \mid \vec{\nu}, \vec{\nu_{\mathrm{th}}})=\sum_{j=1}^{n_{\text{bins}, k}^{\text{gen}}} \mu_j \sigma_j^{\mathrm{SM}}(\vec{\nu_{\mathrm{th}}}) A_{j}^{\mathrm{SM},km}(\vec{\nu_{\mathrm{th}}}) \varepsilon_{ji}^{\mathrm{SM},km}(\vec{\nu}) L(\vec{\nu}) \mathcal{B}^{\mathrm{SM},m}(\vec{\nu_{\mathrm{th}}}) + n^{km}_{\mathrm{OOA}, i}(\vec{\nu}, \vec{\nu_{\mathrm{th}}}),
    \label{eq:nsig}
\end{equation}
where:
\begin{itemize}
    \item $\vec{\mu}$ is the set of signal strength modifiers;
    \item $\vec{\nu}$ is the set of nuisance parameters associated with experimental uncertainties;
    \item $\vec{\nu_{\mathrm{th}}}$ is the set of nuisance parameters associated with theoretical uncertainties;
    \item $j$ is the index of a kinematic bin at the generator level;
    \item $n_{\text{bins}, k}^{\text{gen}}$ is the number of bins at the generator level in analysis category $k$; for some observables (such as \njets) it is the same for all decay channels, while this changes for other observables (such as $\pt^{\PH}$);
    \item $\sigma_j^{\mathrm{SM}}$ is the SM cross section in generator-level bin $j$;
    \item $A_{j}^{\mathrm{SM},km}$ is the SM expectation for the fiducial acceptance (fraction of the events in the fiducial region) in generator-level bin $j$ for decay channel $m$ and analysis category $k$;
    \item $\varepsilon_{ji}^{\mathrm{SM},km}(\vec{\nu})$ is the efficiency with which events in generator-level bin $j$ are reconstructed in the reconstructed bin $i$;
    \item $L$ is the integrated luminosity of the data sets used in the analyses;
    \item $\mathcal{B}^{\mathrm{SM},m}$ is the SM expectation for the branching fraction of the decay channel $m$;
    \item $n^{km}_{\mathrm{OOA}, i}$ is the number of events that originate from outside the fiducial region but are reconstructed within it.
\end{itemize}

Bin-to-bin migrations due to detector resolution effects are taken into account via a folding matrix.
When performing the fits in the individual analyses, extracting the production cross sections in the corresponding fiducial phase space, the signal strength modifiers are defined as
\begin{equation}
    \mu = \frac{\sigma \mathcal{B} A}{\sigma^{\mathrm{SM}} \mathcal{B}^{\mathrm{SM}} A^{\mathrm{SM}}},
    \label{eq:mu_fiducial}
\end{equation}
where the indices of the quantities are omitted for simplicity.
When performing combined fits, however, the cross sections in each decay channel are extrapolated to the full phase space, and the signal strength modifiers are defined as:
\begin{equation}
    \mu = \frac{\sigma}{\sigma^{\mathrm{SM}}}.
    \label{eq:mu_extrapolation}
\end{equation}
It should be noted that the extrapolation procedure introduces theoretical uncertainties and an unavoidable model dependence in the results~\cite{CERN_Report_4}.

Using the terminology introduced in Ref.~\cite{combine}, which we refer to for more details, this combination includes both parametric and template-based statistical models.
Given a statistical model and a data set, a likelihood function is written as the product of the primary and auxiliary likelihoods, where $\mathcal{L}_{\text{primary}}$ is proportional to the probability of observing the event count in data for a given set of model parameters and $\mathcal{L}_{\text{auxiliary}}$ represents external constraints on the nuisance parameters.
For the template-based models, the likelihood is written as
\begin{equation}
    \mathcal{L}=\mathcal{L}_{\text{primary }} \mathcal{L}_{\text{auxiliary }}=\prod_{c=1}^{N_c} \prod_{b=1}^{N_b^c} \operatorname{Pois}\left(n_{c b} ; n_{c b}^\text{exp}(\vec{\mu}, \vec{\nu})\right) \prod_{e=1}^{N_E} p_e\left(y_e ; \nu_{e}\right),
\end{equation}
where $c$ runs over the channels, $b$ over the bins, and $e$ over the auxiliary measurements; $N_{c}$ is the number of channels, $N_{b}^{c}$ is the number of bins in channel $c$, $N_{E}$ is the number of auxiliary measurements, $n_{cb}$ is the observed number of events in bin $b$ of channel $c$, $n_{cb}^{\exp}$ is the expected number of events in the same bin, $y_e$ is the value of the auxiliary measurement $e$, and $\nu_e$ is the corresponding nuisance parameter.
For parametric models, the (unbinned) likelihood is written as
\begin{equation}
    \mathcal{L}=\mathcal{L}_{\text{primary }} \mathcal{L}_{\text{auxiliary }}
    =\left( \prod_c \operatorname{Pois}\left(n_{c, \text{ tot }}^{\text{obs }} ; n_{c, \text{ tot }}^\text{exp}(\vec{\mu}, \vec{\nu})\right) \prod_i^{n_{\text{bs }}^{\text{obs }}} \sum_p f_{c p}^\text{exp} \operatorname{pdf}_{c p}\left(\vec{x}_i ; \vec{\mu}, \vec{\nu}\right) \right) \prod_e^{N_E} p_e\left(y_e ; \PGn_e\right),
\end{equation}
where $c$ runs over the channels (as in the binned case), $i$ runs over the events, $n_{c, \text{ tot }}^\text{exp}$ is the total number of expected events in channel $c$, $\operatorname{pdf}_{c p}$ is the probability density function of the process $p$ in channel $c$, and $f_{c p}^\text{exp}$ is the fraction of the total number of expected events in channel $c$ that originate from process $p$, $f_{c p}=n_{c p}/\sum_p n_{c p}$.
It should be noted that, in the case of parametric models, the likelihood function can be either binned or unbinned.

When combining the likelihoods from different analyses, the likelihood function is the product of the likelihood functions of the individual analyses, both binned and unbinned:
\begin{equation}
    \mathcal{L}_{\text{combined }}=\mathcal{L}_{\text{primary }} \mathcal{L}_{\text{auxiliary }}=\left(\prod_{c_{\text{template }}} \mathcal{L}_{\text{primary }}^{c_{\text{template }}}\right)\left(\prod_{c_{\text{parametric }}} \mathcal{L}_{\text{primary }}^{c_{\text{parametric }}}\right) \mathcal{L}_{\text{auxiliary }},
\end{equation}
where $c_{\text{template }}$ runs over the analyses that adopt template-based models (\ie, \hww, \htt, and boosted \htt), and $c_{\text{parametric }}$ runs over the analyses that adopt parametric models (\ie, \hgamgam\ and \hzz).

The test statistic $q$ is defined as
\begin{equation}
    q(\vec{\mu})=-2 \ln \left(\frac{\mathcal{L}\left(\vec{\mu} \mid \hat{\vec{\nu}}_{\vec{\mu}}\right)}{\mathcal{L}(\hat{\vec{\mu}} \mid \hat{\vec{\nu}})}\right)
    \label{eq:q}
\end{equation}
and is used to set confidence intervals on the signal strength modifiers $\mu$ \cite{combine}.
The quantities $\hat{\vec{\mu}}$ and $\hat{\vec{\nu}}$ are the unconditional maximum likelihood estimates for the parameters $\vec{\mu}$ and $\vec{\nu}$, respectively, while $\hat{\vec{\nu}}_{\vec{\mu}}$ denotes the maximum likelihood estimate for $\vec{\nu}$ conditional on the values of $\vec{\mu}$.

\section{Systematic uncertainties}
\label{sec:systematics}

The experimental systematic uncertainties from the input analyses are incorporated in the combination as nuisance parameters, which are profiled in the maximum likelihood fit.
Detailed descriptions of the systematic uncertainties can be found in the papers describing the individual analyses~\cite{h_gamgam,h_zz,h_ww,h_tt}.
Systematic uncertainties that affect different decay channels are correlated when building the likelihood function: this happens for the integrated luminosity, pileup, jet energy scale and resolution, and b tagging uncertainties.
Some analyses employ a more detailed nuisance parameter scheme (\eg, different nuisance parameters for different eras, final states, etc.); in this case the uncertainties are not correlated between the input analyses. 
This happens in the case of the \PGt energy scales and lepton efficiencies, including electrons, muons, and hadronic taus.

Since the combined spectra are extrapolated to the full phase space, studies have been performed to assess the impact of scale and parton distribution function (PDF) variations on the acceptance of each observable and decay channel.
Scale variations show a nonnegligible impact on the acceptance in most of the observables and decay channels. 
Therefore, two additional nuisance parameters are introduced in the combined fit to account for the renormalization and factorization scale uncertainties in the extrapolation to the full phase space.
The probability density function of the number of expected events as a function of these parameters is an asymmetric log-normal distribution with the $+1\sigma$ ($-1\sigma$) variation obtained by taking the ratio between the acceptance computed with a scale parameter of 2 (0.5) and the nominal acceptance (scale parameter equal to 1). 
When the renormalization scale is varied, the factorization scale is set to 1 and vice versa.
These additional uncertainties are also correlated across all decay channels.
In the case of PDF variations, the impact on the acceptance is negligible and no additional nuisance parameters are introduced.

In the interpretation using the $\kappa$-framework, theoretical uncertainties are implemented following the procedure described in Ref.~\cite{tk_paper}.
Since only the ggH contribution is parametrized, the other contributions are set to their SM predictions. A 2.1\% uncertainty, determined in Ref.~\cite{CERN_Report_4}, is applied to all contributions other than ggH. 

\section{Combination of differential spectra and total cross section measurement}
\label{sec:combination}

In this section we present the combined unfolded differential cross section measurements for the observables \pth, \ptjz, \njets, \yh, \deta, \mjj, and \taujc.
The differential cross section measurements are performed by assigning a parameter of interest $\mu$ to scale predictions for each generator-level bin, as discussed in Section~\ref{sec:statistical_analysis}.
The difference in generator-level binning and, therefore, the difference in the number of parameters of interest across the channels entering the combination, are accounted for with the following procedure. 
First, a set of bins in which measurements are provided is chosen. 
The binning of the \hgamgam\ analysis is employed for this (see Tables~\ref{tab:ptbins}--\ref{tab:taujc}), as this provides better sensitivity to various regions of the differential phase space.
Then, the contributions of processes that have a coarser binning at the generator level are rescaled with a linear combination of the finer parameters of interest contained in the coarser bin. 
The weights used in this rescaling are the ratio of the SM cross section in each bin to the sum over bins.
As an example, one can consider the bins at very low \pth: the choice of parameters of interest is ($\mu_{0-5}$, $\mu_{5-10}$), as in the \hgamgam\ measurement.
In the \hzz analysis, only one generator-level bin is used in the range 0--10\GeV. A weighted sum of the chosen parameters of interest is needed to scale this bin in the \hzz analysis:
\begin{equation}
    \mu^{ZZ}_{0-10} = w_{0-5}\mu_{0-5} + w_{5-10}\mu_{5-10}.
    \label{eq:weights_spectra}
\end{equation}
The weights do not include the contributions of efficiency and acceptance, which are assumed not to vary within the bin.
Tests have been performed to assess the validity of this assumption.
In particular, the deviation from a constant acceptance has been tested in the 0--45\GeV bin of the \pth spectrum in the \htt\ analysis, which includes nine generator-level bins. For all the bins, the deviation from a constant acceptance is less than 2\%.

\begin{figure}[!ht]
    \centering
    \includegraphics[width=0.48\textwidth]{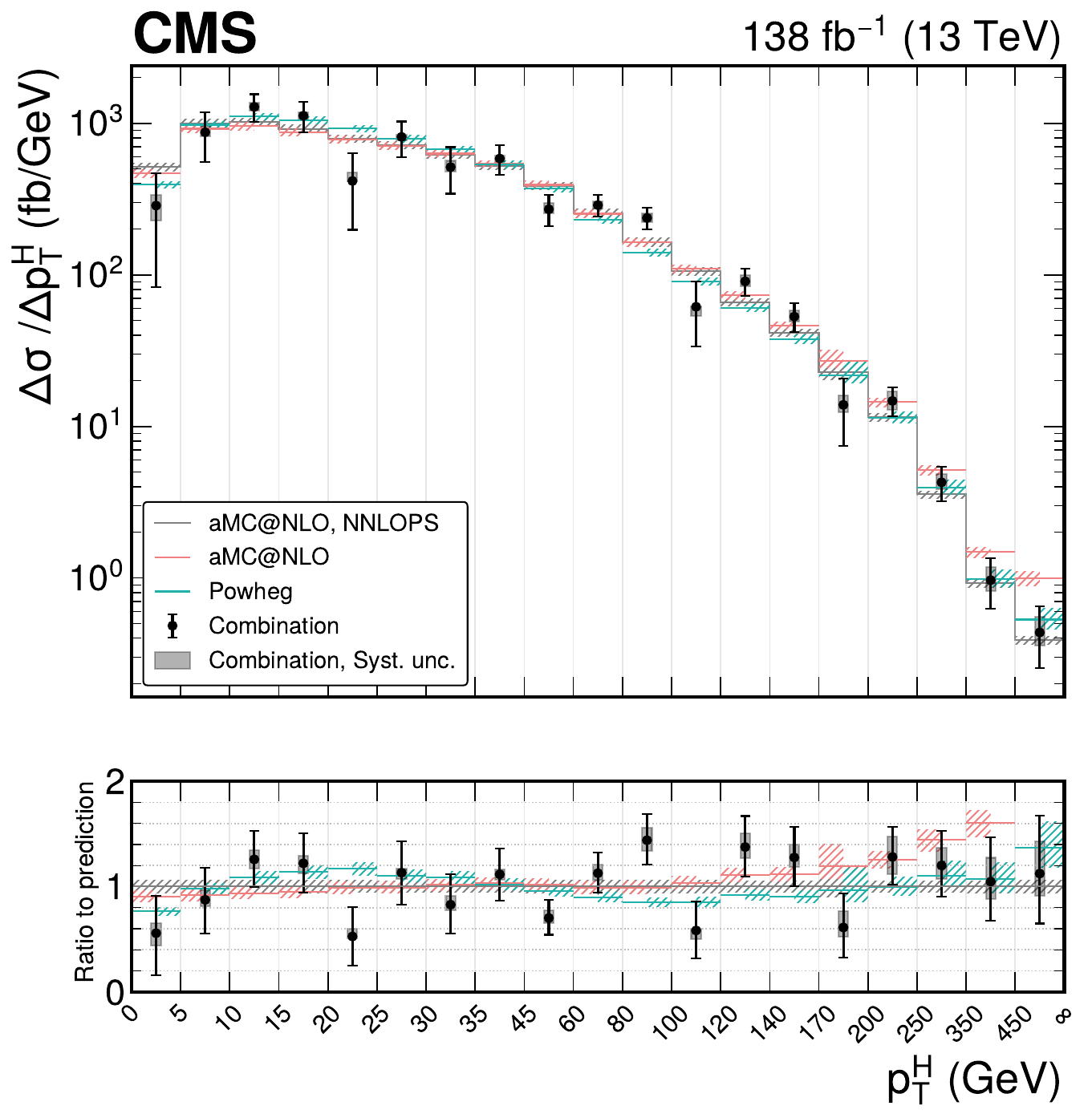}
    \includegraphics[width=0.48\textwidth]{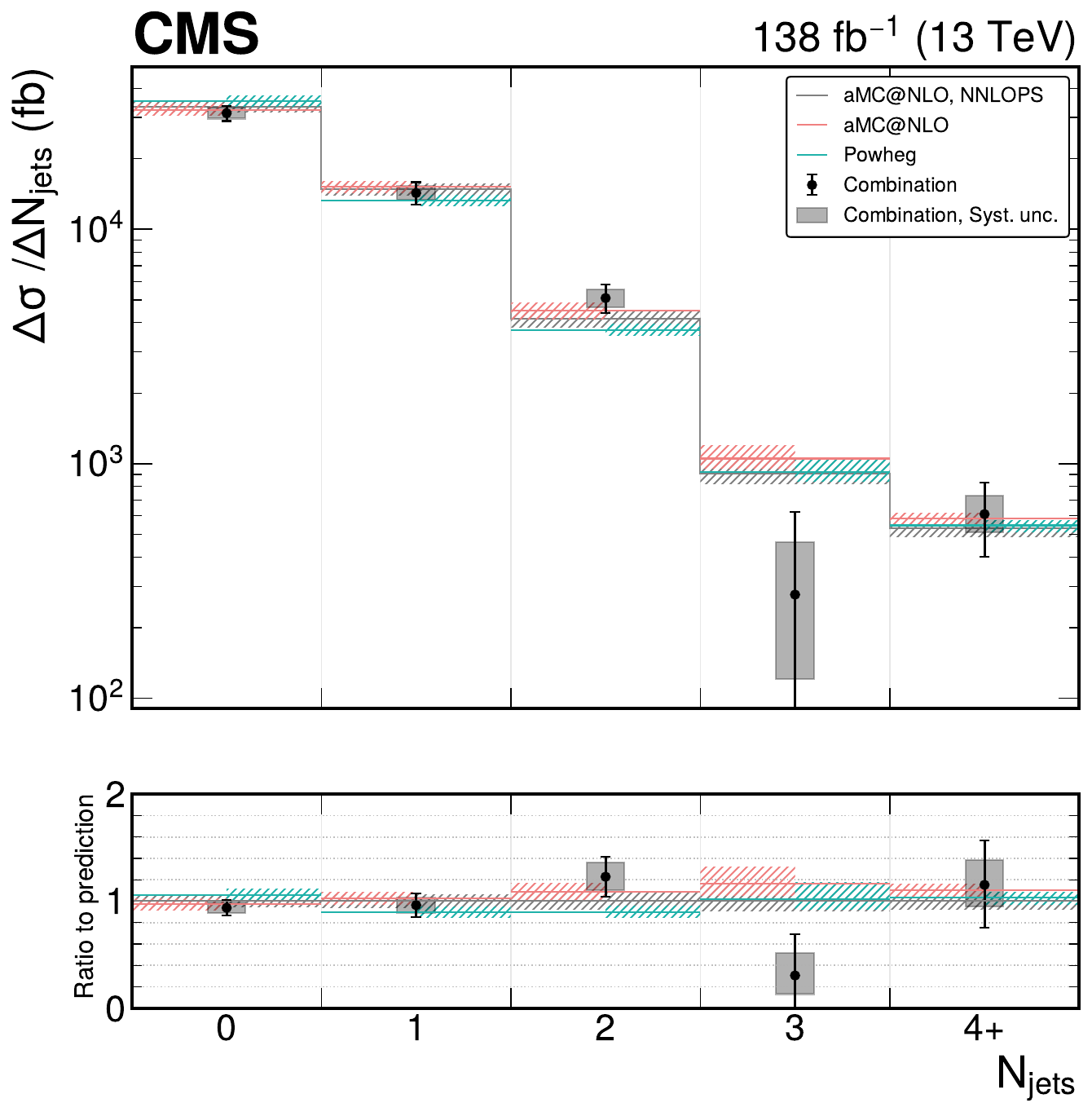} \\
    \includegraphics[width=0.48\textwidth]{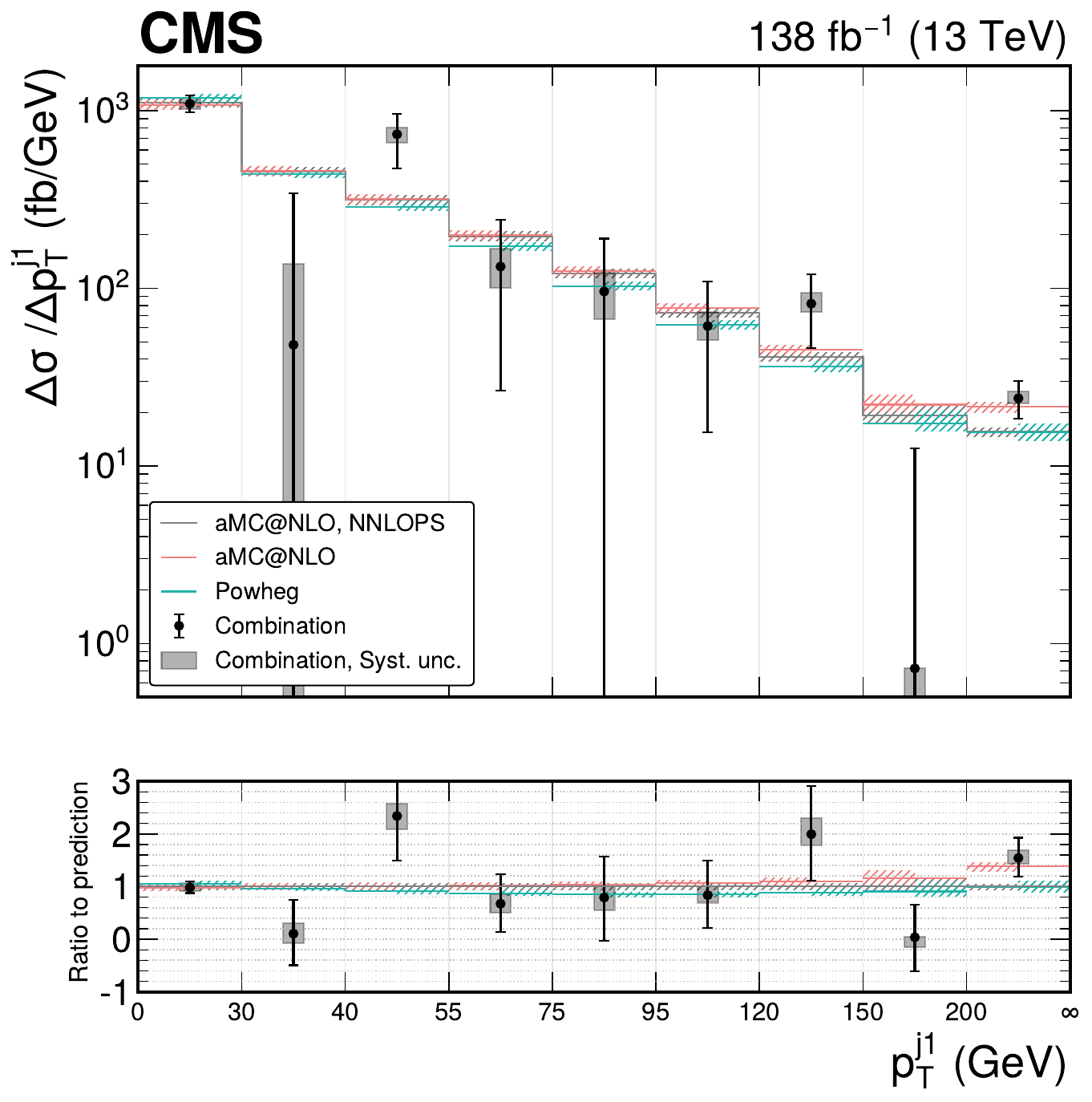}
    \includegraphics[width=0.48\textwidth]{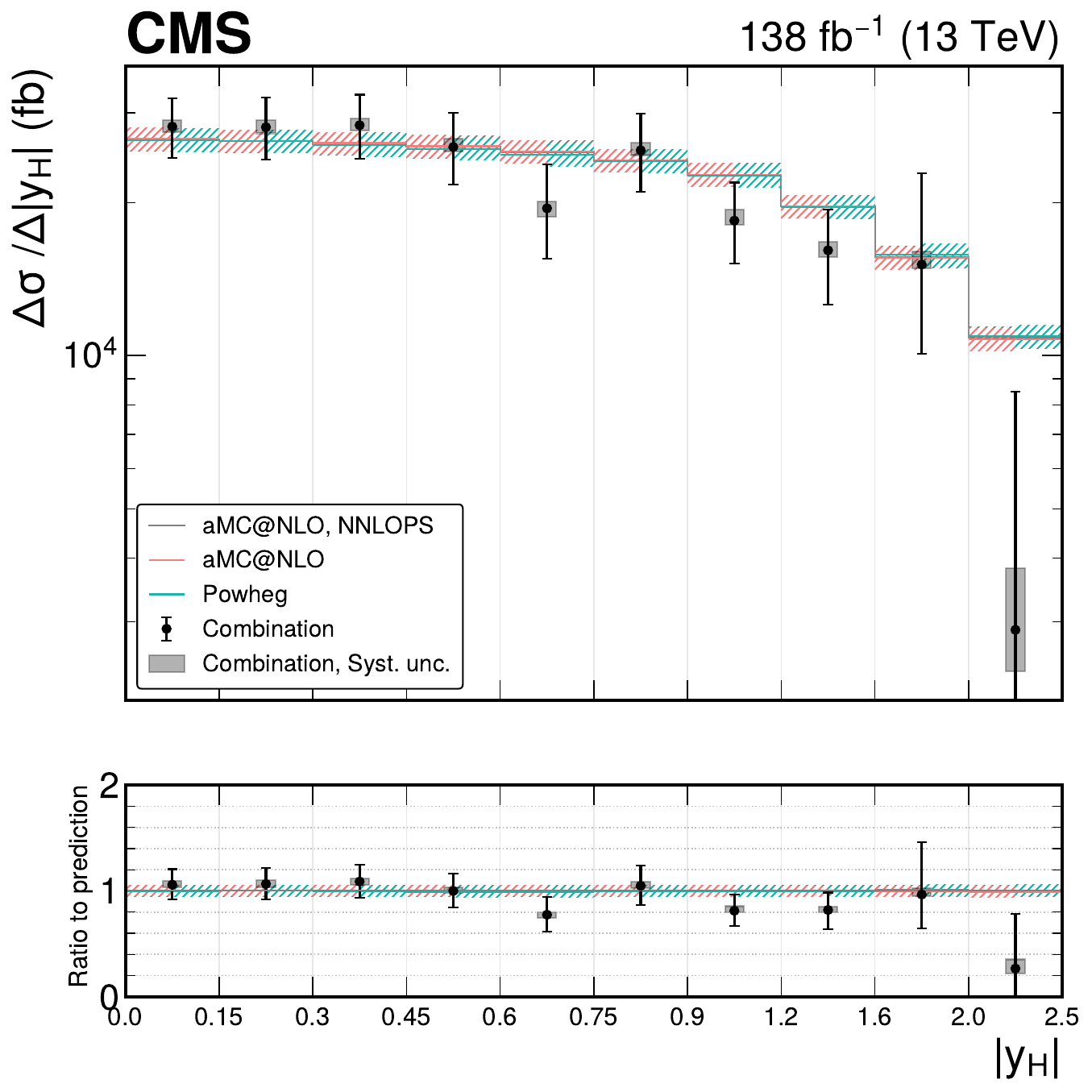}
    \caption{Measurement of the total differential cross section as a function of \pth (upper left), \njets (upper right), \ptjz (lower left), and \yh (lower right). For \ptjz, the first bin comprises all events with less than one jet, for which \ptjz is undefined. The combined spectrum is shown as black points with error bars indicating the $68\%$ confidence interval. The systematic component of the uncertainty is shown in gray. The SM prediction is reported for different generators. The hatched areas show the uncertainties in the shapes of the theoretical predictions from varying the PDFs, the value of $\alpha_{\mathrm{S}}$, and the renormalization and factorization scales. The uncertainty on the total cross section value, taken from Ref.~\cite{CERN_Report_4}, is also included in the hatched areas by summing it in quadrature with the shape uncertainties. In the case of \pth and \ptjz, the rightmost bins of the distributions are overflow bins, and their width is assumed equal to the last but one bin when computing the differential cross section. In cases where the systematic uncertainty band covers only one side of the data point, the systematic uncertainty on the other side is negligible. The ratio between the measurements and the \MGvATNLO NNLOPS-reweighted SM predictions is shown in the lower panel of each plot.}
    \label{fig:spectra_1}
\end{figure}

\begin{figure}[!ht]
    \centering
    \includegraphics[width=0.48\textwidth]{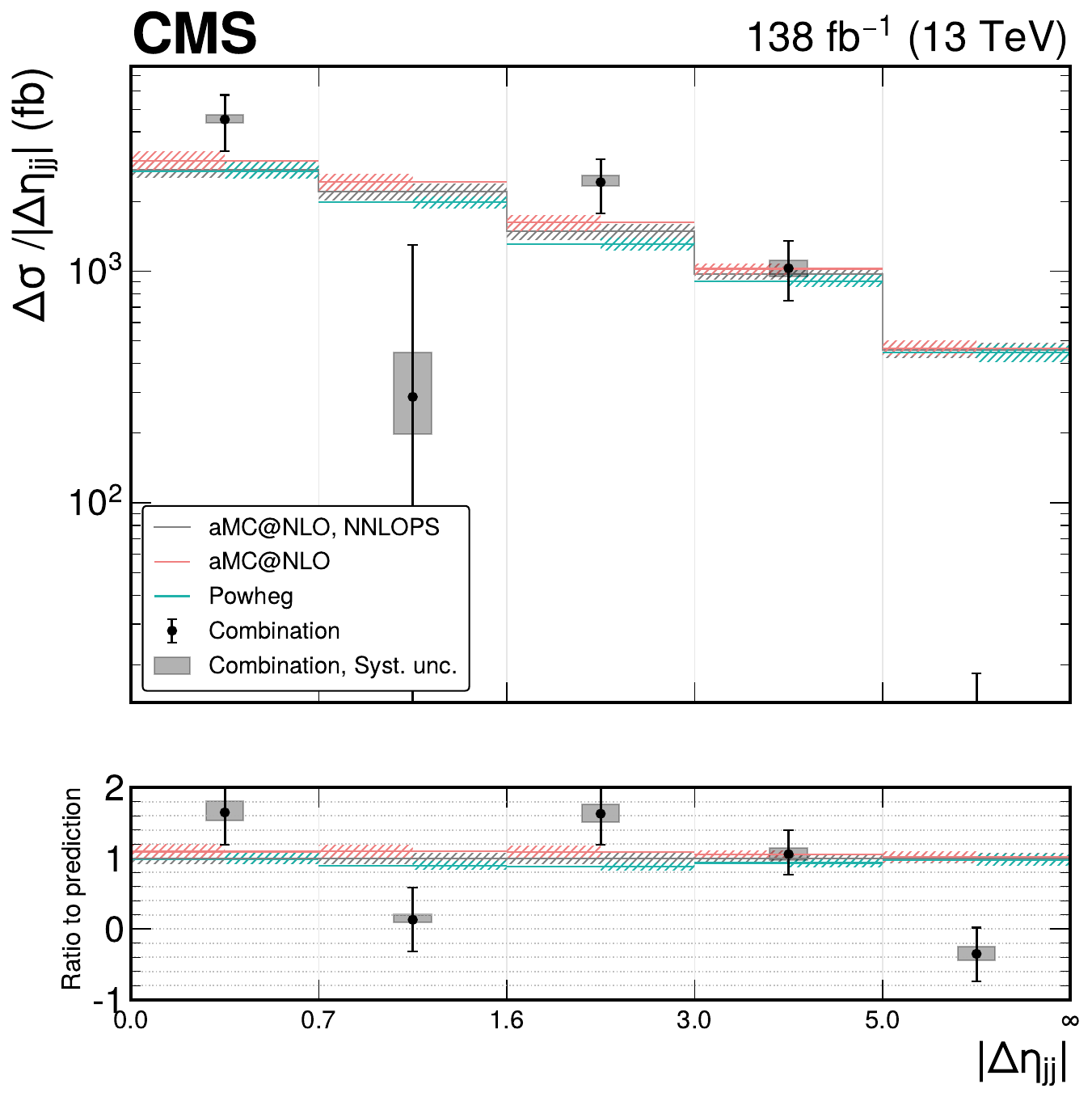}
    \includegraphics[width=0.48\textwidth]{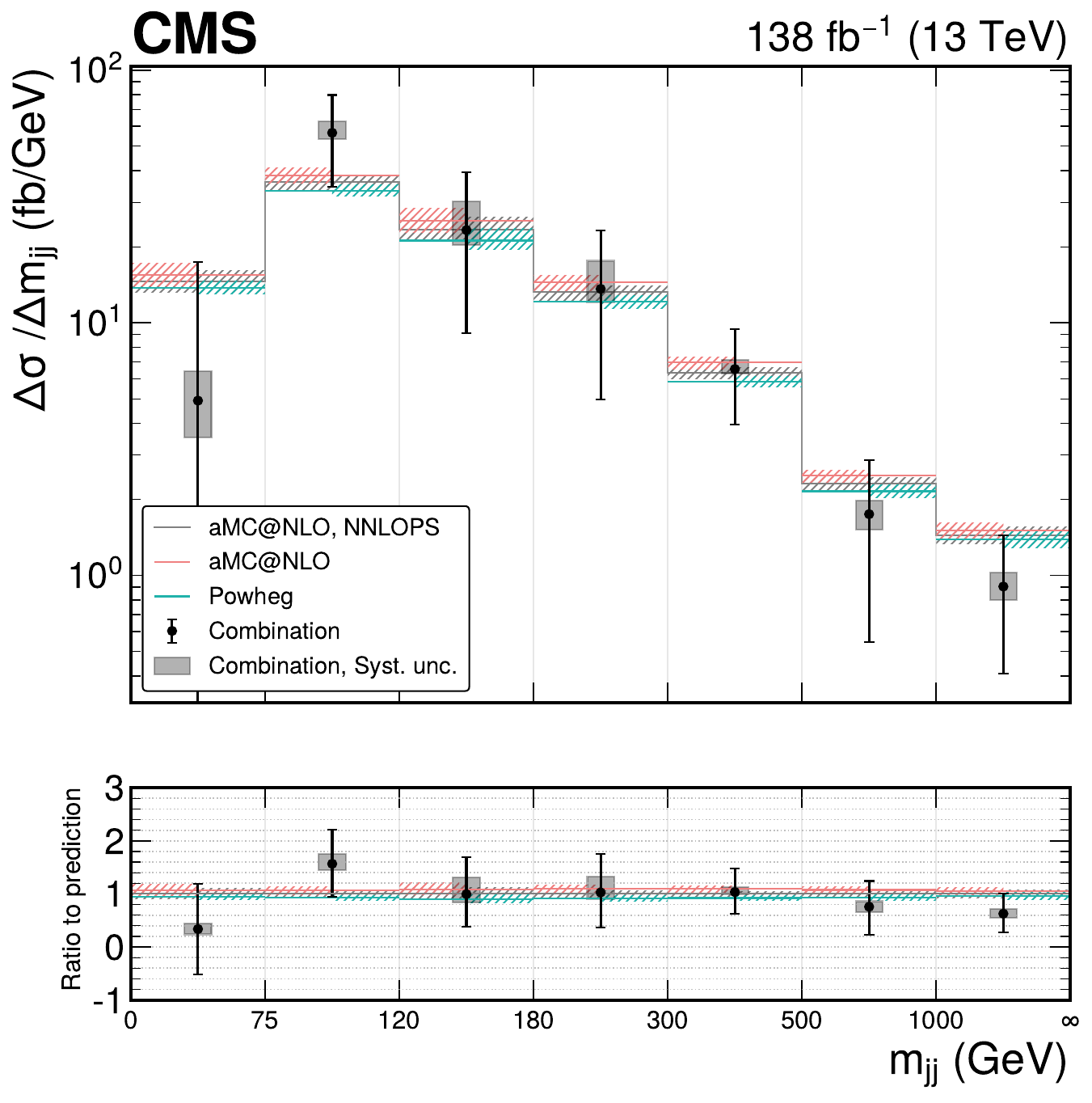} \\
    \includegraphics[width=0.48\textwidth]{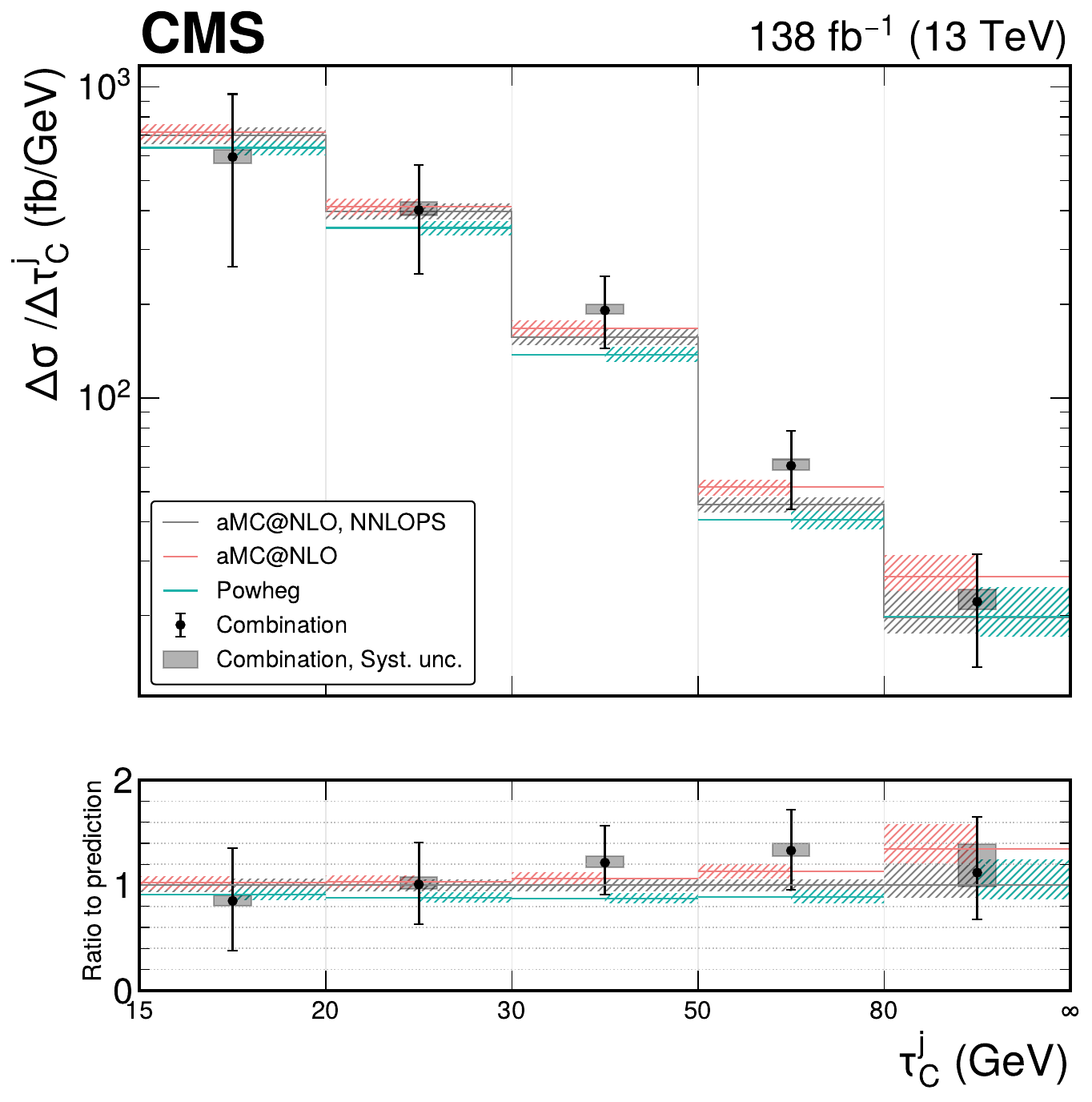}
    \caption{Measurement of the total differential cross section as a function of \deta (upper left), \mjj (upper right), and \taujc (lower). The combined spectrum is shown as black points with error bars indicating the $68\%$ confidence interval. The systematic component of the uncertainty is shown in gray. The SM prediction is reported for different generators. The hatched areas show the uncertainties in the shape of the theoretical predictions from varying the PDFs, the value of $\alpha_{\mathrm{S}}$, and the renormalization and factorization scales. The uncertainty on the total cross section value, taken from Ref.~\cite{CERN_Report_4}, is also included in the hatched areas by summing it in quadrature with the shape uncertainties. The rightmost bins of the distributions are overflow bins, and their width is assumed equal to the last but one bin when computing the differential cross section. The ratio between the measurements and the \MGvATNLO NNLOPS-reweighted SM predictions is shown in the lower panel of each plot.}
    \label{fig:spectra_2}
\end{figure}

The unfolded differential cross sections for the observables \pth, \ptjz, \njets, \yh, \deta, \mjj, and \taujc are shown in Figs.~\ref{fig:spectra_1} and \ref{fig:spectra_2}.
Numerical results are given in Tables~\ref{tab:observed_smH_PTH_xsvalues}--\ref{tab:observed_TauCJ_xsvalues}, available in Appendix~\ref{app:tables_spectra}.
The correlation matrices for the unfolded differential cross sections are shown in Figs.~\ref{fig:corr_matrices_1} and \ref{fig:corr_matrices_2}, Appendix~\ref{app:correlation_matrices}.
Each figure compares the measurement with three theoretical predictions.
The predicted cross section for the ggH production mode is taken from the \MGvATNLO (version 2.6.5) simulation, generated at NLO accuracy, and the \POWHEG (version 2) event generator~\cite{POWHEG1,POWHEG2,POWHEG3,POWHEG:ggH}.
The events produced with \MGvATNLO are weighted as a function of the Higgs boson \pt\ and the number of jets in the event, to match the prediction from the next-to-NLO including parton showering event generator (NNLOPS) program~\cite{Hamilton:2013fea,Hamilton:2012np,Kardos:2014dua}.
The sum of the production cross sections of the other (non-ggH) production modes is taken from the \MGvATNLO (version 2.6.5) simulation; this non-ggH prediction is common to the different SM calculations that are shown.
The uncertainty in the theoretical predictions takes into account variations in the predicted differential cross section spectra from varying the set of PDF replicas, the renormalization and factorization scales, and $\alpha_{\mathrm{S}}$.
The uncertainty on the total cross section value, taken from Ref.~\cite{CERN_Report_4}, is included by summing it in quadrature with the shape uncertainties.
Overall, no significant deviations from the SM predictions are observed.
For all the measurements, nuisance parameters are introduced to account for the scale variations in the acceptance for the extrapolation to the full phase space, as described in Section~\ref{sec:systematics}.
Statistical uncertainties form the dominant source of uncertainty at low \pt in the \pth measurement, while at high \pt the statistical and systematic uncertainties are comparable.
In the case of the \njets measurement, systematic uncertainties dominate, while for the other spectra the statistical uncertainties are the most important.
It should be noted that the differential cross section measurements are not defined to be positive in every bin, and negative values can occur as a result of small event samples and possible anticorrelations between bins.

\begin{figure}[!htb]
    \centering
    \includegraphics[width=0.48\textwidth]{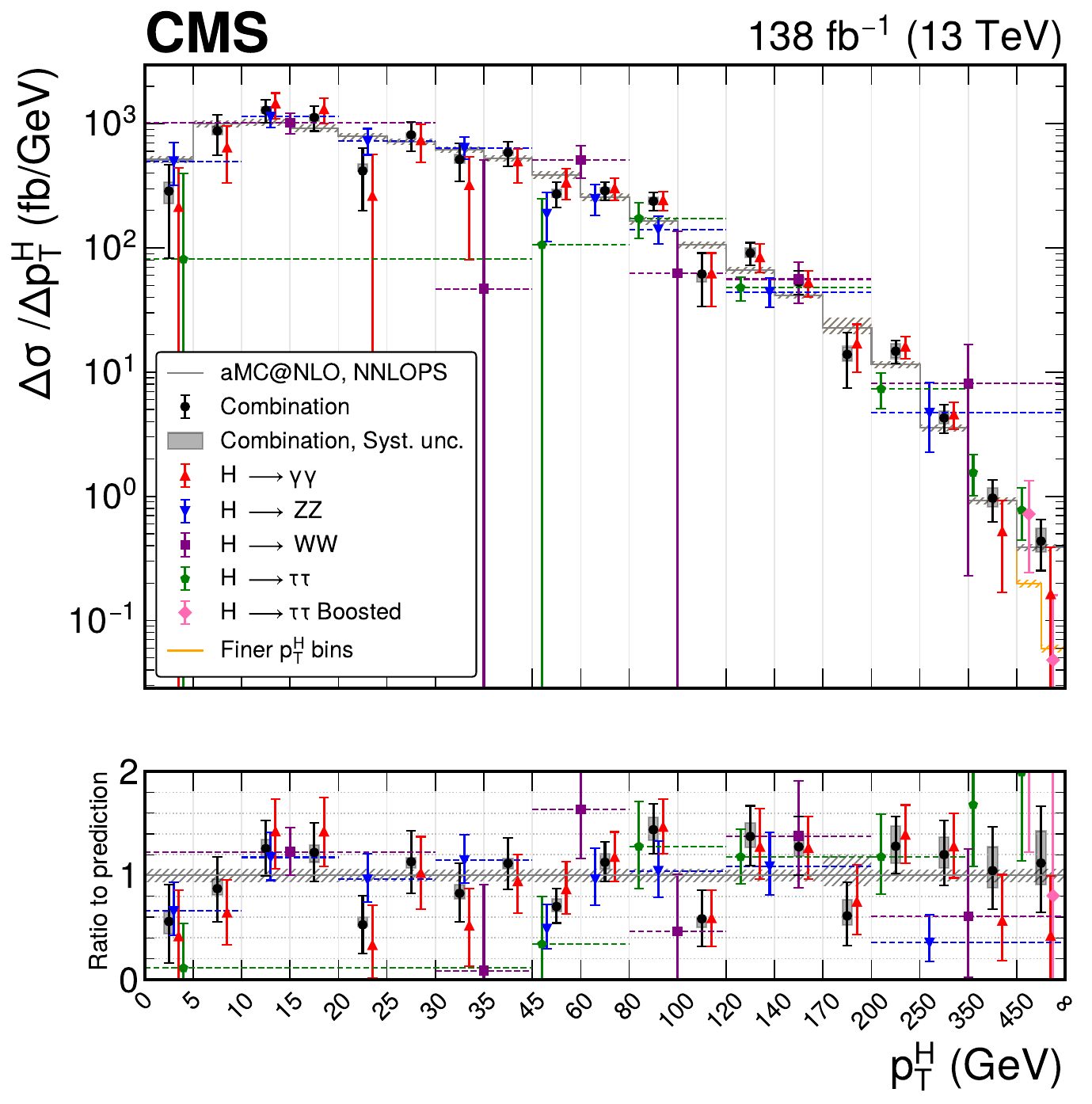}
    \includegraphics[width=0.48\textwidth]{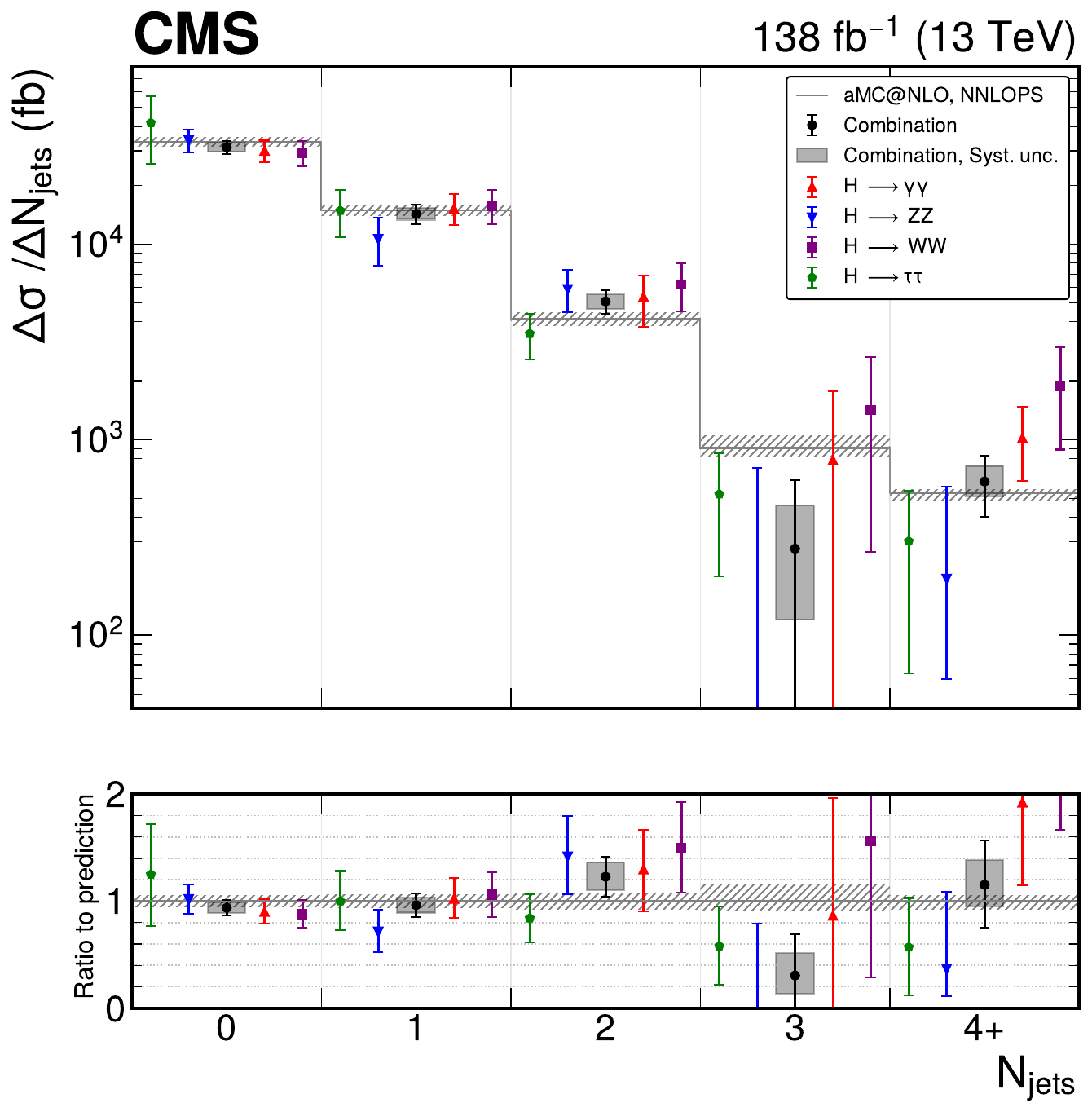} \\
    \includegraphics[width=0.48\textwidth]{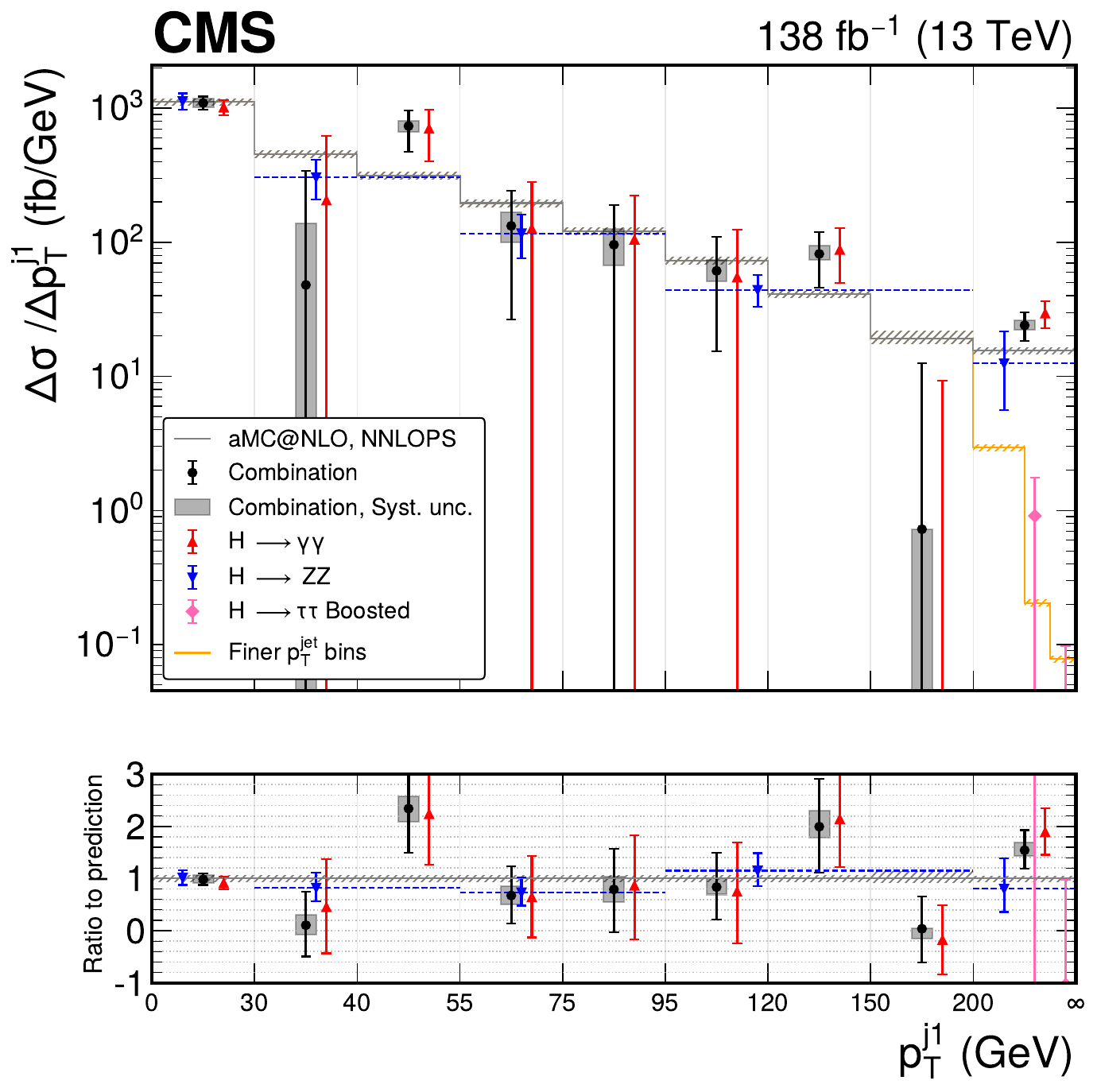}
    \includegraphics[width=0.48\textwidth]{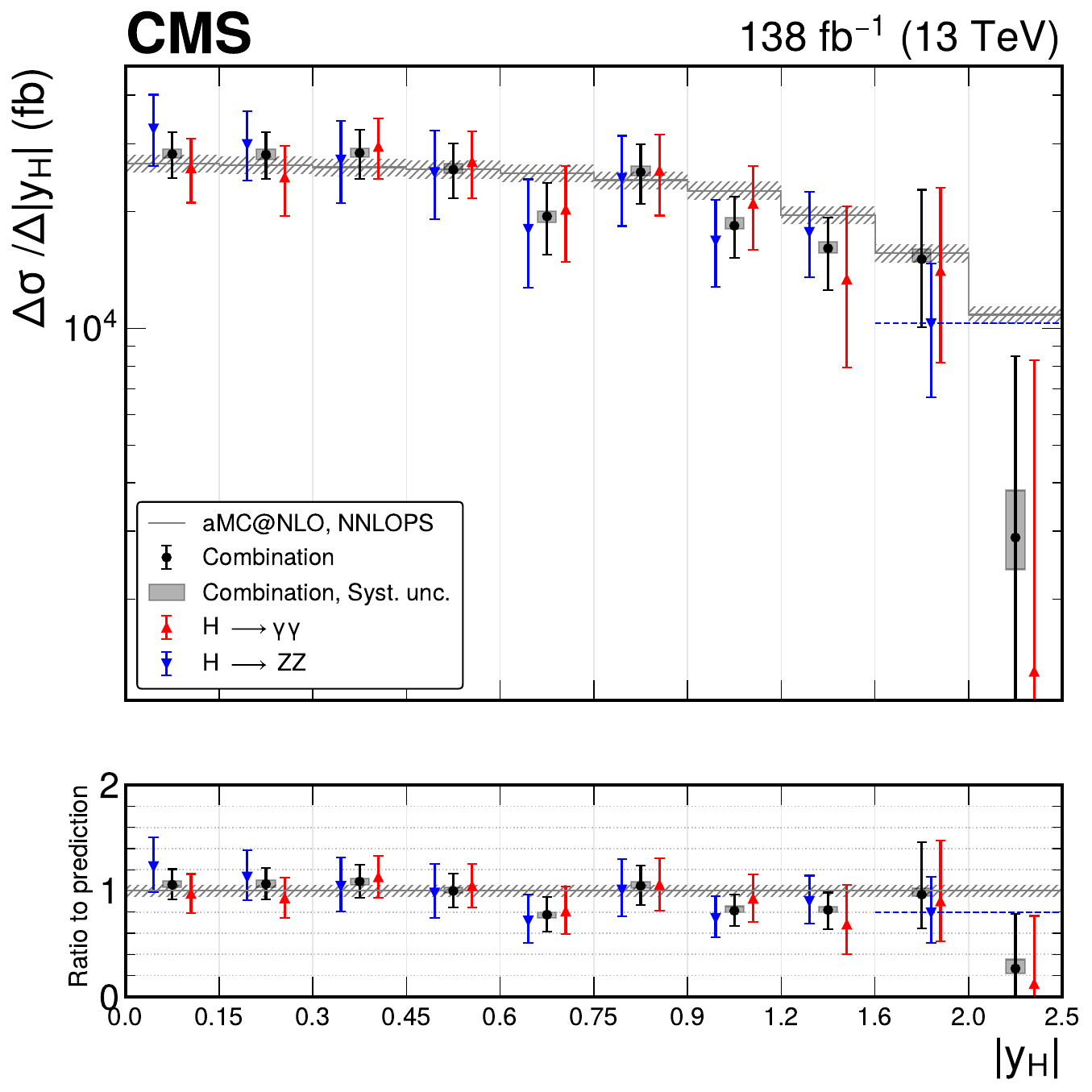}
    \caption{Measurement of the total differential cross section as a function of \pth (upper left), \njets (upper right), \ptjz (lower left), and \yh (lower right). The combined spectrum is shown in black points with error bars indicating the $68\%$ interval. The systematic component of the uncertainty is shown in gray. The spectra for the analyses in \hgamgam, \hzz, \hww, \htt, and \htt boosted are shown in red, blue, purple, green, and pink respectively. The SM prediction is reported in light gray for \MGvATNLO NNLOPS. In the case of \pth and \ptjz, the rightmost bins of the distributions are overflow bins, and their width is assumed equal to the last but one bin when computing the differential cross section. Measurements or predictions with different binnings can be directly compared only in the ratio panels of the figures. In cases where individual contributions have finer bins than the combination, such as the last bin of the \pth and \ptjz\ spectra, a second, finer, SM prediction is shown in the upper panel.}
    \label{fig:spectra_all_1}
\end{figure}

\begin{figure}[!htb]
    \centering
    \includegraphics[width=0.48\textwidth]{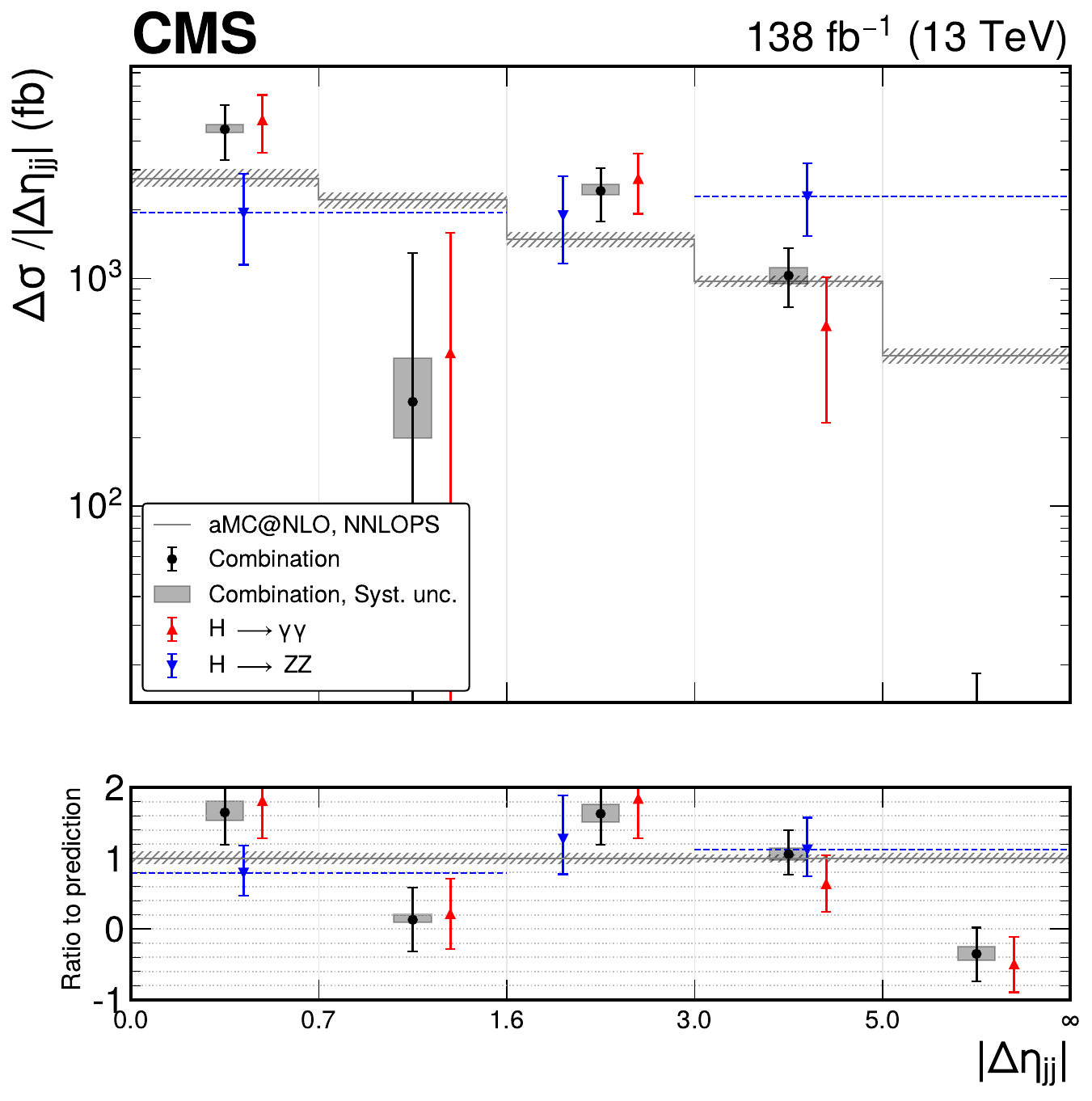}
    \includegraphics[width=0.48\textwidth]{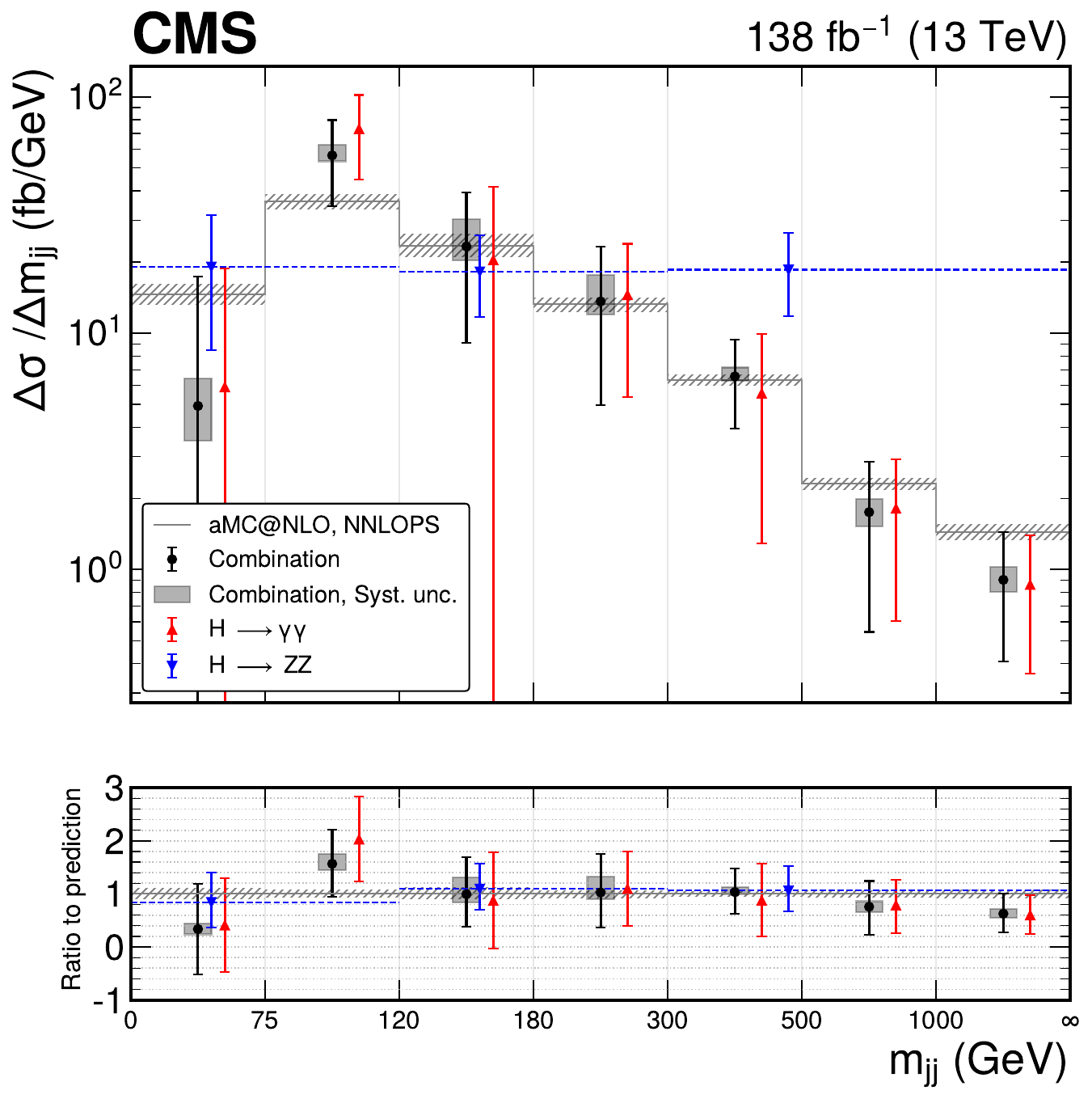} \\
    \includegraphics[width=0.48\textwidth]{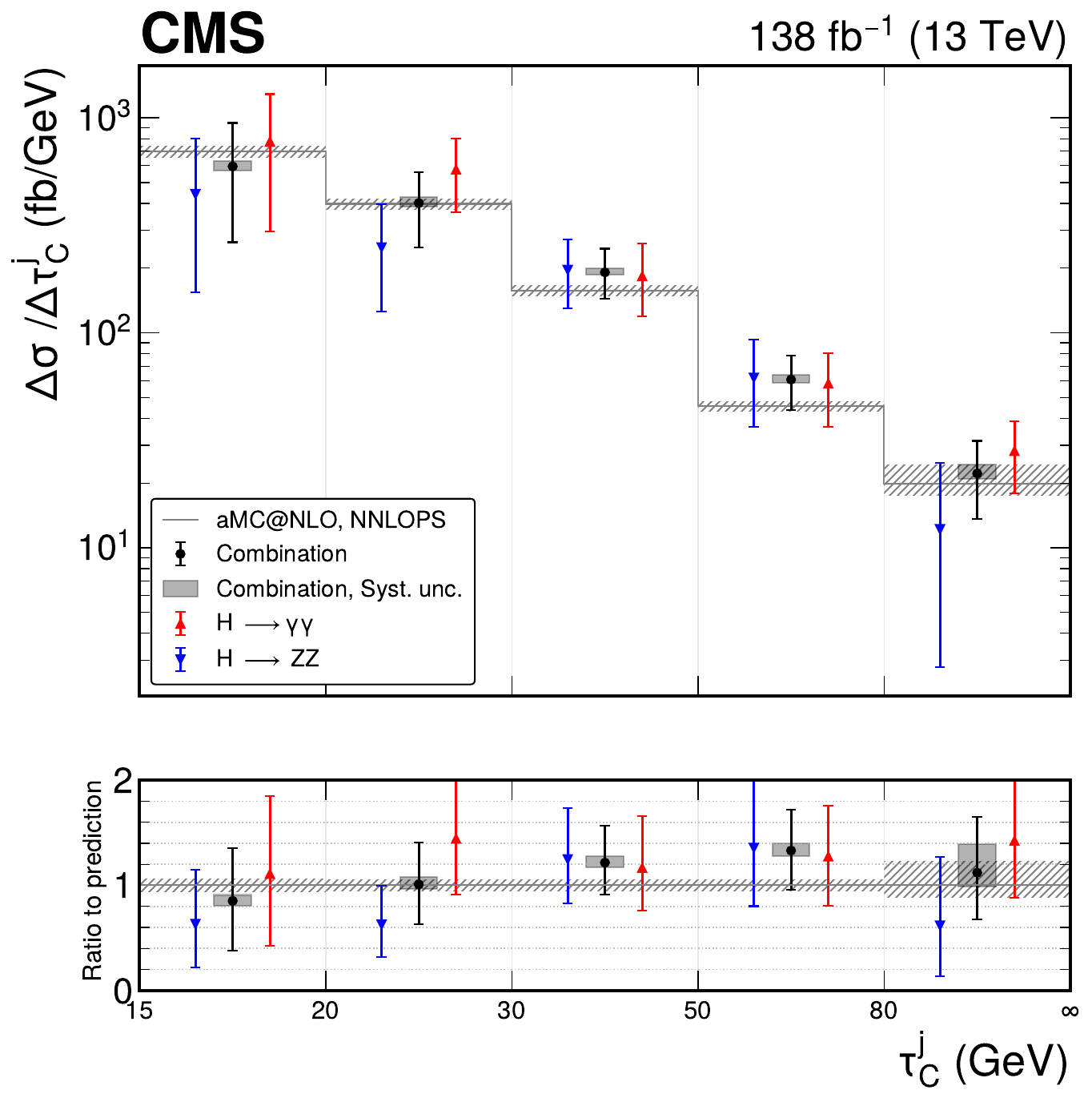}
    \caption{Measurement of the total differential cross section as a function of \deta (upper left), \mjj (upper right), and \taujc (lower). The combined spectrum is shown in black points with error bars indicating the $68\%$ interval. The systematic component of the uncertainty is shown in gray. The spectra for the analyses in \hgamgam and \hzz are shown in red and blue, respectively. The SM prediction is reported in light gray for \MGvATNLO NNLOPS. The rightmost bins of the distributions are overflow bins, and their width is assumed equal to the last but one bin when computing the differential cross section. Measurements or predictions with different binnings can be directly compared only in the ratio panels of the figures.}
    \label{fig:spectra_all_2}
\end{figure}

The same spectra, along with the measurements from the individual channels, are shown in Figs.~\ref{fig:spectra_all_1} and \ref{fig:spectra_all_2}.
In the case of the \pth spectrum~(Fig.~\ref{fig:spectra_all_1}, upper left), the sensitivity is driven by the \hgamgam\ and \hzz\ analyses.
A comparison with the \hgamgam measurement alone shows that the decrease in uncertainty achieved by the combination is most notable in the very low and very high \pth\ regions, with an average reduction of 23\%.
In four out of the nineteen bins, the relative uncertainty does not decrease with respect to the \hgamgam measurement alone.

\begin{figure}[!htb]
    \centering
    \includegraphics[width=0.7\textwidth]{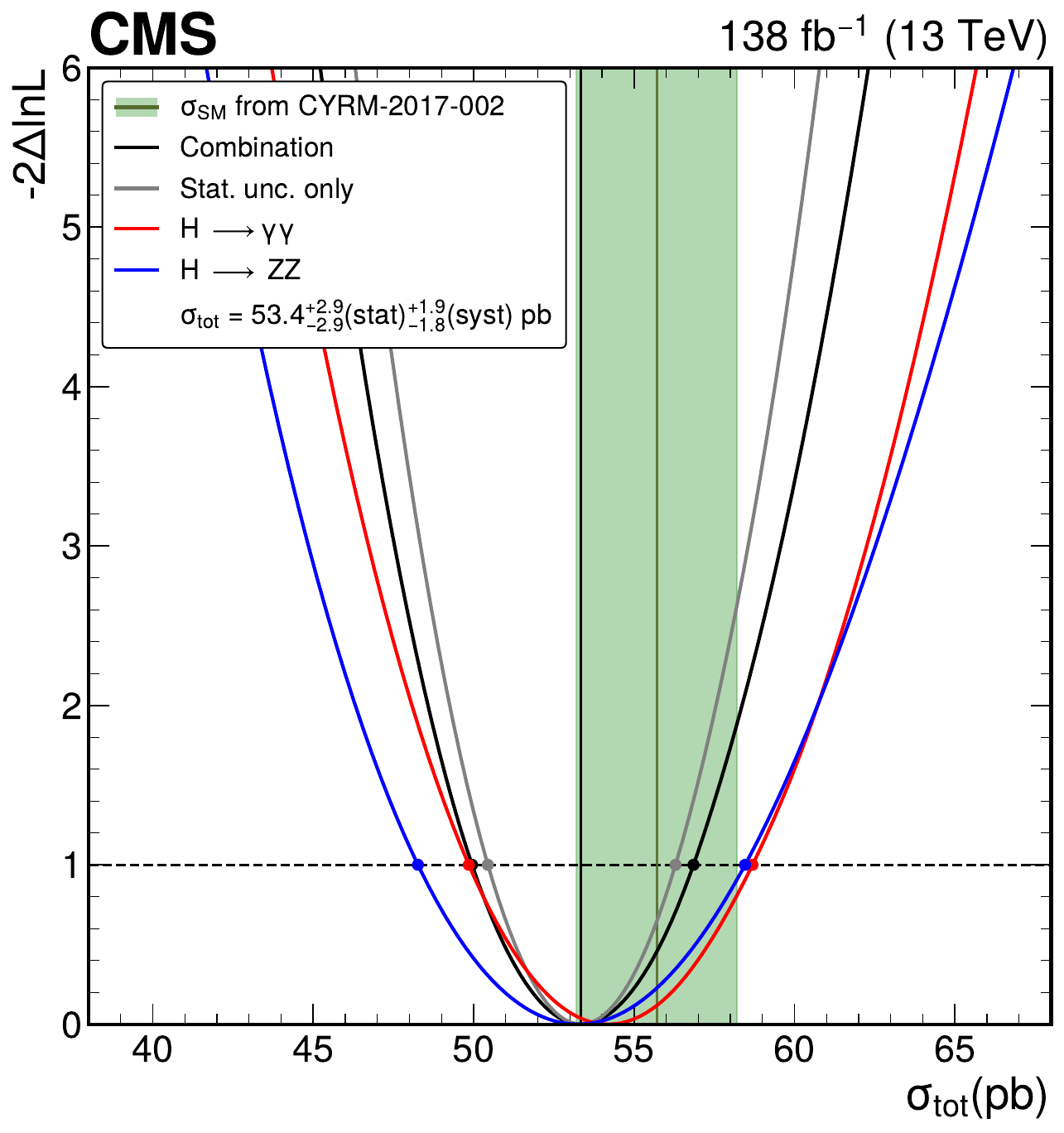}
    \caption{Negative log-likelihood scan of the total Higgs boson production cross section $\sigma_{\mathrm{tot}}$ for the \hgamgam, \hzz, and combined analyses. The markers indicate the 68\% confidence interval. The label \textit{CYRM-2017-002} in the legend denotes Ref.~\cite{CERN_Report_4}.}
    \label{fig:inclusive}
\end{figure}

The total cross section for Higgs boson production, based on a combination of the \hgamgam\ and \hzz\ channels, is measured to be $53.4^{+3.5}_{-3.4}$\unit{pb}, obtained by applying the statistical treatment described in Section~\ref{sec:statistical_analysis} (\ie, with a single bin, both at generator and reconstruction levels).
The measured total cross sections from the individual channels are $54.2^{+4.5}_{-4.3}$\unit{pb} for \hgamgam\ and $53.3^{+5.2}_{-5.0}$\unit{pb} for \hzz; the combination thus improves the precision by 21\% with respect to the \hgamgam\ channel alone.
The likelihood scans for the individual decay channels and their combination are shown in Fig.~\ref{fig:inclusive}.
The combined result agrees with the SM value of $55.6 \pm 2.5$\unit{pb} \cite{CERN_Report_4}.

\clearpage
\section{\texorpdfstring{$\kappa$}{kappa}-framework interpretation}
\label{sec:kappa}
Differential cross section measurements can be used to constrain the couplings of the Higgs boson to other particles.
For Higgs boson production via ggH, variations of the Higgs boson couplings mostly manifest themselves through distortions of the \pth spectrum.
The $\kappa$-framework has been developed~\cite{higgs_handbook} to study the coupling structure of the Higgs boson.
Following the procedure described in Ref.~\cite{tk_paper}, two models are used to interpret the \pth spectrum for ggH: one, referred to as \kappab--\kappac \cite{kbkc}, which takes into account the effects of heavy quarks in the ggH loop, and one, referred to as \kappab--\kappat--\cg \cite{grazzini_1, grazzini_2}, tailored to top and bottom quarks and the effective Higgs boson coupling to gluons, sensitive to effects at high \pth.
The coupling modifiers are defined as:
\begin{equation}
    \kappa_i = \frac{y_i}{y_i^{_{\mathrm{SM}}}},
\end{equation}
where $y_i$ is the Higgs boson coupling to particle $i$. In the SM, the values of all $\kappa_i$ are equal to 1.

\begin{figure}[!ht]
    \centering
    \includegraphics[width=0.48\textwidth]{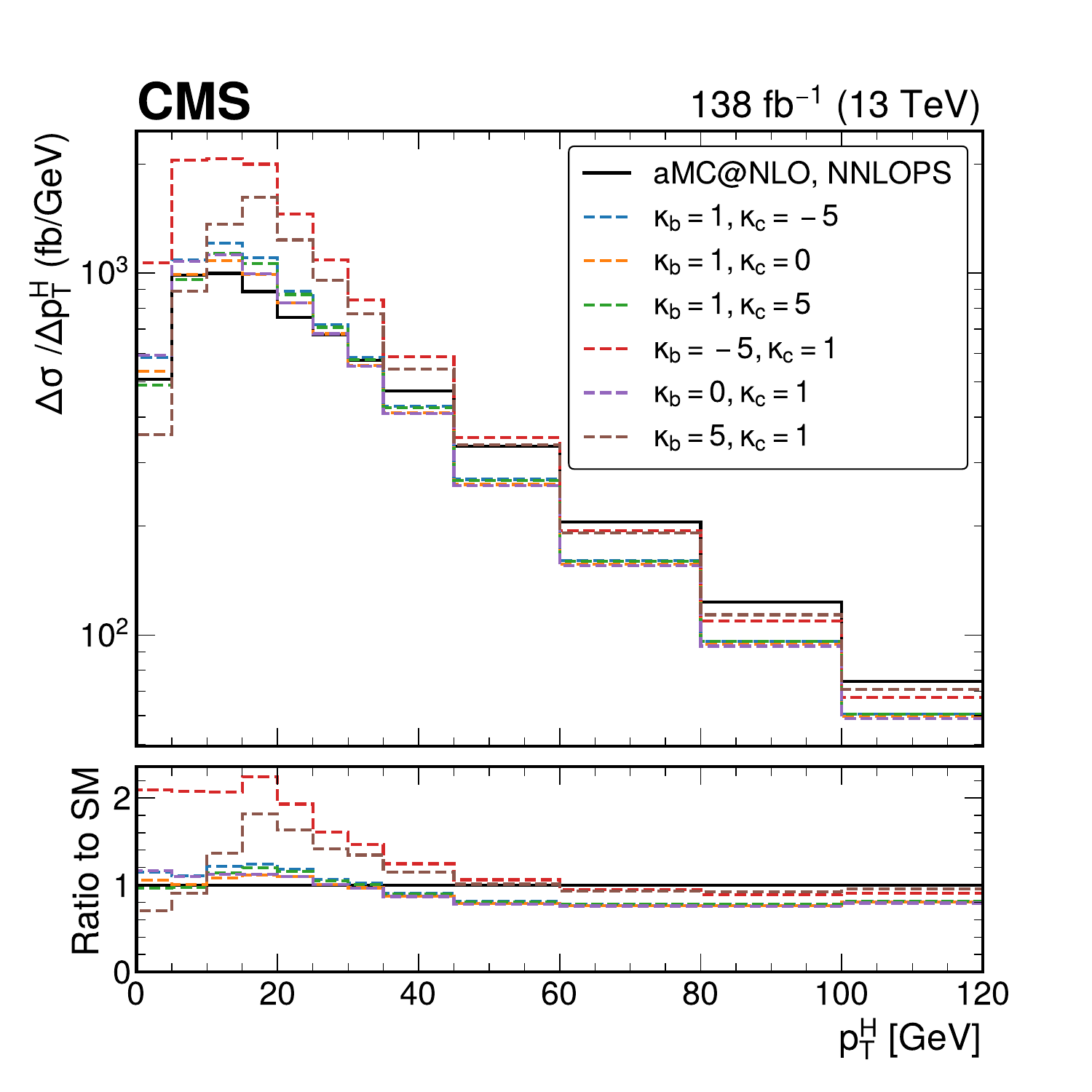}
    \caption{The \pth spectrum, for different values of \kappab and \kappac, as predicted by the \kappab--\kappac model. The \MGvATNLO NNLOPS-reweighted SM prediction is shown in black. The \hgamgam\ binning is used.}
    \label{fig:kappabkappac_model}
\end{figure}

In the \kappab--\kappac model, only the differential ggH cross section is affected by variations of \kappab and \kappac: in this paper, the most recent parametrization~\cite{h_zz} is used.
These variations are parametrized using a quadratic polynomial for each bin of the differential production cross section.
It is important to note that since these parametrizations address the low range of the \pth spectrum, they are available only up to 120\GeV. The \htt\ boosted analysis is therefore not included in this interpretation.
Moreover, the \hww analysis is excluded from this interpretation, as for that analysis the signal predictions are only available inclusively and not separating the ggH contribution from the other production modes.
The \pth spectrum, for different values of \kappab and \kappac, as predicted by this model is shown in Fig.~\ref{fig:kappabkappac_model}.

\begin{figure}[!ht]
    \includegraphics[width=0.48\textwidth]{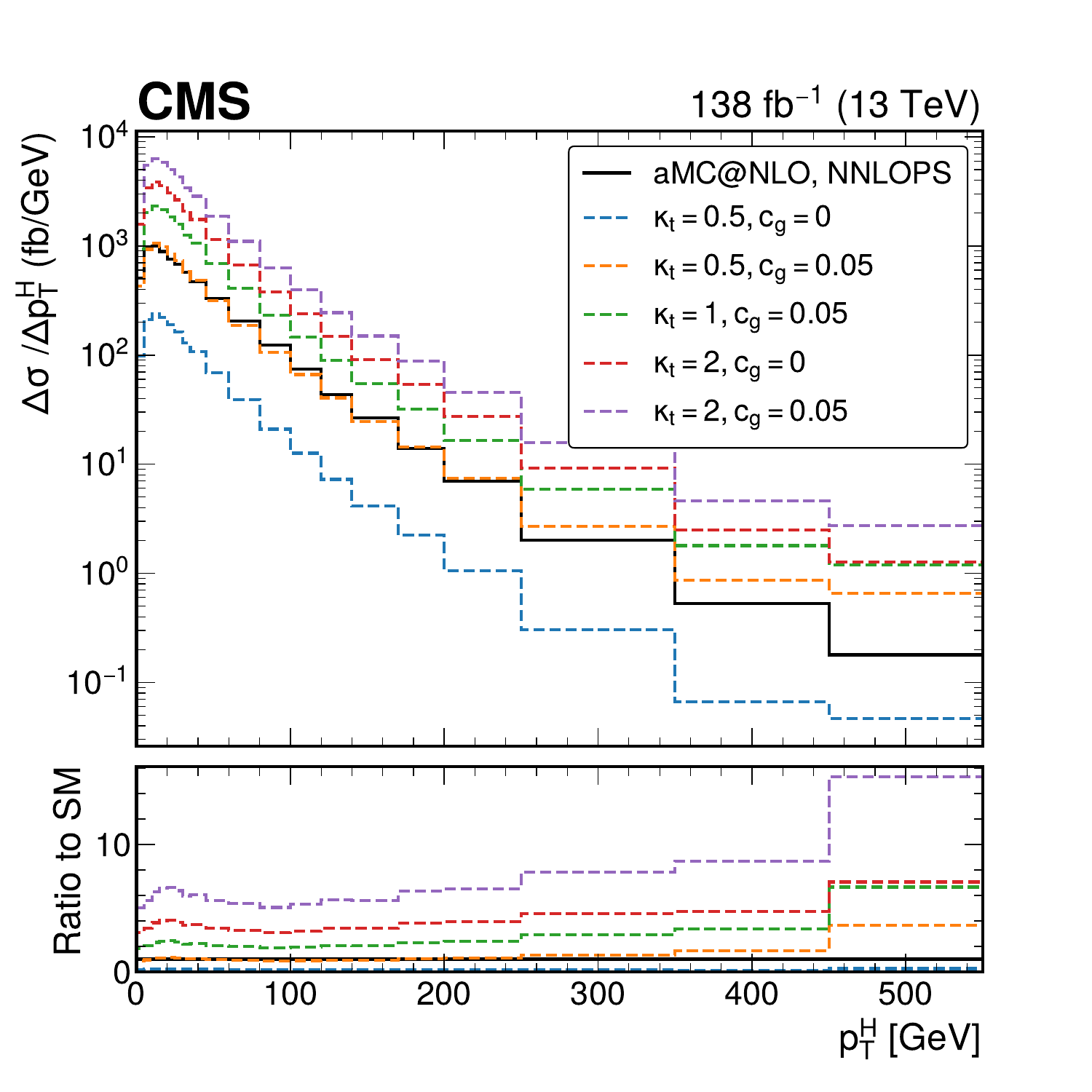}
    \includegraphics[width=0.48\textwidth]{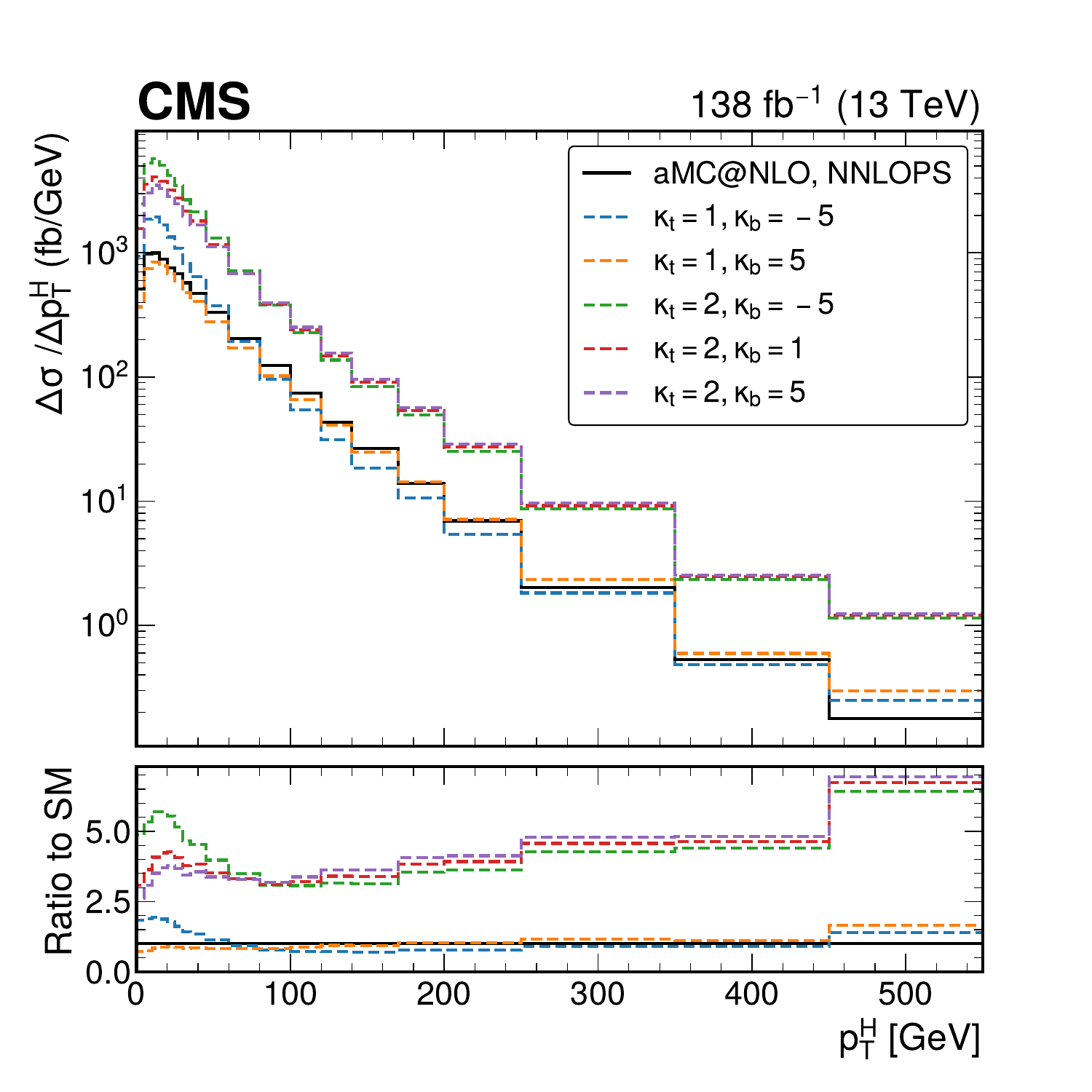}
    \caption{The \pth spectrum, for different values of \kappat and \cg (left) and of \kappat and \kappab (right), as predicted by the \kappab--\kappat--\cg model. The third coupling is set to its SM value. The \MGvATNLO NNLOPS-reweighted SM prediction is shown in black. The \hgamgam\ binning is used. The width of the overflow bin is assumed equal to the last but one bin when computing the differential cross section.}
    \label{fig:kappabkappatcg_model}
\end{figure}

The \kappab--\kappat--\cg model, which produces simultaneous variations of $\kappa_t$, $c_g$, and $\kappa_b$, has been derived in Ref.~\cite{grazzini_1} by adding dimension-6 operators to the SM Lagrangian.
The \pth spectrum is computed at next-to-next-to-leading order (NNLO) accuracy with an analytic resummation performed up to next-to-next-to-leading-logarithmic (NNLL) accuracy.
This is achieved by rescaling the SM prediction at NNLL+NNLO accuracy by a BSM correction factor computed as the ratio of the SMEFT to SM predictions at next-to-leading-logarithmic (NLL)+NLO accuracy; this approach allows bottom and top quark mass effects to be approximately included at NNLL+NNLO.
The dimension-6 operator corresponding to the coefficient \cg\ models a direct coupling of the Higgs field to the gluon field with the same underlying tensor structure as in the heavy top quark limit. The value of \cg\ equals 0 in the SM.
The introduction of \cg\ in the effective Lagrangian is detailed in Ref.~\cite{grazzini_2}.
The inclusive cross section is parametrized as $\sigma \simeq\abs{12 c_{\mathrm{g}}+\kappa_{\mathrm{t}}}^2 \sigma^{\mathrm{SM}}$.
Two other operators are included in the Lagrangian to describe modifications of the top quark and bottom quark Yukawa couplings, having coefficients \kappat and \kappab, respectively.
Simultaneous variations of \kappat\ and \cg\ and of \kappat\ and \kappab\ are considered.
In this result, all the decay channels apart from \hww are included. 
The \pth spectrum, for different values of \kappat and \cg (left) and of \kappat and \kappab (right), as predicted by this model is shown in Fig.~\ref{fig:kappabkappatcg_model}. In both cases, the third coupling is set to its SM value.

\begin{figure}[!ht]
    \includegraphics[width=0.45\textwidth]{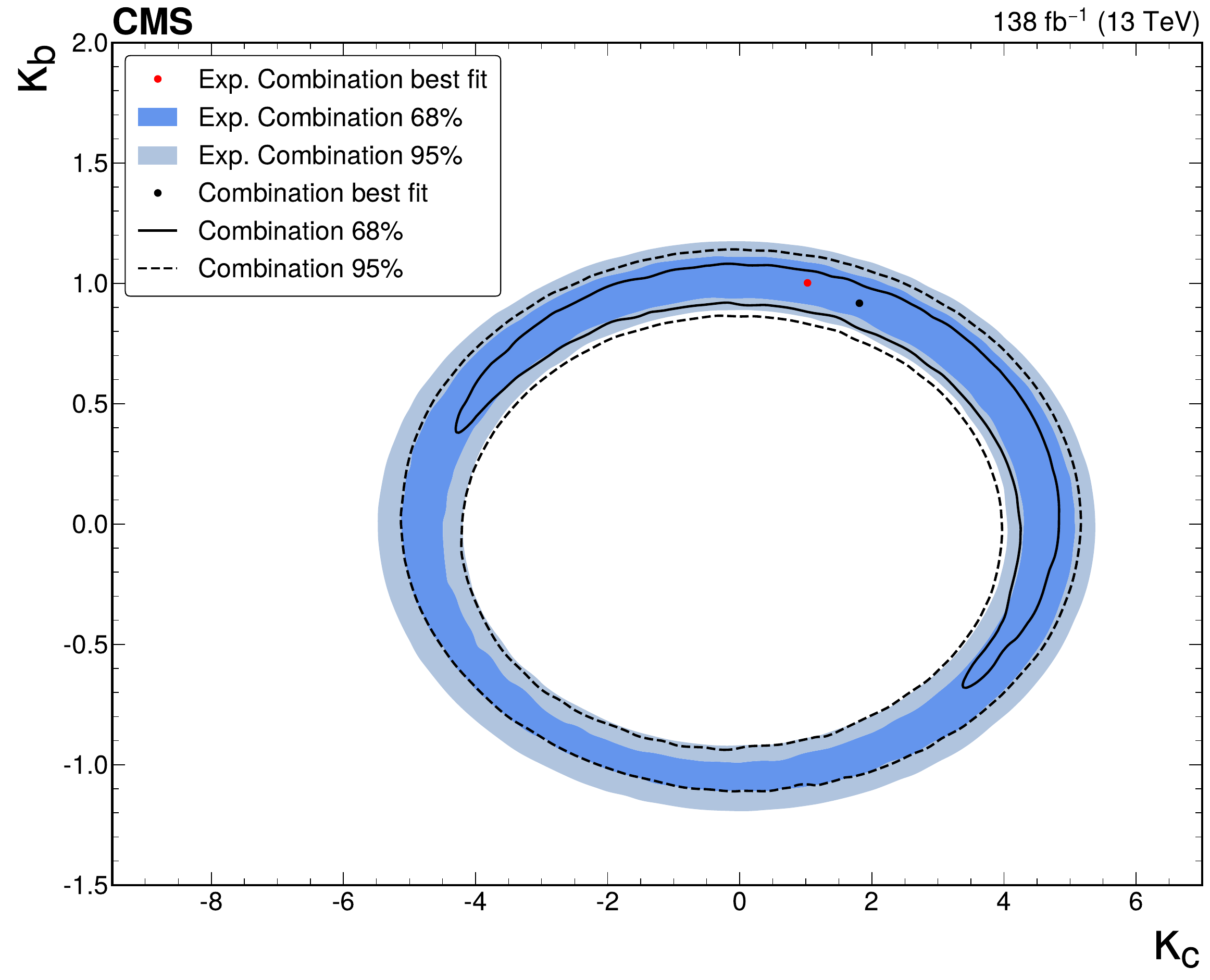}
    \includegraphics[width=0.45\textwidth]{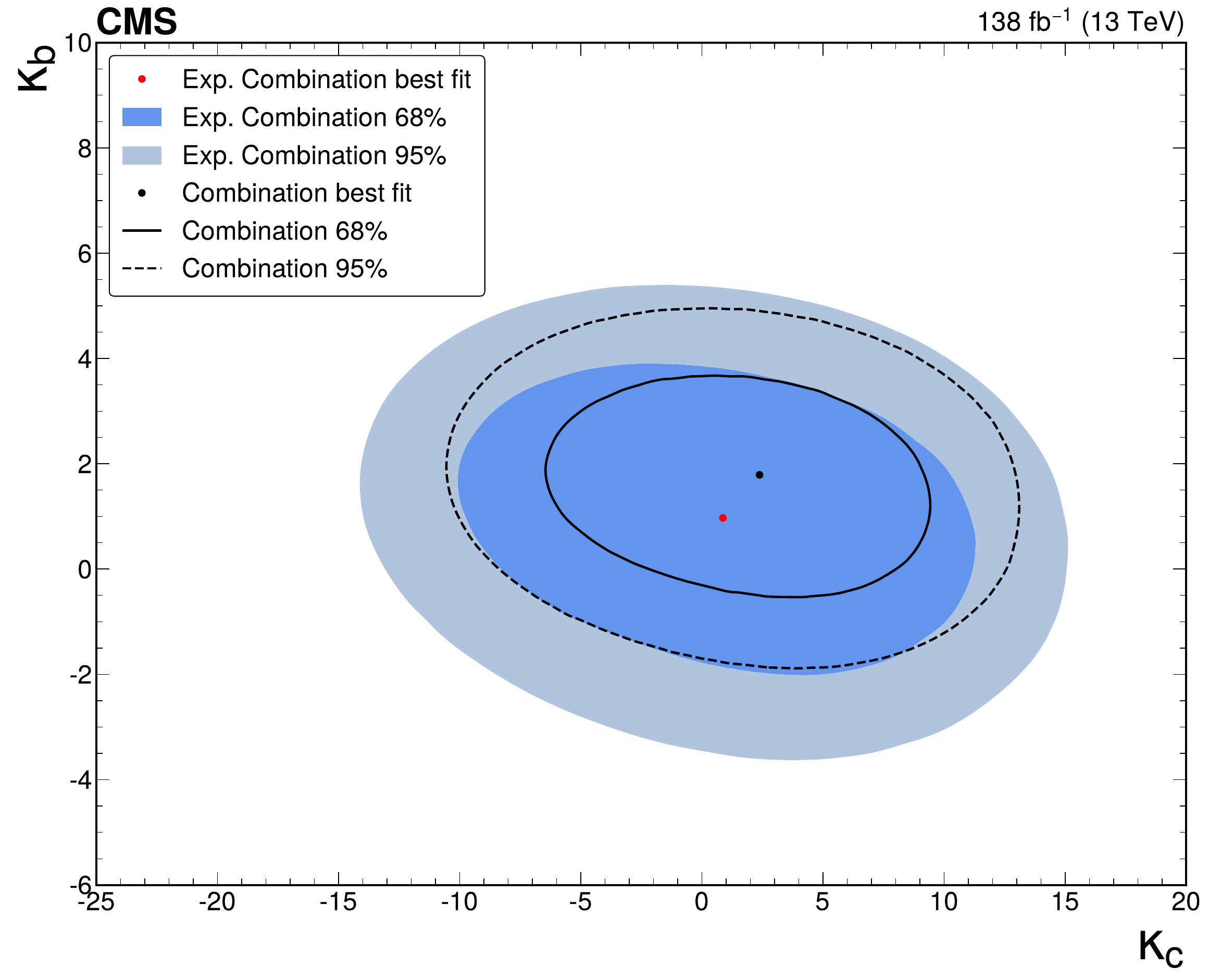}
    \caption{Observed and expected simultaneous fits for \kappab\ and \kappac, including the coupling dependence of the branching fractions (left) and with the branching fractions of the decay channels entering the combination implemented as nuisance parameters with no dependence on the couplings (right). The $68\%$ and $95\%$ \CL contours are shown in solid and dashed lines for the observed data, with the expected contours indicated in blue.}
    \label{fig:yukawa_2D_superimposed}
\end{figure}

It should be noted that the vector of signal strength modifiers includes one parameter per bin per decay channel, hence the procedure exemplified in Eq.~(\ref{eq:weights_spectra}) is not necessary in this case.
Since the parametrizations are derived in the full phase space, the acceptance term is not included in the signal strength modifiers.
These considerations lead to the following form for the vector of signal strength modifiers:
\begin{equation}
    \begin{aligned}
        \vec{\mu} &= \left(\vec{\mu}_{\hgamgam}, \vec{\mu}_{\hzz}, \dots \right) \\
                &= \left(\mu_{\hgamgam, 0-5}, \dots, \mu_{\hzz, 0-10}, \dots \right) \\
                &= \left(
                    \frac{\sigma_{0-5}(\vec{\kappa}) \mathcal{B}_{\hgamgam}(\vec{\kappa})}{\sigma_{0-5}^{\mathrm{SM}} \mathcal{B}_{\hgamgam}^{\mathrm{SM}}},
                    \dots,
                    \frac{\sigma_{0-10}(\vec{\kappa}) \mathcal{B}_{\hzz}(\vec{\kappa})}{\sigma_{0-10}^{\mathrm{SM}} \mathcal{B}_{\hzz}^{\mathrm{SM}}},
                    \dots
                \right).
    \end{aligned}
\end{equation}
Figure~\ref{fig:yukawa_2D_superimposed} shows the constraints on \kappab and \kappac when including the coupling dependence of the branching fraction (left) and when implemented as nuisance parameters with no branching-fraction dependence on the couplings and no prior constraint, \ie, floating (right).
The shapes of the constraints are similar to the ones obtained in Ref.~\cite{tk_paper}. They are in agreement with the SM at 68\% confidence level (\CL).

\begin{figure}[!ht]
    \includegraphics[width=0.45\textwidth]{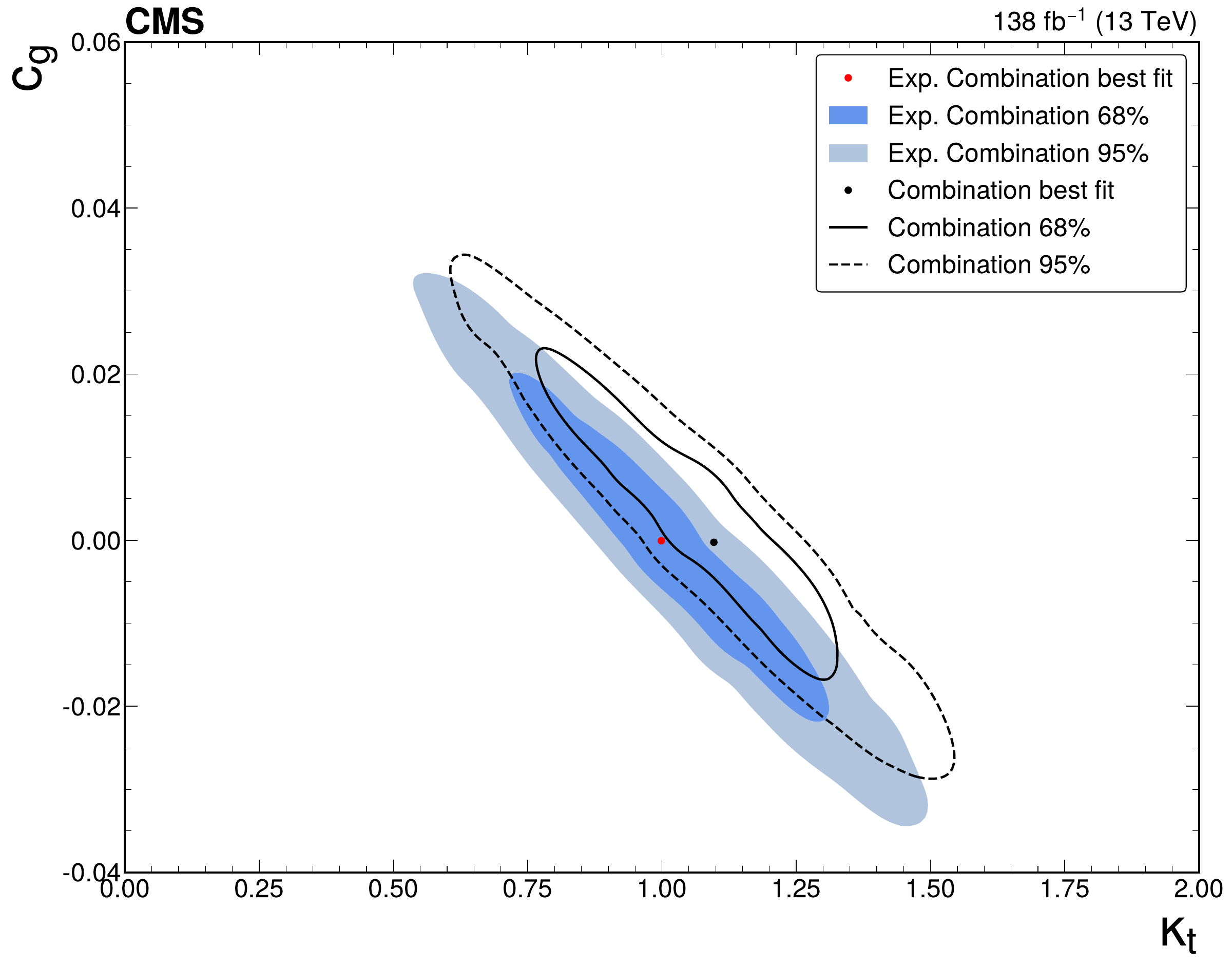}
    \includegraphics[width=0.45\textwidth]{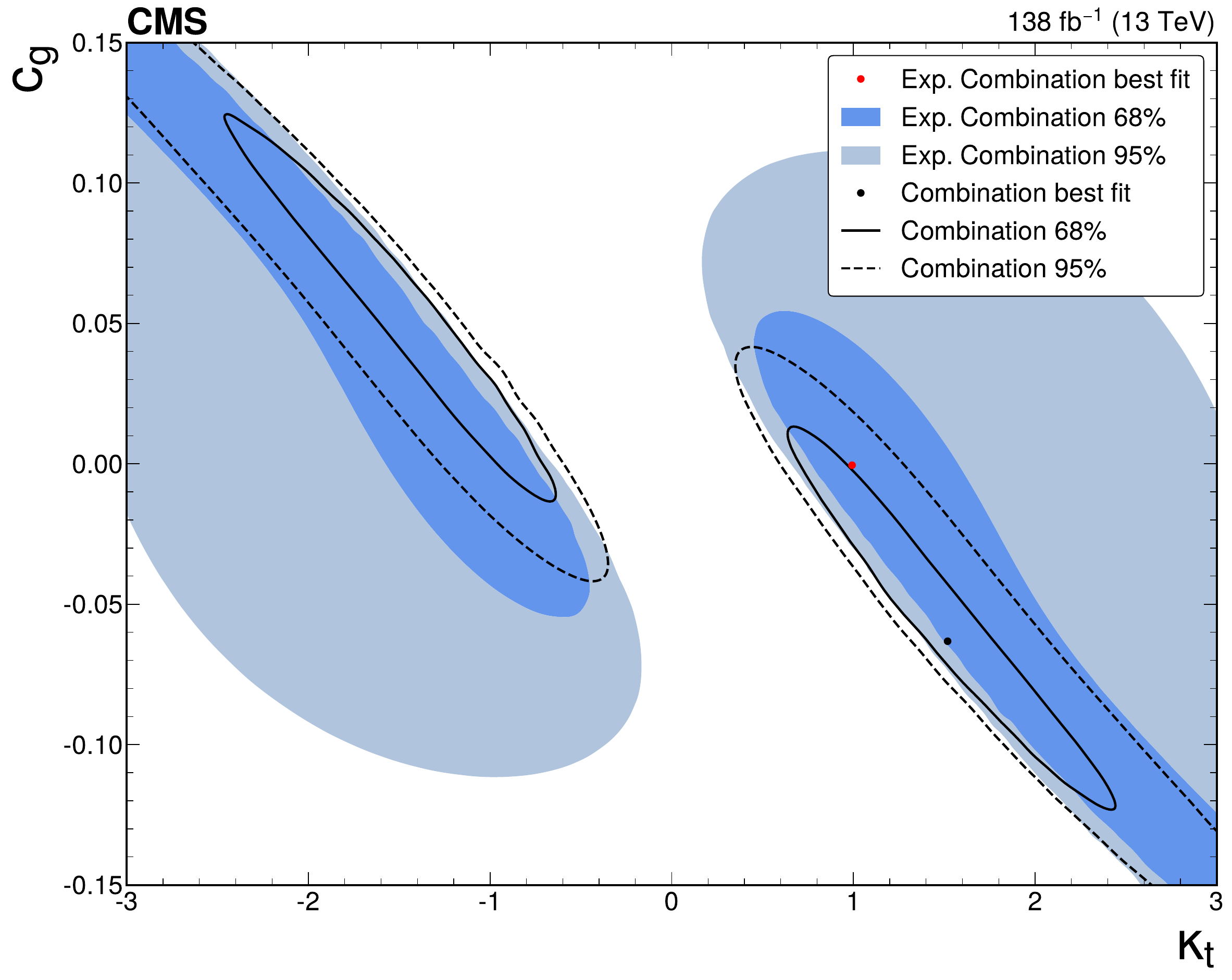}
    \caption{Simultaneous fit for \kappat\ and \cg, observed and expected, including the coupling dependence of the branching fractions (left) and with the branching fractions of the decay channels entering the combination implemented as nuisance parameters with no dependence on the couplings (right). The $68\%$ and $95\%$ CL contours are shown in solid and dashed lines for observed data, the expected contours are indicated by the blue shaded areas. In the fit with the branching fractions implemented as nuisance parameters, the best fit point is shown in the region with positive \kappat, but a second degenerate best fit point exists for negative \kappat.}
    \label{fig:top_cgkt_2D}
\end{figure}

The observed and expected two-dimensional confidence intervals for \kappat\ and \cg\ are shown in Fig.~\ref{fig:top_cgkt_2D}.
For the case of coupling dependence of the branching fractions, the normalization of the spectrum is, by construction, equal to the SM normalization for the set of coefficients satisfying \\ $12 \cg + \kappat \simeq 1$.
The shape of the parametrized spectrum, $s$, is calculated by normalizing the differential cross section to 1:
\begin{equation}
    s_i(\kappa_{\mathrm{t}}, c_{\mathrm{g}})=\frac{\sigma_i(\kappa_{\mathrm{t}}, c_{\mathrm{g}})}{\sum_j \sigma_j(\kappa_{\mathrm{t}}, c_{\mathrm{g}})} ,
\end{equation}
where $\sigma_i$ is the parametrization in bin $i$.
Inserting the expected parabolic dependence of $\sigma_i(\kappat,\cg)$ reveals that the shape of the parametrization for \kappat and \cg\ variations becomes a function only of the ratio of the two couplings, $s_i(\cg/\kappat)$.
Thus, the dependence of the likelihood on the radial distance $\sqrt{\smash[b]{\kappat^2 + \cg^2}}$ stems from the constraints on the overall normalization, while the dependence on the slope $\cg/\kappat$ is due to the shape of the distribution.
The dependence of the likelihood on the slope becomes apparent in Fig. \ref{fig:top_cgkt_2D} (right), where the branching fractions are implemented as nuisance parameters with no prior constraint.
Except at small values of the couplings, the constraint on the couplings comes from their ratio.
The two symmetric sets of contours are due to a symmetry of the parametrization under $(\kappat, \cg) \to (-\kappat, -\cg)$.
In both scenarios, the results are consistent with the SM at the 68\% \CL.
The shapes of the constrained regions are in agreement with those in Ref.~\cite{tk_paper}, once scaled for the increase in the data sample size and the number of decay channels included in the combination.

\begin{figure}[!ht]
    \includegraphics[width=0.45\textwidth]{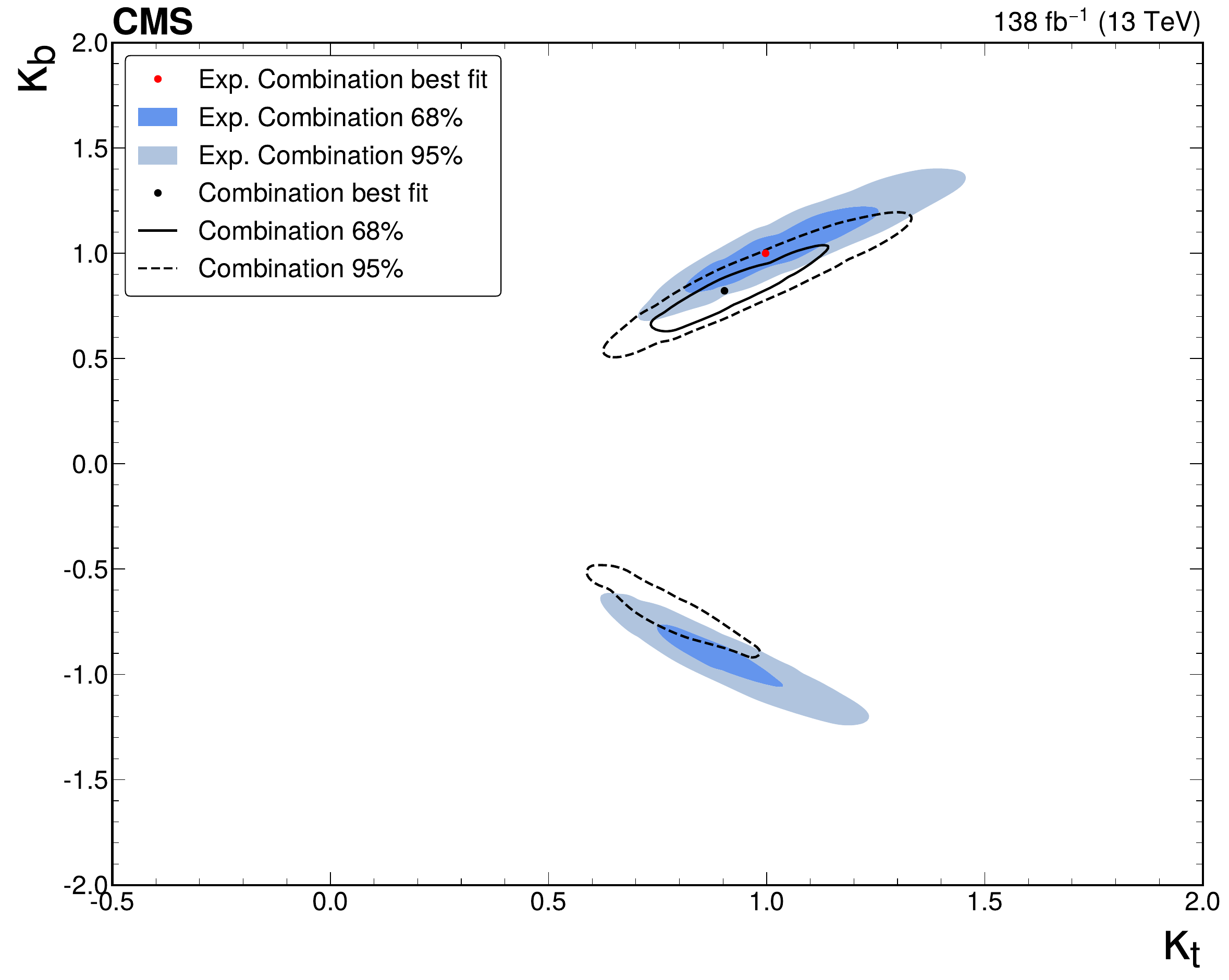}
    \includegraphics[width=0.45\textwidth]{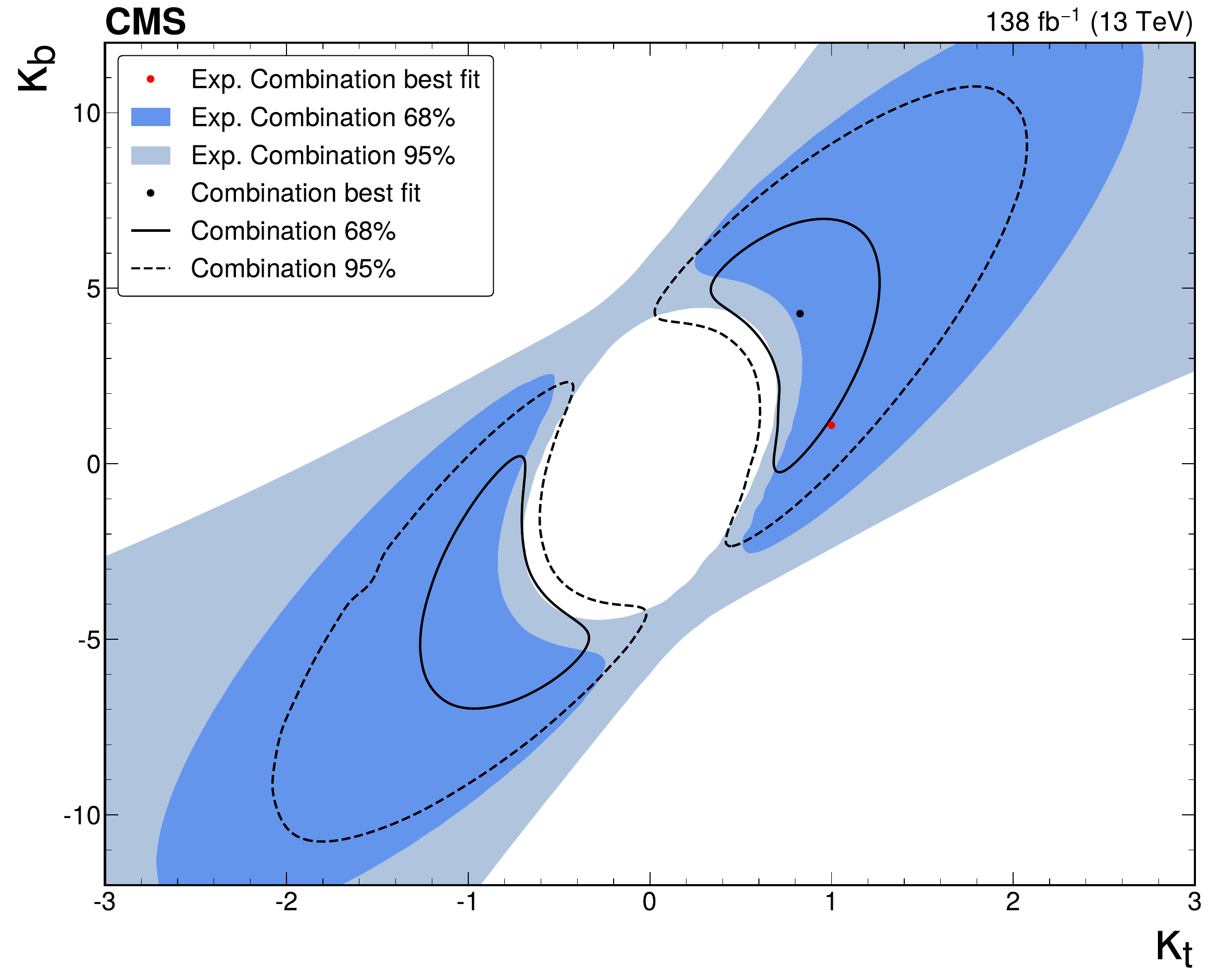}
    \caption{Simultaneous fit for \kappat\ and \kappab, observed and expected, including the coupling dependence of the branching fractions (left) and with the branching fractions of the decay channels entering the combination implemented as nuisance parameters with no dependence on the couplings (right). The $68\%$ and $95\%$ \CL contours are indicated by the solid and dashed lines for the observed data, the expected contours are indicated by the blue shaded regions. In the fit with the branching fractions implemented as nuisance parameters, the best fit point is shown in the region with positive \kappat, but a second degenerate best fit point exists for negative \kappat.}
    \label{fig:top_kbkt_2D}
\end{figure}

The observed and expected two-dimensional likelihood scans for \kappat\ and \kappab\ are shown in Fig.~\ref{fig:top_kbkt_2D}.
For the branching fractions implemented as nuisance parameters with no prior constraint, the parametrization is symmetric under $(\kappat, \kappab) \to (-\kappat, -\kappab)$, hence the two sets of contours are symmetric.
In both scenarios, the results are consistent with the SM at the 68\% confidence level.
The shapes of the likelihoods are in agreement with those in Ref.~\cite{tk_paper}, once scaled for the increase in the size of the data sample and the number of decay channels included.

\clearpage
\section{SMEFT interpretation}
\label{sec:smeft}

The effective field theory (EFT) approach aims to constrain BSM physics in a model-agnostic way.
Assuming the existence of a yet unknown phenomenon at an energy scale $\Lambda$ above the energy scale that our experiment can reach directly, effects of BSM physics may manifest themselves through effective interactions between SM fields.
The effective Lagrangian is written as:
\begin{equation}
    \mathcal{L}_{\mathrm{SMEFT}}=\mathcal{L}_{\mathrm{SM}}+\sum_{i=5}^{\infty} \sum_{j=0}^{N_i} \frac{c_j^{(i)}}{\Lambda^{i - 4}} O_j^{(i)},
    \label{eq:main_lagrangian}
\end{equation}
where $i$ runs over the number of dimensions, $j$ runs over the number of operators of dimension $i$, $\mathcal{L}_{\mathrm{SM}}$ is of dimension 4, the operators $O_j^{(i)}$ have dimensions of $i$, $c_j^{(i)}$ are dimensionless Wilson coefficients (WCs) that correspond to the strength of the interaction, and $\Lambda$ is the above-mentioned energy scale.
Dimension-five operators are related to the neutrino sector and lepton number violation, and are not studied in this paper.
The leading contributions are then from dimension-six operators and Eq.~(\ref{eq:main_lagrangian}) can be written as:
\begin{equation}
    \mathcal{L}_{\mathrm{SMEFT}}=\mathcal{L}_{\mathrm{SM}}+ \sum_{j=0}^{N_{(6)}} \frac{c_j^{(6)}}{\Lambda^2} O_j^{(6)},
\end{equation}
where the number of independent operators $N_{(6)}$ in this basis can be reduced to 59 (barring flavor structure and Hermitian conjugations), divided into eight classes, following the scheme introduced in Ref. \cite{Grzadkowski_2010}.

\subsection{SMEFT model details}
The two main packages used to generate events within \MGvATNLO\ are SMEFTsim3.0 \cite{smeftsim} and SMEFT@NLO \cite{smeftatnlo}.
The \textsc{FeynRules} model used is \textsc{topU3l}, which implements $U(2)^3_{\PQq,\PQu,\PQd}$ and $U(3)^2_{\ell,\Pe}$ flavor symmetry.
Light quarks (\PQu, \PQd, \PQs, \PQc) and heavy quarks (\PQb, \PQt) have separate WCs, while all leptons share the same WCs.
The input parameter scheme is $\{G_F, m_Z, m_W\}$.
The value of the parameter $\Lambda$ is set to 1\TeV, which is the scale at which the SMEFT operators are assumed to be valid.
Several differences exist between SMEFTsim3.0 and SMEFT@NLO.
SMEFTsim3.0 includes leading order (LO) corrections, while SMEFT@NLO includes NLO corrections in QCD: this translates to a higher accuracy in the calculation of coefficients affecting ggH production in the latter case.
Another important difference is that SMEFT@NLO does not include $CP$-odd operators, while SMEFTsim3.0 does.
For the different interpretations provided in this paper, three input configurations are used to derive the parametrizations, exploiting the strengths of both packages.
The differences are related to the model used to derive the parameterizations of ggH production and of the \hgamgam\ decay width: 
the first configuration, adopted to produce the results for (\chg, \chgt) in Section~\ref{subsec:cpeven_cpodd}, uses SMEFTsim3.0 to derive both the ggH parametrization and the \hgamgam decay width parameterization; 
the second configuration, adopted to produce the results for the pairs (\chb, \chbt), (\chw, \chwt) and (\chwb, \chwtb) in Section~\ref{subsec:cpeven_cpodd}, uses SMEFT@NLO to derive the ggH parametrization and SMEFTsim3.0 to derive the \hgamgam decay width parameterization; 
the last configuration, adopted to produce the results shown in Section~\ref{subsec:pca}, uses SMEFT@NLO to derive the ggH parametrization and the NLO theoretical predictions provided in Ref.~\cite{gammagamma_eft_parametrization} for the parametrization of the \hgamgam decay width.
Production modes other than ggH and decay modes other than \hgamgam are parametrized using SMEFTsim3.0 and are the same for all the configurations.

\subsection{Derivation of the parametrizations}
When working within an EFT framework, the amplitude for each Higgs boson production and decay process can be described as:
\begin{equation}
    \left|\mathcal{M}_{\mathrm{SMEFT}}\right|^2=\left|\mathcal{M}_{\mathrm{SM}}+\mathcal{M}_{\mathrm{BSM}}\right|^2=\left|\mathcal{M}_{\mathrm{SM}}\right|^2+2 \operatorname{Re}\left\{\mathcal{M}_{\mathrm{SM}} \mathcal{M}_{\mathrm{BSM}}^{\dagger}\right\}+\left|\mathcal{M}_{\mathrm{BSM}}\right|^2,
    \label{eq:full_matrix_element}
\end{equation}
where $\mathcal{M}_{\mathrm{SM}}$ and $\mathcal{M}_{\mathrm{BSM}}$ are the matrix elements originating from the SM and BSM Lagrangians, respectively.
The SM-BSM interference term is suppressed by a factor of $1/\Lambda^2$ and the purely-BSM term is suppressed by a factor of $1/\Lambda^4$.
If the BSM contributions are restricted to diagrams with a single insertion of a BSM vertex, then $\mathcal{M}_{\mathrm{BSM}}$ is linear in the WCs $c_i$.
Thus, using the fact that $\sigma \propto |\mathcal{M}|^2$, the production cross section in a bin $i$ can be written as:
\begin{equation}
    \sigma_{\mathrm{SMEFT}}^i=\sigma_{\mathrm{SM}}^i+\sigma_{\mathrm{int}}^i+\sigma_{\mathrm{BSM}}^i,
\end{equation}
and a scaling function quadratic in the WCs can be derived as:
\begin{equation}
    \mu_{\mathrm{prod }}^i(\vec{c})=\frac{\sigma_{\mathrm{SMEFT}}^i}{\sigma_{\mathrm{SM}}^i}=1+\sum_j A_j^i c_i+\sum_{j k} B_{j k}^i c_j c_k.
\end{equation}
The $A_j^i$ and $B_{jk}^i$ constants encode the impact of the WCs on the production cross section in bin $i$: $A_j^i$ are the linear terms, $B_{jj}^i$ are the quadratic terms, and $B_{jk}^i$ for $j \neq k$ are the cross terms.
This procedure also applies to both partial and total decay widths, meaning the scaling function of the branching fraction to a given final state $f$ can be written as:
\begin{equation}
    \mu_{\text{decay}}^f(\vec{c})=\frac{\mathcal{B}_{\mathrm{SMEFT}}^f}{\mathcal{B}_{\mathrm{SM}}^f}=\frac{\Gamma_{\mathrm{SMEFT}}^f / \Gamma_{\mathrm{SM}}^f}{\Gamma_{\mathrm{SMEFT}}^H / \Gamma_{\mathrm{SM}}^H}=\frac{1+\sum_j A_j^f c_j+\sum_{j k} B_{j k}^f c_j c_k}{1+\sum_j A_j^H c_j+\sum_{j k} B_{j k}^H c_j c_k}.
\end{equation}
Introducing the narrow width approximation, the total scaling function for a given bin $i$ and final state $f$ is then:
\begin{equation}
    \mu^{i, f}(\vec{c})=
    \frac{(\sigma \mathcal{B})^{i, \PH \to f}}{(\sigma \mathcal{B})_{\mathrm{SM}}^{i, \PH \to f}}=
    \mu_{\text{prod }}^i(\vec{c}) \mu_{\text{decay }}^f(\vec{c}) .
    \label{eq:full_scaling_function}
\end{equation}
The terms $A_j$ and $B_{jk}$ are derived either analytically at higher orders than possible with current Monte Carlo tools, or using the \textsc{EFT2Obs} tool~\cite{leshouches_2019}, which wraps and interfaces widely-used packages to facilitate the process of deriving an EFT parametrization.
It utilizes \MGvATNLO \cite{madgraph} (version 2.6.7) for simulation and \PYTHIA8 \cite{Pythia82} (version 8.2) for parton showering and hadronization.
Fiducial selections and histograms of observables are defined in the \textsc{Rivet} \cite{rivet} (version 3.0.1) framework. 
Through the use of \textsc{Rivet}, acceptance effects, which are crucial in the case of differential fiducial cross sections, are taken into account.

An important note concerning the parametrization of the production terms is that most of the analyses are not sufficiently sensitive to measure different production modes separately, hence a weighted parametrization, scaling the expected inclusive cross section in each observable bin, is applied:
\begin{equation}
    \mu_{i} = \sum_{j} \frac{\sigma^{\mathrm{MG5}}_{ij} \frac{\sigma^{\mathrm{YR}}_{j}}{\sigma^{\mathrm{MG5}}_{j}}}{\sum_{k} \sigma^{\mathrm{MG5}}_{ik} \frac{\sigma^{\mathrm{YR}}_{k}}{\sigma^{\mathrm{MG5}}_{k}}} \mu_{ij},
\end{equation}
where:
\begin{itemize}
    \item $i$ and $j$ refer to observable bin and production mode, respectively;
    \item $\sigma^{YR}_{j}$ is the SM cross section for production mode $j$ (taken from Ref.~\cite{higgs_handbook});
    \item $\sigma^{MG5}_{j}$ is the full cross section for production mode $j$ before the application of fiducial selections;
    \item $\sigma^{MG5}_{ij}$ is the cross section in bin $i$ for production mode $j$ after the application of fiducial selections.
\end{itemize}

\subsection{Constraints on CP-even and CP-odd pairs of Wilson coefficients}
\label{subsec:cpeven_cpodd}

A particularly relevant group of operators is the group $X^2H^2$, listed in Table~\ref{tab:wilson_coeffs_x2h2}.
For each process shown, the first operator conserves $CP$ while the second violates it. The coefficients \chg\ and \chgt\ mainly affect ggH production, while the others affect qqH and VH production along with the Higgs boson decay.

\renewcommand{\arraystretch}{1.7}
\begin{table}[!htb]
   \centering
   \topcaption{List of $X^2H^2$ operators and corresponding Wilson coefficients. Example Feynman diagrams of the processes affected by the operators are shown in the rightmost column. The notation used is based on Ref.~\cite{smeftsim}.}
   \begin{tabular}{lccc}
    Class & Operator & Wilson coefficient & Example process\\
    \hline
    \multirow{8}{*}{$\mathcal{L}_6^{(4)}-X^2 H^2$} & $H^{\dagger} H G_{\mu \nu}^a G^{a \mu \nu}$ & $c_{\mathrm{HG}}$ & \multirow{2}{*}{\includegraphics[width=0.15\linewidth]{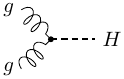}} \\
    & $H^{\dagger} H \tilde{G}_{\mu \nu}^a G^{a \mu \nu}$ & $\tilde{c}_{\mathrm{HG}}$ & \\
    \cline{2-4}
    & $H^{\dagger} H B_{\mu \nu} B^{\mu \nu}$ & $c_{\mathrm{HB}}$ & \multirow{2}{*}{\includegraphics[width=0.15\linewidth]{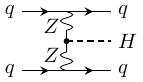} \includegraphics[width=0.1\linewidth]{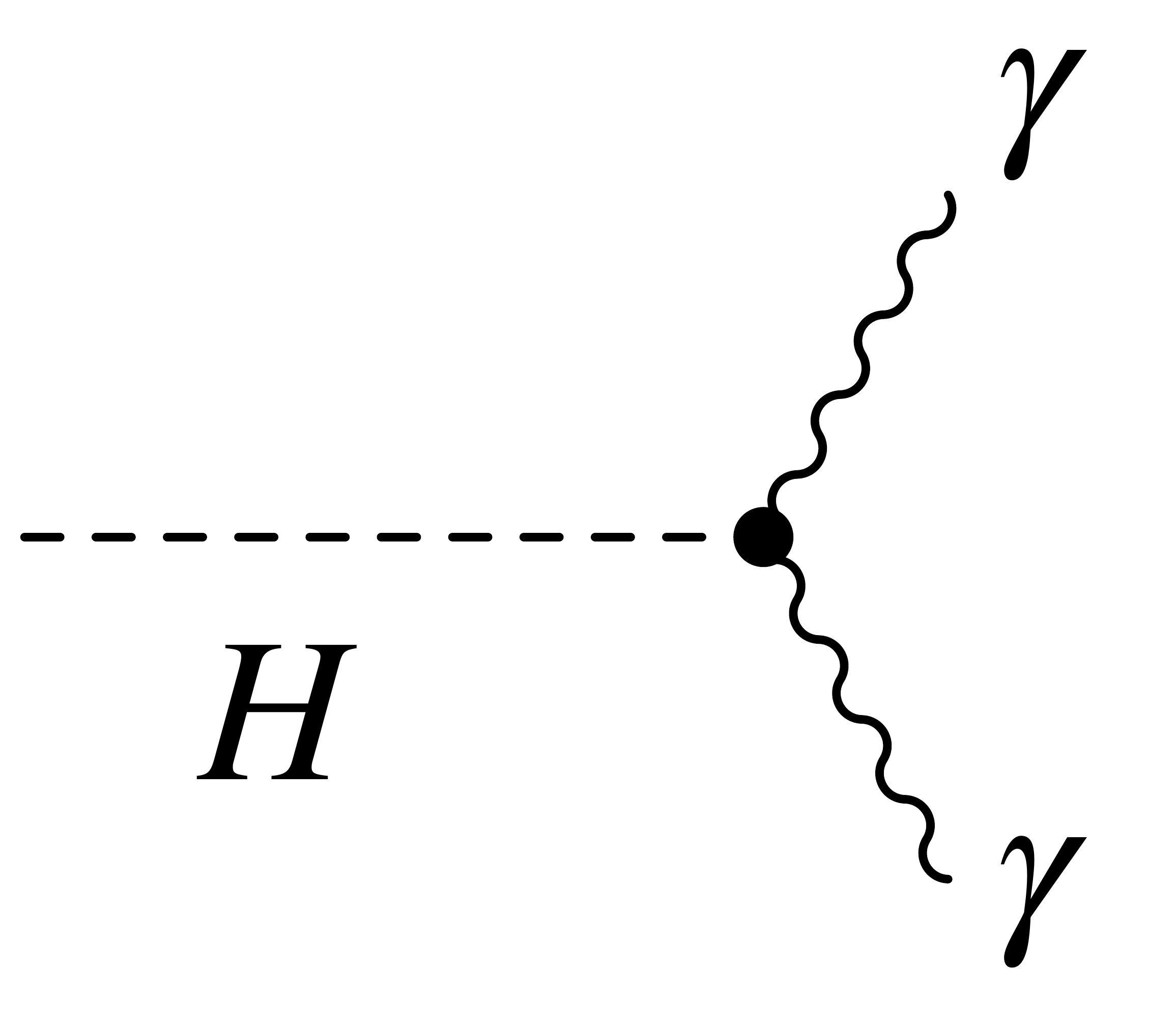}}\\
    & $H^{\dagger} H \tilde{B}_{\mu \nu} B^{\mu \nu}$ & $\tilde{c}_{\mathrm{HB}}$ & \\
    & & & \includegraphics[width=0.1\linewidth]{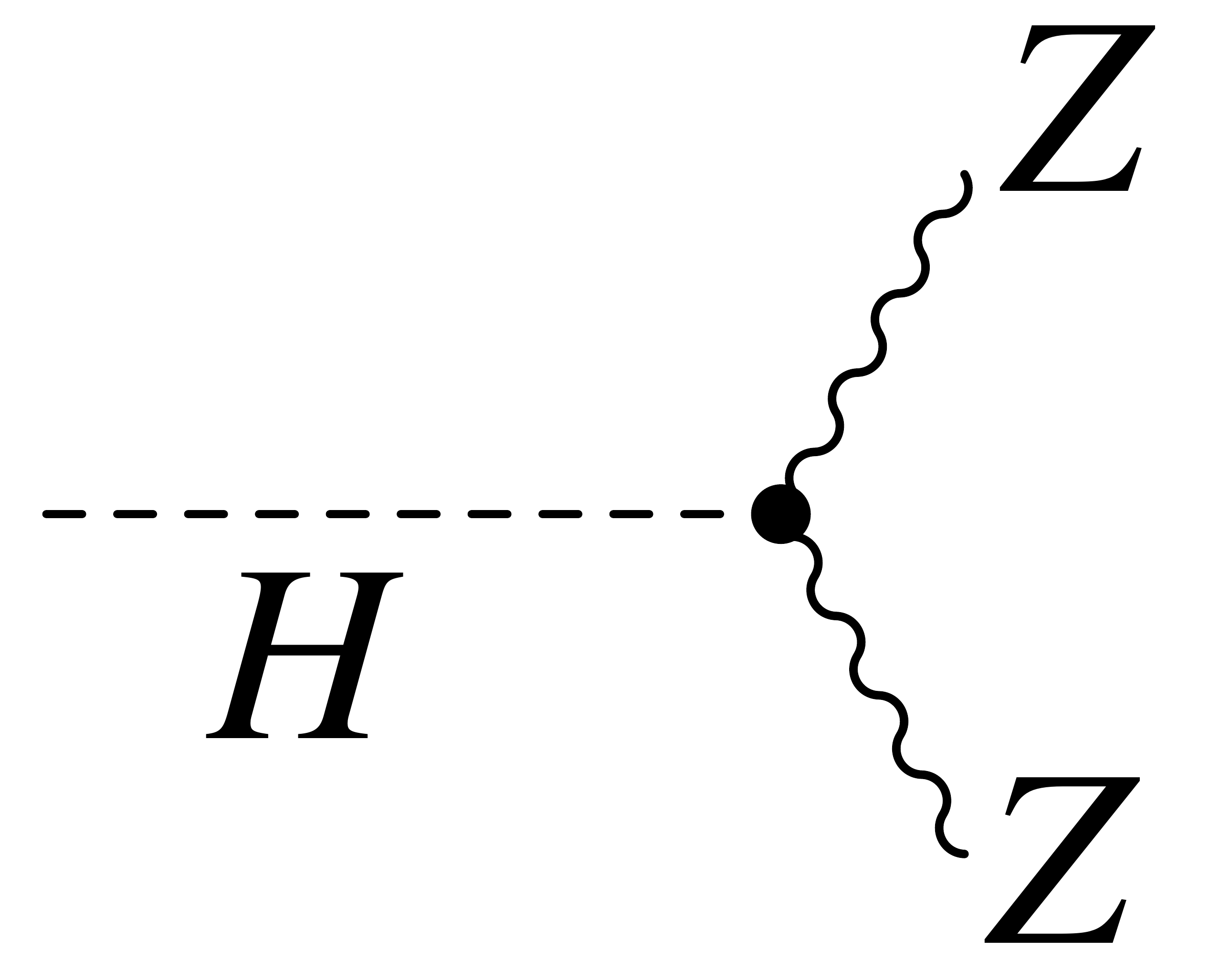} \\
    \cline{2-4}
    & $H^{\dagger} H W_{\mu \nu}^i W^{i \mu \nu}$ & $c_{\mathrm{HW}}$ & \multirow{2}{*}{\includegraphics[width=0.15\linewidth]{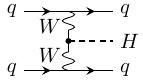} \includegraphics[width=0.1\linewidth]{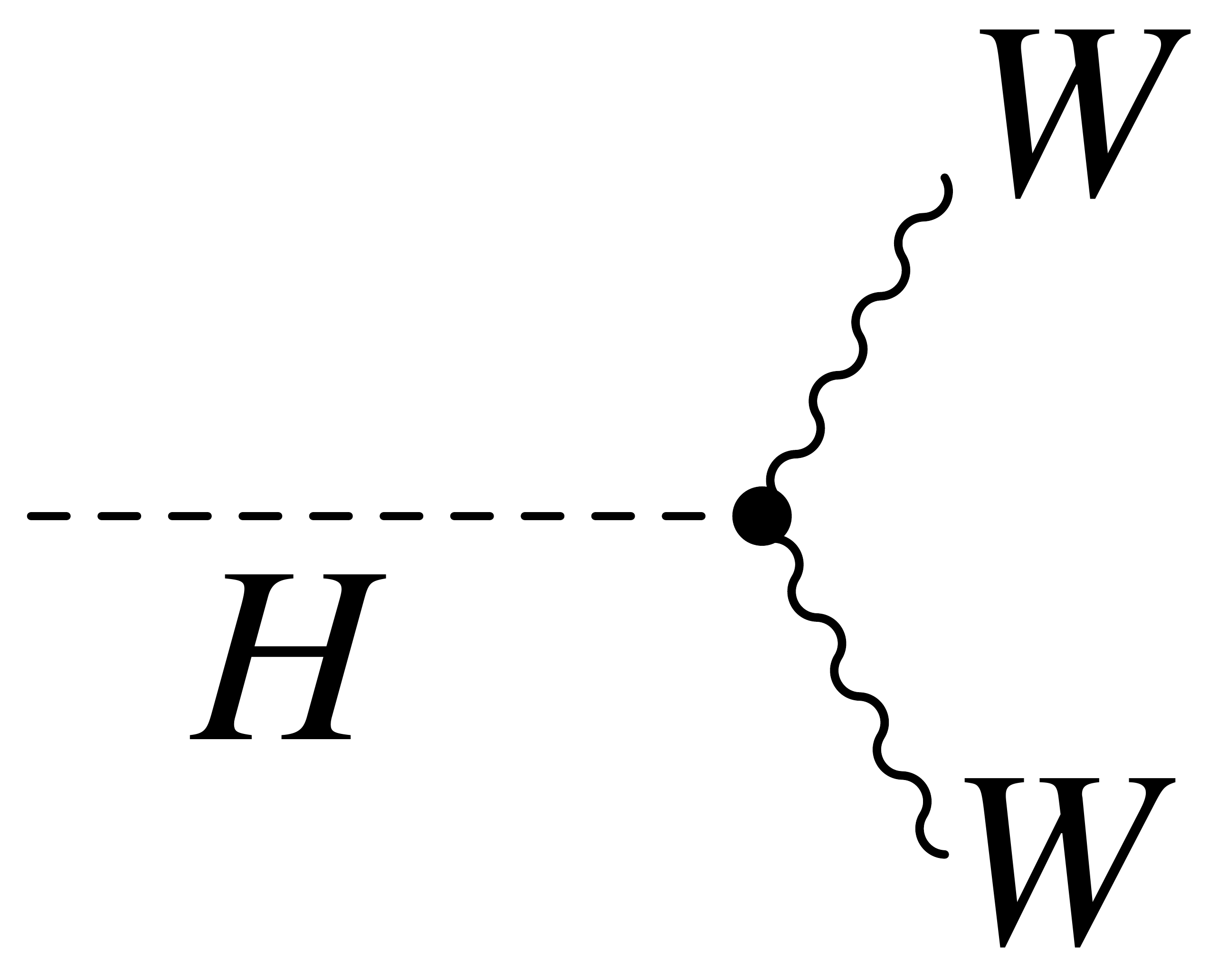}} \\
    & $H^{\dagger} H \tilde{W}_{\mu \nu}^i W^{i \mu \nu}$ & $\tilde{c}_{\mathrm{HW}}$ & \\
    \cline{2-4}
    & $H^{\dagger} \sigma^i H W_{\mu \nu}^i B^{i \mu \nu}$ & $c_{\mathrm{HWB}}$ & \multirow{2}{*}{\includegraphics[width=0.15\linewidth]{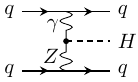} \includegraphics[width=0.1\linewidth]{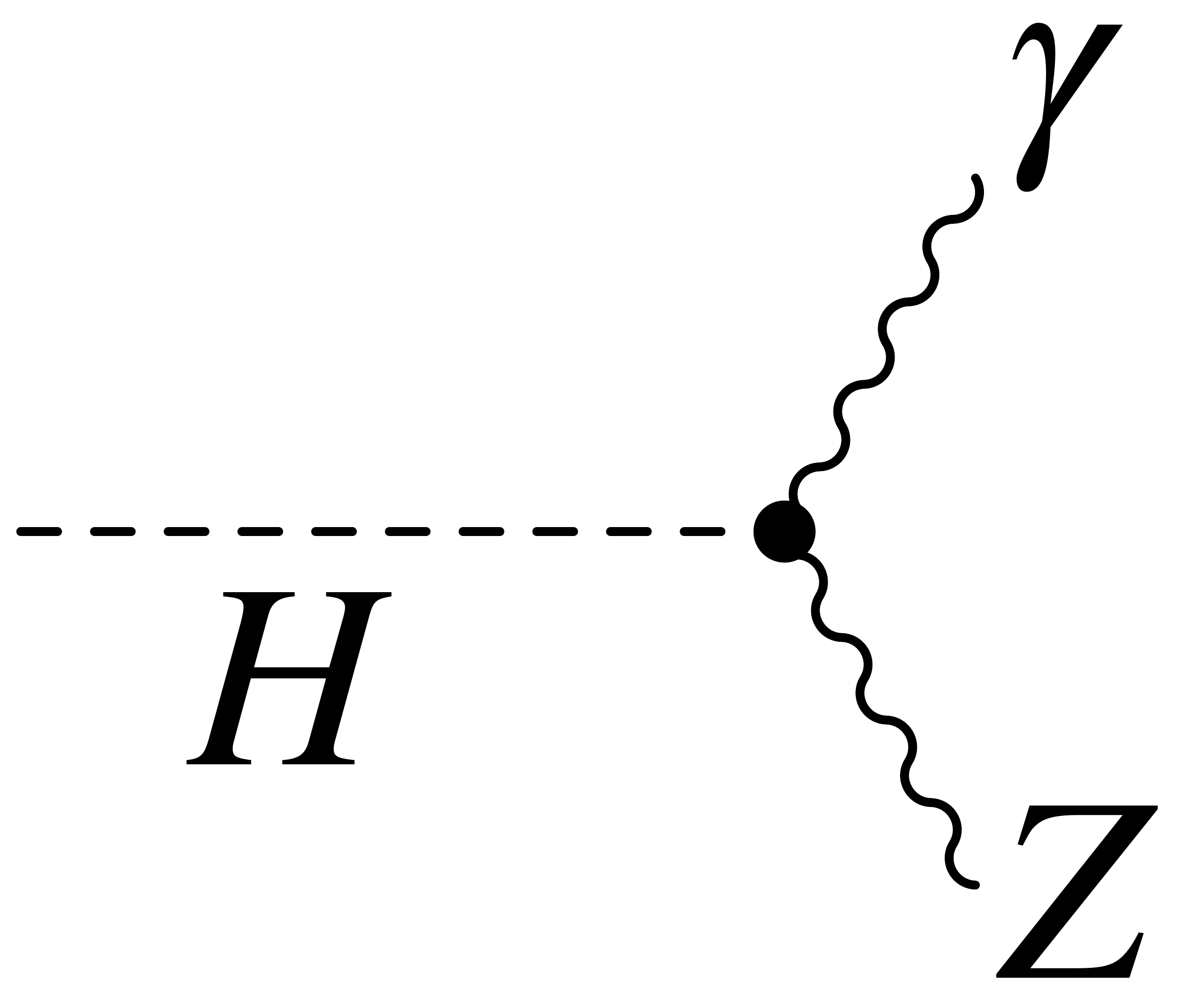}} \\
    & $H^{\dagger} \sigma^i H \tilde{W}_{\mu \nu}^i B^{i \mu \nu}$ & $\tilde{c}_{\mathrm{HWB}}$ & \\
   \end{tabular}
   \label{tab:wilson_coeffs_x2h2}
\end{table}

\begin{figure}[!ht]
    \centering
    \includegraphics[width=0.49\textwidth]{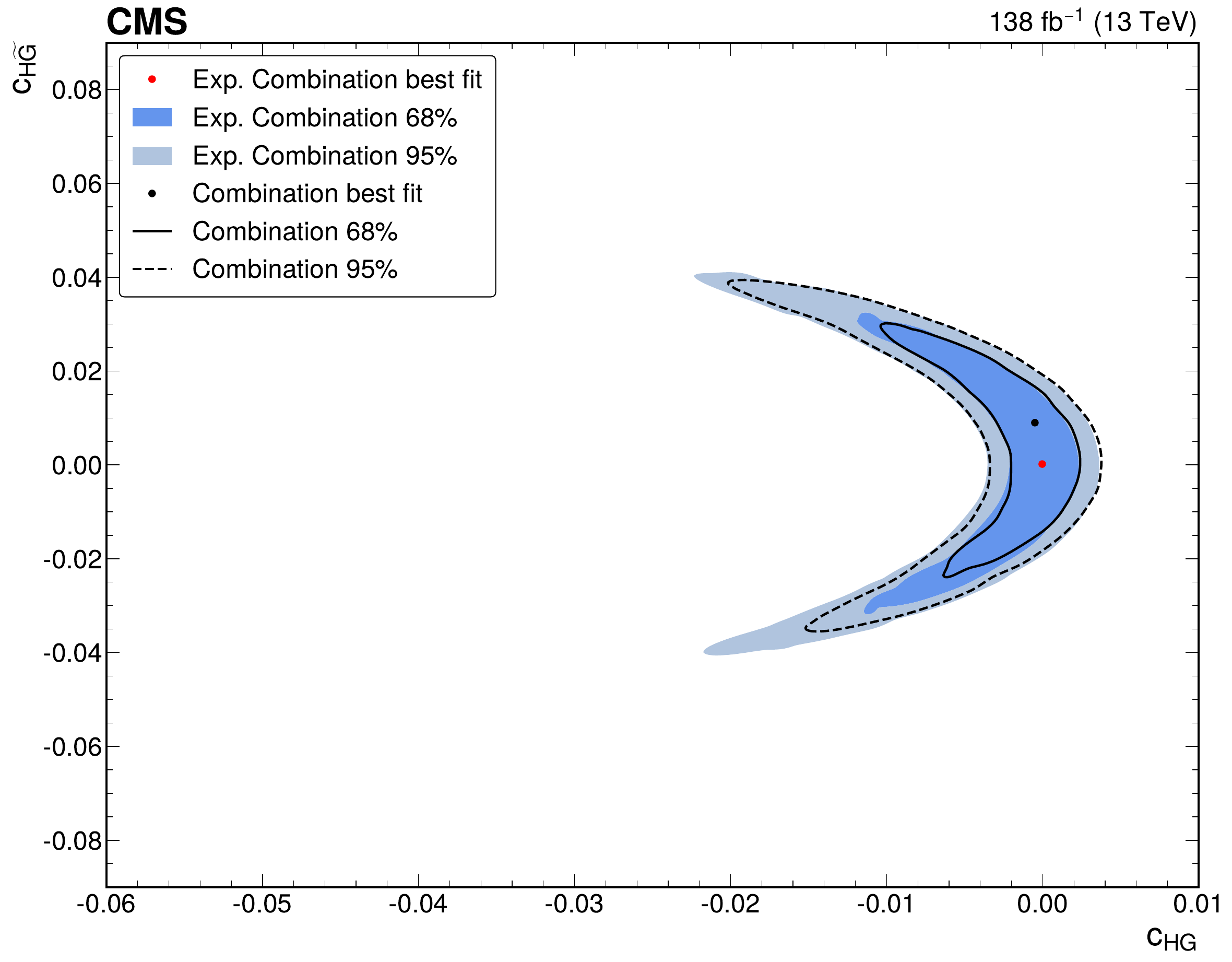}
    \includegraphics[width=0.49\textwidth]{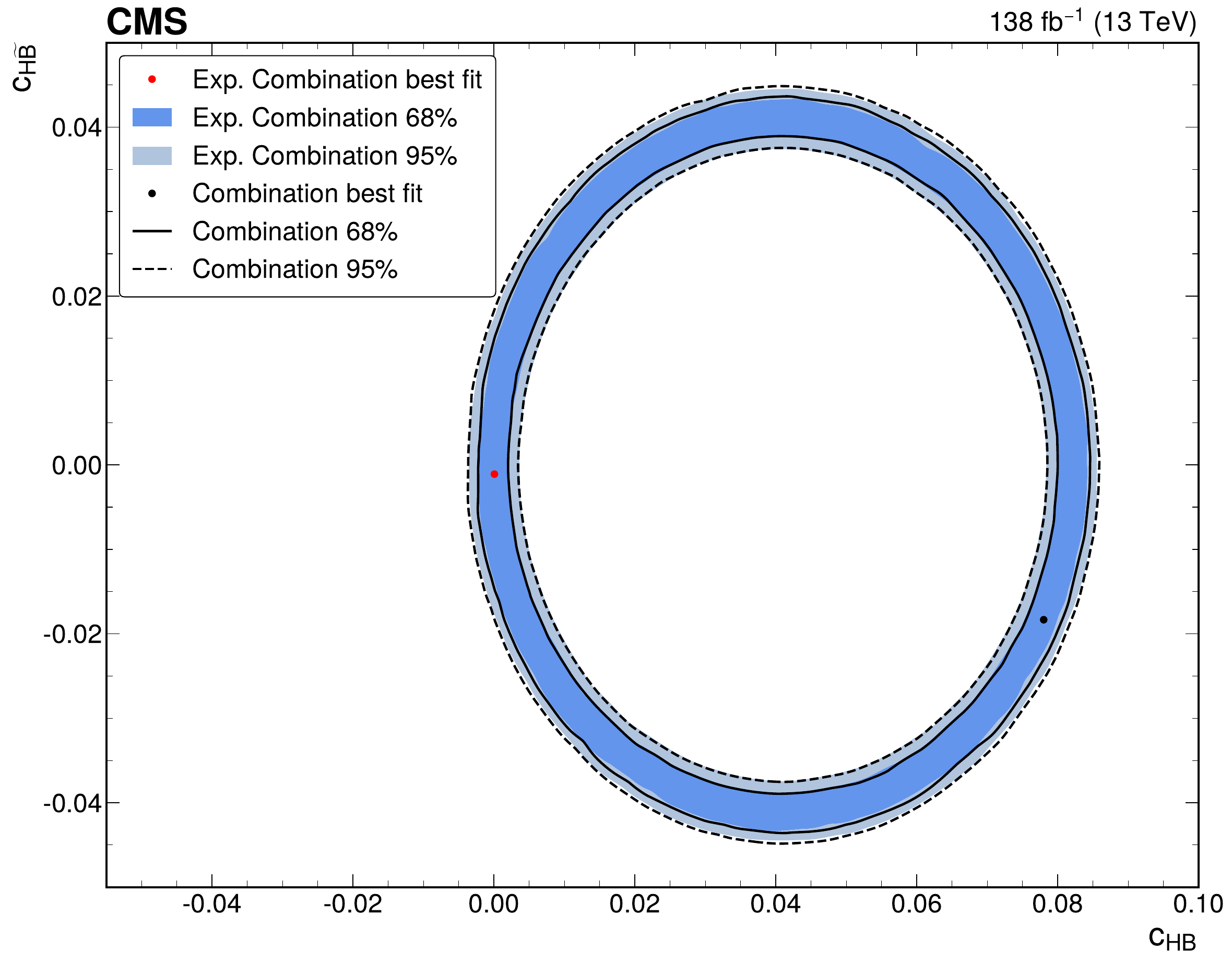} \\
    \includegraphics[width=0.48\textwidth]{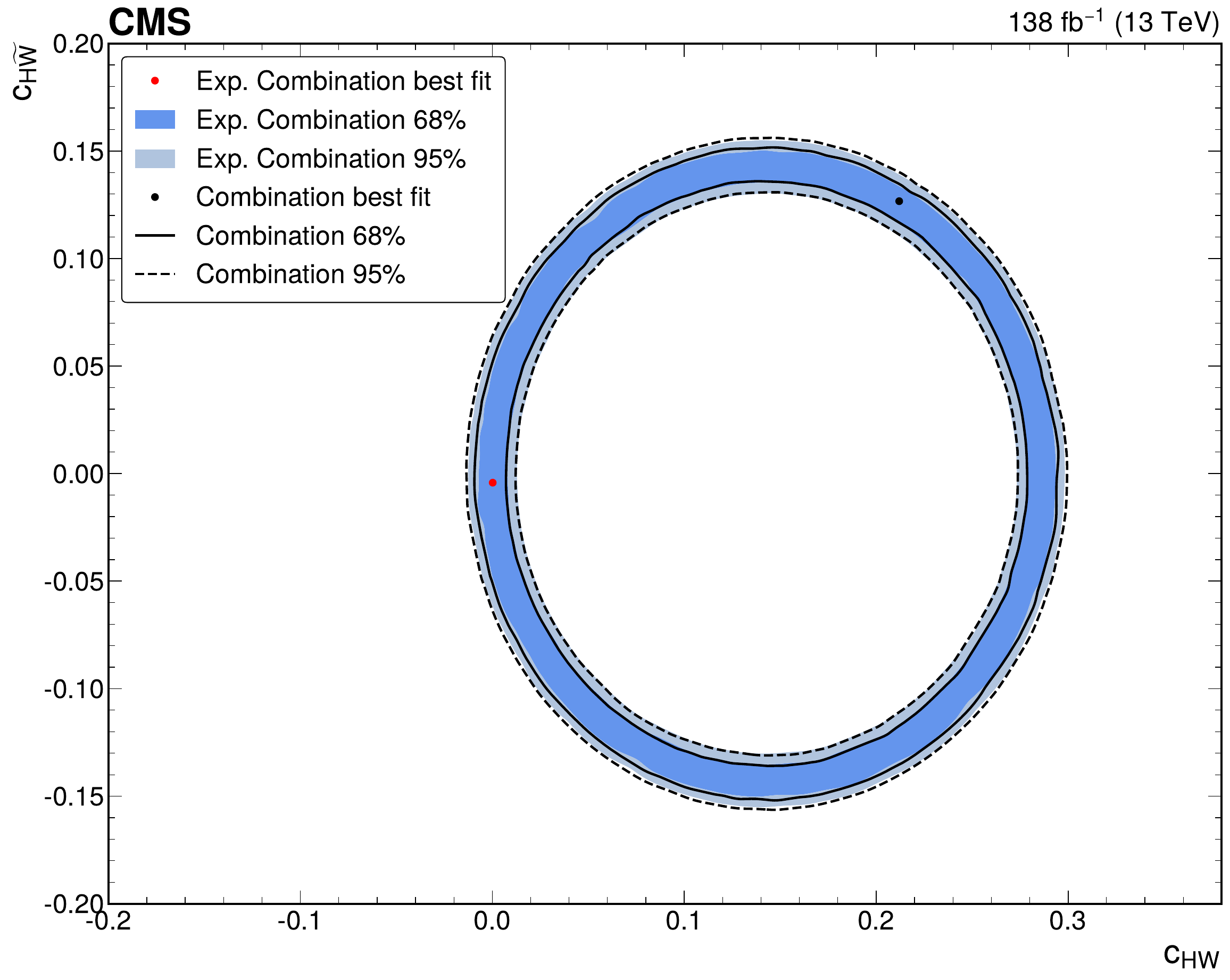}
    \includegraphics[width=0.5\textwidth]{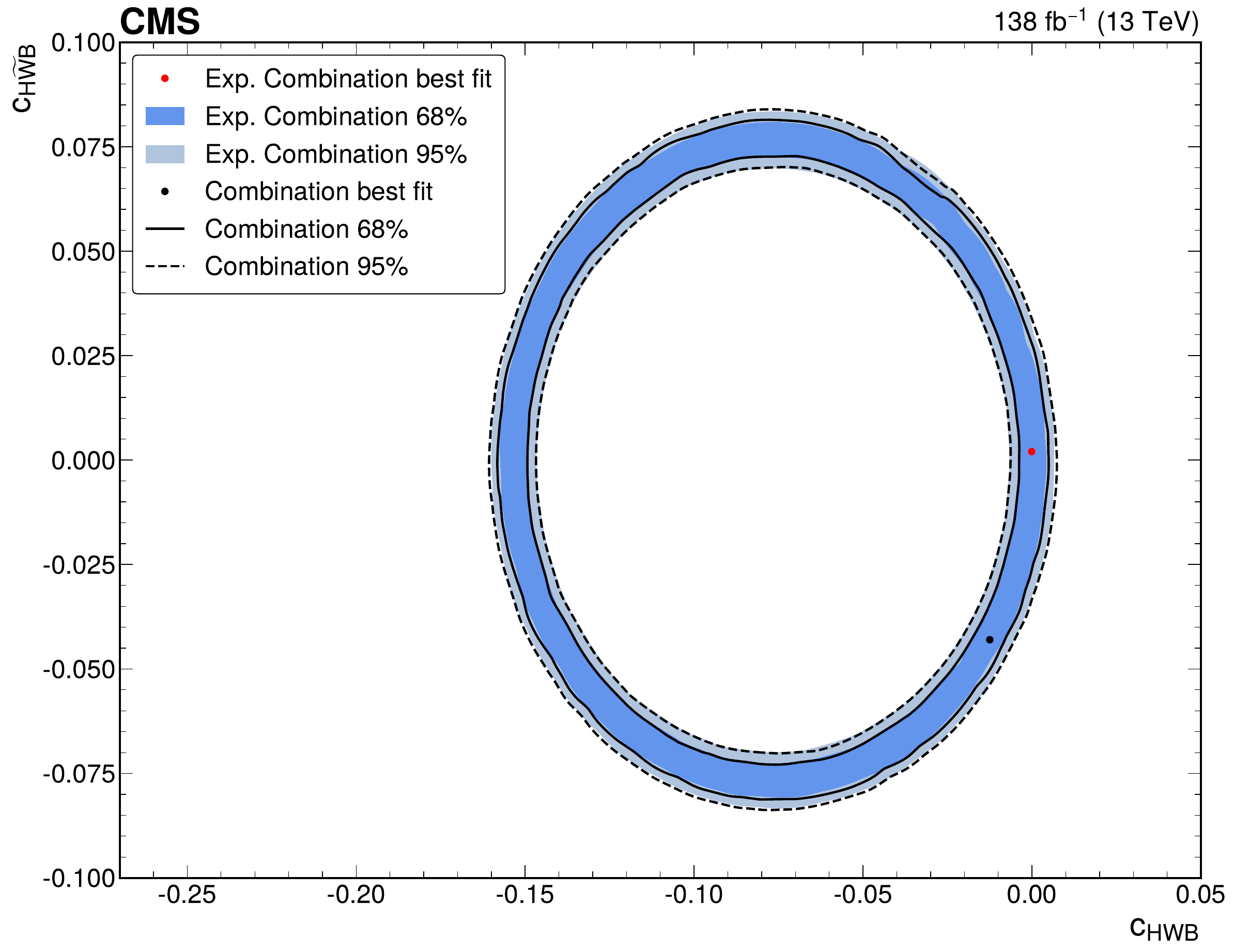}
    \caption{Observed and expected two-dimensional scans for the \chg-\chgt\ (upper left), \chb-\chbt\ (upper right), \chw-\chwt\ (lower left), and \chwb-\chwtb\ (lower right) pairs with \pth spectra in all decay channels.}
    \label{fig:pth_scans_2d_expbkg}
\end{figure}

Confidence regions are obtained for the following pairs:
\begin{itemize}
    \item \chg-\chgt;
    \item \chb-\chbt;
    \item \chw-\chwt;
    \item \chwb-\chwtb.
\end{itemize}
When a pair is studied, all the other WCs are set to their SM values (\ie, 0).
Two-dimensional constraints, obtained using Wilks theorem and by combining the \pth\ spectra of all the input analyses, are shown in Fig.~\ref{fig:pth_scans_2d_expbkg}. The results are consistent with the SM at the 68\% \CL.
Constraints in this configuration have also been set by the ATLAS Collaboration~\cite{hgg_atlas}, using only the \hgamgam\ decay channel. The contour plots presented in this paper are in agreement with the results obtained by the ATLAS Collaboration, but the constraints presented in this paper are tighter because of the use of a larger number of decay channels.

Constraints on the same coefficients are set using the \dpjj spectra of the \hgamgam\ and \hzz\ decay channels.
The results, shown in Fig.~\ref{fig:deltaphijj_scans_2d_expbkg} in Appendix \ref{app:smeft_deltaphijj_scans}, are consistent with the SM at the 68\% \CL and provide less stringent constraints than the ones obtained using the \pth spectra.
However, since the observable is CP-even, it can only be used to constrain CP-odd operators in conjunction with CP-even operators, and not to determine if the Wilson coefficients of CP-odd operators are nonzero.

\subsection{Constraints on linear combinations of Wilson coefficients}
\label{subsec:pca}

The available data do not contain enough information to simultaneously constrain all coefficients $c_i$.
Many degrees of freedom are left unconstrained by the data, manifesting as flat directions of the likelihood in the parameter space.
One way to get insights into the values of the WCs is to use techniques such as principal component analysis.
Performing an eigenvector decomposition of the Fisher information matrix provides linear combinations of the original coefficients $c_i$ with an indication of their constraining power (the eigenvalues): the ones with the largest constraining power are left floating in the fit, while the remaining directions are fixed to their SM value (\ie, 0).

In the case of a single measurement, taking as example \hgamgam, the PCA is performed as follows. Under the Gaussian approximation, the following equality holds:
\begin{equation}
    \mathcal{I}_{\gamma\gamma, \text{diff}} = \mathcal{H}_{\gamma\gamma, \text{diff}} = \mathrel{C}^{-1}_{\gamma\gamma, \text{diff}},
    \label{eq:fisher_hessian_one_measurement}
\end{equation}
where, referring to the \hgamgam differential cross section measurement, $\mathcal{I}_{\gamma\gamma, \text{diff}}$ and $\mathcal{H}_{\gamma\gamma, \text{diff}}$ are the Fisher information matrix and the Hessian of the likelihood built with parameters of interest and nuisances of the analysis parametrized with the cross section modifiers $\mu_i$, while $\mathrel{C}^{-1}_{\gamma\gamma, \text{diff}}$ is the inverse of the covariance matrix between the parameters of interest.
To obtain $\mathrel{C}^{-1}_{\gamma\gamma, \text{diff}}$, a fit to the Asimov data set~\cite{Cowan:2010js} is performed in the original analysis.
To move to a WC basis, one can build a matrix $P^{\gamma\gamma}$ by expanding Eq.~\eqref{eq:full_scaling_function} to only include terms linear in the WCs:
\begin{equation}
    P_{ij}^{\PGg\PGg}=A_{ij}^{\Pg\Pg \to \PH}+A_{j}^{\PH \to \PGg\PGg}-A_j^{H},
    \label{eq:p_matrix}
\end{equation}
where the index $i$ runs over the generator level bins in \hgamgam, the index $j$ runs over the WCs, and $A_j^{\PH}$ refers to the linear term for the total Higgs boson decay width. The inverse of the covariance matrix in the new basis is then obtained through:
\begin{equation}
    \mathrel{C}^{-1}_{\gamma\gamma, \mathrm{SMEFT}} = \mathrel{P^{\gamma\gamma T}} \mathrel{C}^{-1}_{\gamma\gamma, \text{diff}} \mathrel{P^{\gamma\gamma}}.
    \label{eq:inverse_covariance_matrix}
\end{equation}
By performing the eigenvector decomposition of $\mathrel{C}^{-1}_{\gamma\gamma, \mathrm{SMEFT}}$ one can obtain a matrix $EV_{\gamma\gamma}$ whose columns are the eigenvectors of $\mathrel{C}^{-1}_{\gamma\gamma, \mathrm{SMEFT}}$ and a diagonal matrix $\Lambda_{\gamma\gamma}$ whose elements are the eigenvalues of $\mathrel{C}^{-1}_{\gamma\gamma, \mathrm{SMEFT}}$ so that:
\begin{equation}
    \mathrel{C}^{-1}_{\gamma\gamma, \mathrm{SMEFT}} = (EV_{\gamma\gamma})\Lambda_{\gamma\gamma} (EV_{\gamma\gamma})^{-1}.
    \label{eq:eigenvector_decomposition}
\end{equation}
The coefficients in $(EV)^{-1}$ are used to write linear combinations of the WCs. 
The last step consists of rewriting the scaling equations in the newly defined linear combinations.
From the description above and in particular from Eq.~\eqref{eq:p_matrix}, one can see that only linear terms (up to $1/\Lambda^2$) are used to build the new basis.

The procedure to perform a PCA in the case of a combination of a set of differential cross section measurements follows the same steps as in the case of a single measurement, with two main differences:
\begin{itemize}
    \item the Fisher information matrix $\mathcal{I}_{\text{diff}}$ is built as a block-diagonal matrix by concatenating the matrices from individual measurements, since no correlation is assumed between the measurements;
    \item the new $P$ matrix is built by concatenating the $P$ matrices from individual measurements along the rows, since each bin at generator level is considered independently and the number of WCs (\ie, the number of columns of $P$) remains the same.
\end{itemize}

\begin{figure}[!htb]
    \centering
    \includegraphics[width=0.9\textwidth]{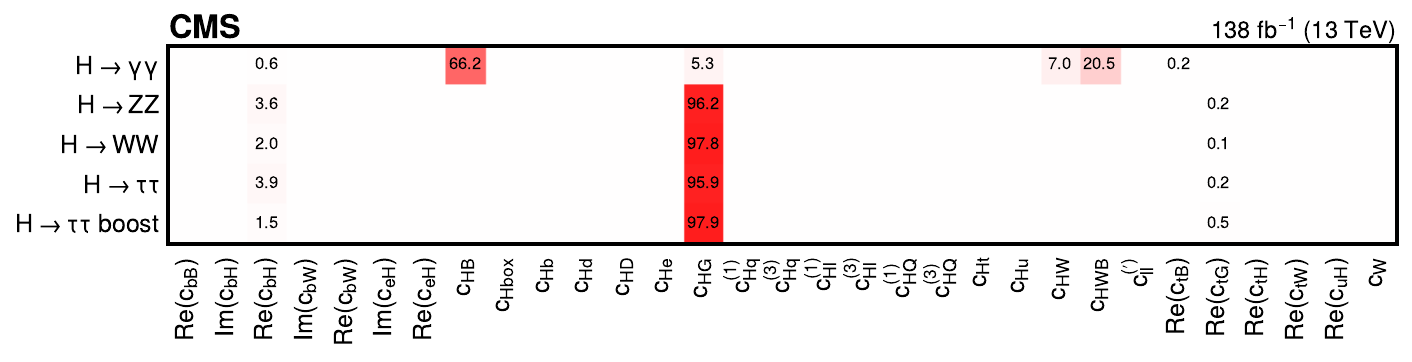}
    \caption{The values of the diagonal entries of the Fisher information matrix, presented as rows, for each decay channel. The normalization is such that the sum of the entries associated with each decay channel is equal to 100.}
    \label{fig:fisher}
\end{figure}

Figure~\ref{fig:fisher} displays the values of the diagonal entries of the Fisher information matrix, separated by decay channel.
The normalization is such that the sum of the entries associated with each decay channel is equal to 100.
In this way, it is possible to see, within each decay channel, which WCs are the most constrained (the higher the value, the more constrained the coefficient is by the data).
In all the decay channels except \hgamgam, the sensitivity is dominated by \chg: this is because, in these cases, the correction in production is two orders of magnitude larger than the correction in the decay.
This, combined with the fact that differential fiducial cross section measurements are mostly sensitive to ggH production, leads to these channels being sensitive almost exclusively to \chg.
The case of \hgamgam\ is different: the production and decay contributions are comparable (both at one SM loop), hence the sensitivity is also high for \chb, \chw, and \chwb, which affect the decay.
To justify the proportions between \chg, \chb, \chw, and \chwb, as defined in SMEFTsim3.0, one has to consider the linear terms $A$ entering the scaling equations, since these terms are used to build the rotation matrix $P$.
The linear terms for \chb, \chw, and \chwb in the decay scaling equation are higher than the ones for \chg in the production scaling formula, leading to the observed proportions. This comes from the fact that the linear terms are proportional to the ratio between the interference term and the SM term, and the SM term contains $\alpha_{\text{S}}$ while the interference term contains $\alpha_{\text{EM}}$.

\begin{figure}[!htb]
    \centering
    \includegraphics[width=0.99\textwidth]{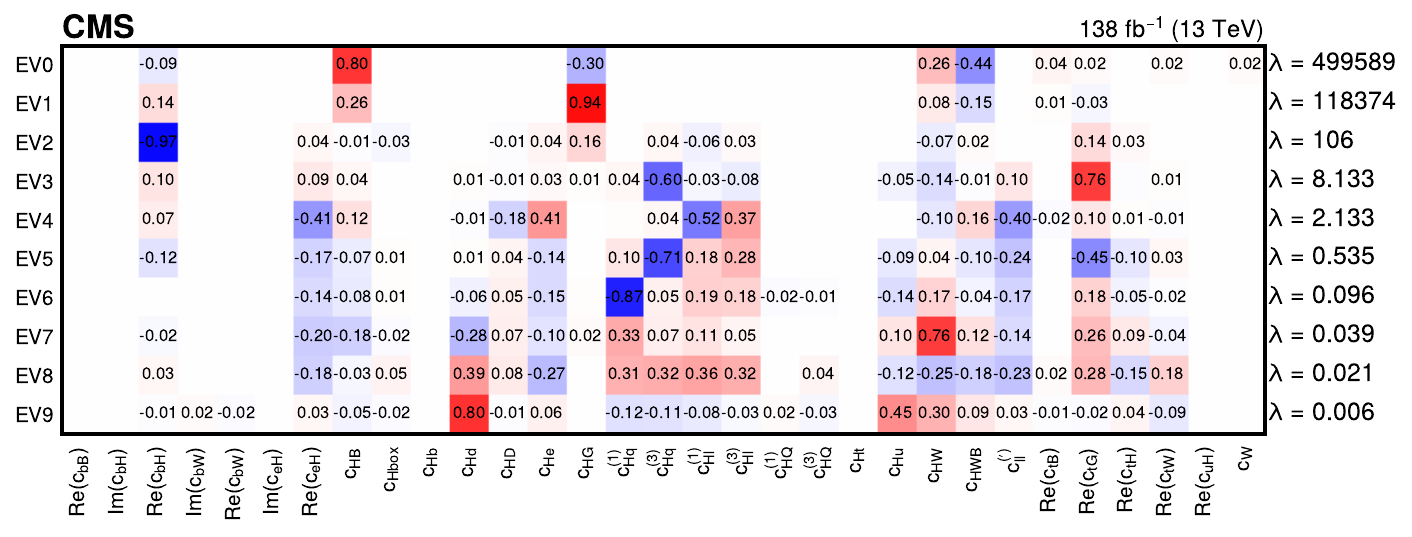}
    \caption{Graphical representation of the ten eigenvectors with the highest eigenvalues $\lambda$ of the expected combined Fisher information matrix in the SMEFT basis. Values lower than $10^{-3}$ are not shown. The intensity of the color represents the absolute value of the coefficient, going from $-1$ (blue) to $1$ (red).}
    \label{fig:pca_result}
\end{figure}

The WCs chosen as input to the PCA are selected from the initial set used to derive the parametrizations, based on a threshold on the $A$ coefficients.
The selected subset is reported in Table~\ref{tab:wilson_coeffs}.
The rotation matrix $EV^{-1}$, shown in Fig.~\ref{fig:pca_result}, is then used to derive linear combinations of the WCs.
The absolute values of the coefficients shown in Fig.~\ref{fig:pca_result} provide an idea of the weight that each WC has in the linear combination.
The larger the weight of a WC in a linear combination with large eigenvalues, the more constrained it is by the data.
As an example, one can see that the first two linear combinations are dominated by \chg, \chb, \chw, and \chwb, which are the most constrained WCs in the analysis and also dominate in Fig.~\ref{fig:fisher}.

\renewcommand{\arraystretch}{1.}
\begin{table}[!htb]
   \centering
   \topcaption{Wilson coefficients used as input to the SMEFT interpretation (right column). In the left and center columns the class they belong to and the corresponding operator are reported. The notation used is based on Ref.~\cite{smeftsim}.}
   \begin{tabular}{l c c}
    Class & Operator & Wilson coefficient \\
    \hline
    $\mathcal{L}_6^{(1)}-X^3$ & $\varepsilon^{i j k} W_\mu^{i \nu} W_\nu^{j \rho} W_\rho^{k \mu}$ & $c_{\mathrm{W}}$ \\
    \multirow{2}{*}{$\mathcal{L}_6^{(3)}-H^4 D^2$} & $(D^\mu H^{\dagger} H)(H^{\dagger} D_\mu H)$ & $c_{\mathrm{HD}}$ \\
    & $(H^{\dagger} H)\Box(H^{\dagger} H)$ & $c_{\mathrm{H\Box}}$ \\
    \multirow{4}{*}{$\mathcal{L}_6^{(4)}-X^2 H^2$} & $H^{\dagger} H G_{\mu \nu}^a G^{a \mu \nu}$ & $c_{\mathrm{HG}}$ \\
    & $H^{\dagger} H B_{\mu \nu} B^{\mu \nu}$ & $c_{\mathrm{HB}}$ \\
    & $H^{\dagger} H W_{\mu \nu}^i W^{i \mu \nu}$ & $c_{\mathrm{HW}}$ \\
    & $H^{\dagger} \sigma^i H W_{\mu \nu}^i B^{i \mu \nu}$ & $c_{\mathrm{HWB}}$ \\
    \multirow{7}{*}{$\mathcal{L}_6^{(5)}-\psi^2 H^3$} & \multirow{2}{*}{$(H^{\dagger}H)(\bar{Q} H b)$} & $\mathrm{Re}(c_{\mathrm{bH}})$ \\
    & & $\mathrm{Im}(c_{\mathrm{bH}})$ \\
    & $(H^{\dagger}H)(\bar{Q} H t)$ & $\mathrm{Re}(c_{\mathrm{tH}})$ \\
    & \multirow{2}{*}{$(H^{\dagger}H)(\bar{l}_p e_r H)$} & $\mathrm{Re}(c_{\mathrm{eH}})$ \\
    & & $\mathrm{Im}(c_{\mathrm{eH}})$ \\
    & $(H^{\dagger}H)(\bar{q} Y^{\dagger}_u u \tilde{H})$ & $\mathrm{Re}(c_{\mathrm{uH}})$ \\
    \multirow{6}{*}{$\mathcal{L}_6^{(6)}-\psi^2 X H$} & $(\bar{Q} \sigma^{\mu \nu} T^a t) \tilde{H} G_{\mu \nu}^a$ & $\mathrm{Re}(c_{\mathrm{tG}})$ \\
    & $(\bar{Q} \sigma^{\mu\nu} b) H B_{\mu \nu}$ & $\mathrm{Re}(c_{\mathrm{bB}})$ \\
    & $(\bar{Q} \sigma^{\mu\nu} t) H B_{\mu \nu}$ & $\mathrm{Re}(c_{\mathrm{tB}})$ \\
    & \multirow{2}{*}{$(\bar{Q} \sigma^{\mu\nu} b) \sigma^i H W_{\mu \nu}^i$} & $\mathrm{Re}(c_{\mathrm{bW}})$ \\
    & & $\mathrm{Im}(c_{\mathrm{bW}})$ \\
    & $(\bar{Q} \sigma^{\mu\nu} t) \sigma^i \tilde{H} W_{\mu \nu}^i$ & $\mathrm{Re}(c_{\mathrm{tW}})$ \\
    \multirow{11}{*}{$\mathcal{L}_6^{(7)}-\psi^2 H^2 D$} & $(H^{\dagger} i \overleftrightarrow{D}_\mu H)(\bar{l}_p \gamma^\mu l_r)$ & $c_{\mathrm{Hl}}^{\mathrm{(1)}}$ \\
    & $(H^{\dagger} i \overleftrightarrow{D}_\mu^i H)(\bar{l}_p \sigma^i \gamma^\mu l_r)$ & $c_{\mathrm{Hl}}^{\mathrm{(3)}}$ \\
    & $(H^{\dagger} i \overleftrightarrow{D}_\mu H)(\bar{q}_p \gamma^\mu q_r)$ & $c_{\mathrm{Hq}}^{\mathrm{(1)}}$ \\
    & $(H^{\dagger} i \overleftrightarrow{D}_\mu^i H)(\bar{q}_p \sigma^i \gamma^\mu q_r)$ & $c_{\mathrm{Hq}}^{\mathrm{(3)}}$ \\
    & $(H^{\dagger} i \overleftrightarrow{D}_\mu H)(\bar{Q}_p \gamma^\mu Q_r)$ & $c_{\mathrm{HQ}}^{\mathrm{(1)}}$ \\
    & $(H^{\dagger} i \overleftrightarrow{D}_\mu^i H)(\bar{Q}_p \sigma^i \gamma^\mu Q_r)$ & $c_{\mathrm{HQ}}^{\mathrm{(3)}}$ \\
    & $(H^{\dagger} i \overleftrightarrow{D}_\mu H)(\bar{u}_p \gamma^\mu u_r)$ & $c_{\mathrm{Hu}}$ \\
    & $(H^{\dagger} i \overleftrightarrow{D}_\mu H)(\bar{d}_p \gamma^\mu d_r)$ & $c_{\mathrm{Hd}}$ \\
    & $(H^{\dagger} i \overleftrightarrow{D}_\mu H)(\bar{e}_p \gamma^\mu e_r)$ & $c_{\mathrm{He}}$ \\
    & $(H^{\dagger} i \overleftrightarrow{D}_\mu H)(\bar{b} \gamma^\mu b)$ & $c_{\mathrm{Hb}}$ \\
    & $(H^{\dagger} i \overleftrightarrow{D}_\mu H)(\bar{t} \gamma^\mu t)$ & $c_{\mathrm{Ht}}$ \\
    $\mathcal{L}_6^{(8a)}-(\bar{L}L)(\bar{L}L)$ & $ (\bar{l}_p \gamma_{\mu} l_r)(\bar{l}_s \gamma^{\mu} l_t) $ & $c^{\prime}_{\mathrm{ll}}$ \\
   \end{tabular}
   \label{tab:wilson_coeffs}
\end{table}

\begin{figure}[!htb]
    \centering
    \includegraphics[width=0.5\textwidth]{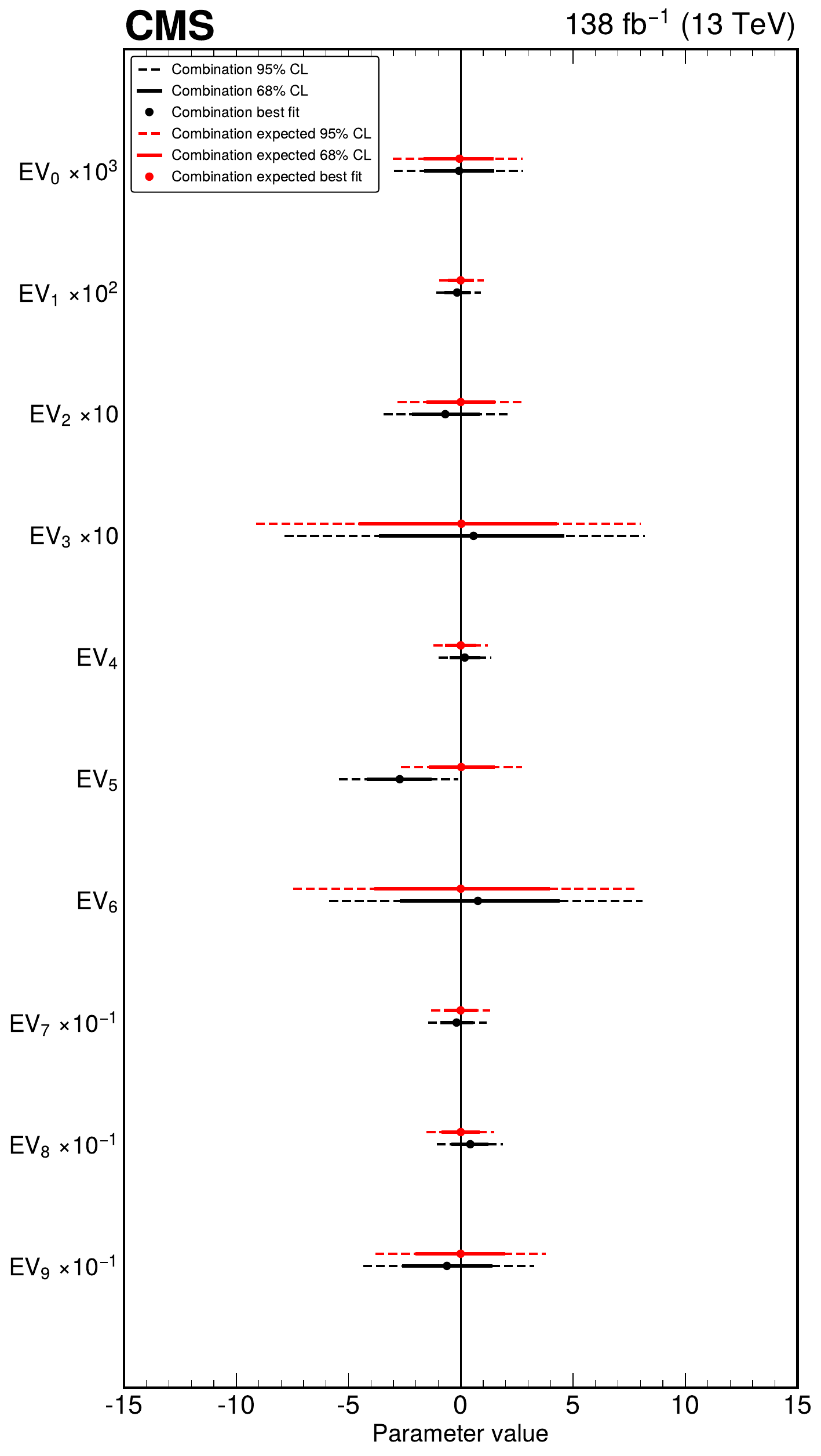}
    \caption{Summary of observed and expected confidence intervals at 68\% and 95\% confidence level for the first ten eigenvectors. On the y-axis, the quantity being displayed is multiplied by the corresponding power of ten. The eigenvectors are ordered by decreasing eigenvalue.}
    \label{fig:fits_summary}
\end{figure}

Observed and expected results for the ten eigenvectors with the largest eigenvalues are shown in Fig.~\ref{fig:fits_summary}.
The corresponding one-dimensional scans, obtained profiling the other nine eigenvectors, are shown in Figs.~\ref{fig:fits_linear_1}, \ref{fig:fits_linear_2}, and \ref{fig:fits_linear_3} in Appendix \ref{app:smeft_eigenvectors}.
The results are consistent with the SM within two standard deviations.
The largest tension with the SM is observed in the sixth eigenvector, where the best fit value is $\mathrm{EV}_5 = 2.71^{+1.33}_{-1.39}$.
The p-value corresponding to this tension is 3.6\%.
As shown in Fig. \ref{fig:pca_result}, the WC with the highest weight in $\mathrm{EV}_5$ is $c_{\mathrm{Hq}}^{(3)}$.

\begin{figure}[!htb]
    \centering
    \includegraphics[width=0.7\textwidth]{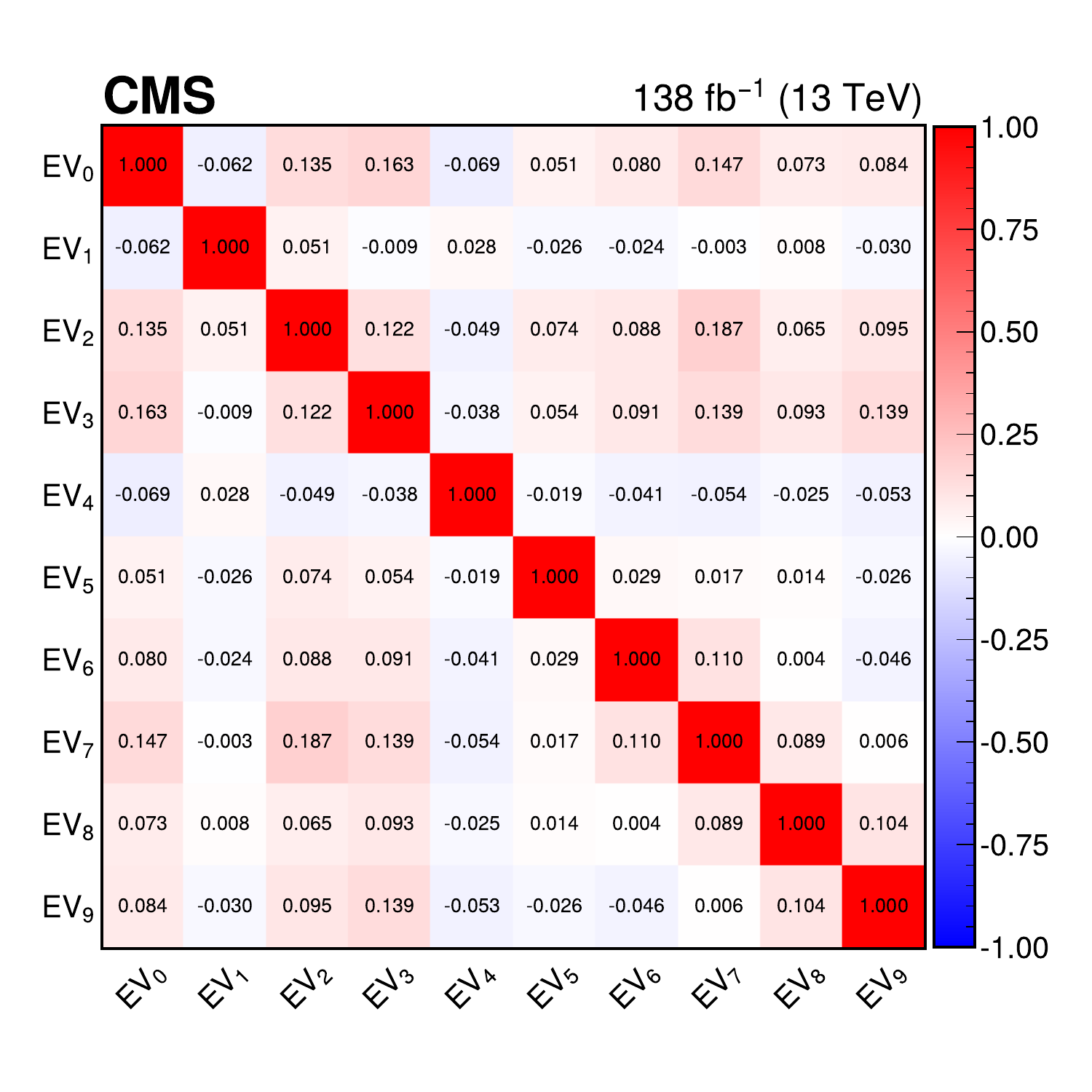}
    \caption{Correlation matrix of the linear combinations of Wilson coefficients obtained from the PCA, obtained by fitting the observed data to the \pth spectra of all decay channels.}
    \label{fig:fits_pca_corrmat}
\end{figure}

The correlation matrix of the linear combinations of WCs is shown in Fig.~\ref{fig:fits_pca_corrmat}. Some level of correlation is present between the eigenvectors (up to 18$\%$ in the worst cases).
This can be explained by the fact that the eigenvectors are obtained by diagonalizing the expected Fisher information matrix, and not the observed one, hence a perfect level of decorrelation is not to be expected.

\section{Summary}
\label{sec:conclusions}

Combined measurements of differential Higgs boson production cross sections for the observables \pth, \njets, \yh, \ptjz, \mjj, \deta, and \taujc are presented, using proton-proton collision data collected at \com and corresponding to an integrated luminosity of $138\fbinv$.
The spectra are obtained with data from the \hgamgam, \hzz, \hww, and \htt\ (both in the small and large Lorentz-boost regimes) decay channels.
The precision of the combined measurement of the \pth differential cross section is improved by about 23\% with respect to the \hgamgam\ channel alone.
The improvement is particularly significant in the low- and high-\pth regions.
No significant deviations from the SM predictions are observed in the differential distributions.
Additionally, the total cross section for Higgs boson production based on a combination of the \hgamgam\ and \hzz\ channels is measured to be $53.4^{+2.9}_{-2.9}\stat^{+1.9}_{-1.8}\syst$\unit{pb}, consistent with the SM prediction.

The obtained \pth spectra are interpreted using the $\kappa$ and SM effective field theory frameworks.
In the former, multiple couplings are varied using the models provided in Refs. \cite{kbkc,grazzini_1,grazzini_2}.
In the latter, two-dimensional constraints are obtained for pairs of Wilson coefficients.
A principal component analysis is then performed to identify nonflat directions of the likelihood.
The studies performed in this context highlight that the differential fiducial cross section measurements are sensitive to a limited set of operators and related Wilson coefficients, with the most constrained ones being \chg, \chb, \chw, and \chwb.
No significant deviations from the SM are observed in the results obtained with either framework.

\begin{acknowledgments}
We congratulate our colleagues in the CERN accelerator departments for the excellent performance of the LHC and thank the technical and administrative staffs at CERN and at other CMS institutes for their contributions to the success of the CMS effort. In addition, we gratefully acknowledge the computing centers and personnel of the Worldwide LHC Computing Grid and other centers for delivering so effectively the computing infrastructure essential to our analyses. Finally, we acknowledge the enduring support for the construction and operation of the LHC, the CMS detector, and the supporting computing infrastructure provided by the following funding agencies: SC (Armenia), BMBWF and FWF (Austria); FNRS and FWO (Belgium); CNPq, CAPES, FAPERJ, FAPERGS, and FAPESP (Brazil); MES and BNSF (Bulgaria); CERN; CAS, MoST, and NSFC (China); MINCIENCIAS (Colombia); MSES and CSF (Croatia); RIF (Cyprus); SENESCYT (Ecuador); ERC PRG, RVTT3 and MoER TK202 (Estonia); Academy of Finland, MEC, and HIP (Finland); CEA and CNRS/IN2P3 (France); SRNSF (Georgia); BMBF, DFG, and HGF (Germany); GSRI (Greece); NKFIH (Hungary); DAE and DST (India); IPM (Iran); SFI (Ireland); INFN (Italy); MSIP and NRF (Republic of Korea); MES (Latvia); LMTLT (Lithuania); MOE and UM (Malaysia); BUAP, CINVESTAV, CONACYT, LNS, SEP, and UASLP-FAI (Mexico); MOS (Montenegro); MBIE (New Zealand); PAEC (Pakistan); MES and NSC (Poland); FCT (Portugal); MESTD (Serbia); MICIU/AEI and PCTI (Spain); MOSTR (Sri Lanka); Swiss Funding Agencies (Switzerland); MST (Taipei); MHESI and NSTDA (Thailand); TUBITAK and TENMAK (Turkey); NASU (Ukraine); STFC (United Kingdom); DOE and NSF (USA).
   
\hyphenation{Rachada-pisek} Individuals have received support from the Marie-Curie program and the European Research Council and Horizon 2020 Grant, contract Nos.\ 675440, 724704, 752730, 758316, 765710, 824093, 101115353, 101002207, and COST Action CA16108 (European Union); the Leventis Foundation; the Alfred P.\ Sloan Foundation; the Alexander von Humboldt Foundation; the Science Committee, project no. 22rl-037 (Armenia); the Fonds pour la Formation \`a la Recherche dans l'Industrie et dans l'Agriculture (FRIA-Belgium); the Beijing Municipal Science \& Technology Commission, No. Z191100007219010 and Fundamental Research Funds for the Central Universities (China); the Ministry of Education, Youth and Sports (MEYS) of the Czech Republic; the Shota Rustaveli National Science Foundation, grant FR-22-985 (Georgia); the Deutsche Forschungsgemeinschaft (DFG), among others, under Germany's Excellence Strategy -- EXC 2121 ``Quantum Universe" -- 390833306, and under project number 400140256 - GRK2497; the Hellenic Foundation for Research and Innovation (HFRI), Project Number 2288 (Greece); the Hungarian Academy of Sciences, the New National Excellence Program - \'UNKP, the NKFIH research grants K 131991, K 133046, K 138136, K 143460, K 143477, K 146913, K 146914, K 147048, 2020-2.2.1-ED-2021-00181, TKP2021-NKTA-64, and 2021-4.1.2-NEMZ\_KI-2024-00036 (Hungary); the Council of Science and Industrial Research, India; ICSC -- National Research Center for High Performance Computing, Big Data and Quantum Computing and FAIR -- Future Artificial Intelligence Research, funded by the NextGenerationEU program (Italy); the Latvian Council of Science; the Ministry of Education and Science, project no. 2022/WK/14, and the National Science Center, contracts Opus 2021/41/B/ST2/01369 and 2021/43/B/ST2/01552 (Poland); the Funda\c{c}\~ao para a Ci\^encia e a Tecnologia, grant CEECIND/01334/2018 (Portugal); the National Priorities Research Program by Qatar National Research Fund; MICIU/AEI/10.13039/501100011033, ERDF/EU, "European Union NextGenerationEU/PRTR", and Programa Severo Ochoa del Principado de Asturias (Spain); the Chulalongkorn Academic into Its 2nd Century Project Advancement Project, and the National Science, Research and Innovation Fund via the Program Management Unit for Human Resources \& Institutional Development, Research and Innovation, grant B39G670016 (Thailand); the Kavli Foundation; the Nvidia Corporation; the SuperMicro Corporation; the Welch Foundation, contract C-1845; and the Weston Havens Foundation (USA).
\end{acknowledgments}
\bibliography{auto_generated} 

\appendix

\numberwithin{figure}{section}
\numberwithin{table}{section}

\section{Tables for the differential cross section measurements}
\label{app:tables_spectra}
Tables~\ref{tab:observed_smH_PTH_xsvalues}--\ref{tab:observed_TauCJ_xsvalues} show the measured combined differential cross sections for the considered observables.

\begin{table}[!ht]
    \centering
    \topcaption{Observed best fit differential cross section for the \pth (\GeVns{}) observable. For its calculation, the last bin is assumed to have a bin width equal to that of the last but one bin.}
    \begin{tabular}{lc}
    \pth (\GeVns{})&Best fit (fb/\GeVns{})\\
    \hline
    $\sigma_{0 - 5}$&$2.86^{+0.49}_{-0.58}\syst^{+1.75}_{-1.95}\stat \times 10^{2}$\\
    $\sigma_{5 - 10}$&$8.73^{+0.89}_{-0.54}\syst^{+2.90}_{-3.13}\stat \times 10^{2}$\\
    $\sigma_{10 - 15}$&$1.28^{+0.08}_{-0.08}\syst^{+0.26}_{-0.25}\stat \times 10^{3}$\\
    $\sigma_{15 - 20}$&$1.12^{+0.06}_{-0.08}\syst^{+0.26}_{-0.24}\stat \times 10^{3}$\\
    $\sigma_{20 - 25}$&$4.16^{+0.56}_{-0.00}\syst^{+2.14}_{-2.19}\stat \times 10^{2}$\\
    $\sigma_{25 - 30}$&$8.13^{+0.25}_{-0.43}\syst^{+2.14}_{-2.11}\stat \times 10^{2}$\\
    $\sigma_{30 - 35}$&$5.14^{+0.52}_{-0.32}\syst^{+1.74}_{-1.68}\stat \times 10^{2}$\\
    $\sigma_{35 - 45}$&$5.85^{+0.23}_{-0.24}\syst^{+1.28}_{-1.29}\stat \times 10^{2}$\\
    $\sigma_{45 - 60}$&$2.71^{+0.26}_{-0.16}\syst^{+0.62}_{-0.59}\stat \times 10^{2}$\\
    $\sigma_{60 - 80}$&$2.88^{+0.17}_{-0.13}\syst^{+0.46}_{-0.45}\stat \times 10^{2}$\\
    $\sigma_{80 - 100}$&$2.37^{+0.18}_{-0.14}\syst^{+0.36}_{-0.35}\stat \times 10^{2}$\\
    $\sigma_{100 - 120}$&$6.16^{+0.00}_{-0.83}\syst^{+2.97}_{-2.65}\stat \times 10^{1}$\\
    $\sigma_{120 - 140}$&$9.07^{+0.86}_{-0.66}\syst^{+1.73}_{-1.70}\stat \times 10^{1}$\\
    $\sigma_{140 - 170}$&$5.31^{+0.50}_{-0.37}\syst^{+1.08}_{-1.06}\stat \times 10^{1}$\\
    $\sigma_{170 - 200}$&$1.39^{+0.22}_{-0.15}\syst^{+0.65}_{-0.63}\stat \times 10^{1}$\\
    $\sigma_{200 - 250}$&$1.47^{+0.22}_{-0.18}\syst^{+0.25}_{-0.24}\stat \times 10^{1}$\\
    $\sigma_{250 - 350}$&$4.28^{+0.59}_{-0.43}\syst^{+1.00}_{-0.97}\stat \times 10^{0}$\\
    $\sigma_{350 - 450}$&$9.67^{+2.09}_{-1.49}\syst^{+3.27}_{-3.08}\stat \times 10^{-1}$\\
    $\sigma_{> 450}$&$4.37^{+1.19}_{-0.80}\syst^{+1.77}_{-1.66}\stat \times 10^{-1}$\\
    \end{tabular}%
    \label{tab:observed_smH_PTH_xsvalues}
    \end{table}

    \begin{table}[!ht]
    \centering
    \topcaption{Observed best fit differential cross section for the \njets observable.}
    \begin{tabular}{lc}
    \njets&Best fit (fb)\\
    \hline
    $\sigma_{0}$&$3.13^{+0.17}_{-0.16}\syst^{+0.17}_{-0.17}\stat \times 10^{4}$\\
    $\sigma_{1}$&$1.43^{+0.10}_{-0.09}\syst^{+0.13}_{-0.13}\stat \times 10^{4}$\\
    $\sigma_{2}$&$5.09^{+0.47}_{-0.44}\syst^{+0.56}_{-0.56}\stat \times 10^{3}$\\
    $\sigma_{3}$&$2.77^{+1.86}_{-1.56}\syst^{+2.91}_{-2.83}\stat \times 10^{2}$\\
    $\sigma_{>=4}$&$6.10^{+1.24}_{-1.00}\syst^{+1.82}_{-1.82}\stat \times 10^{2}$\\
    \end{tabular}
    \label{tab:observed_Njets_xsvalues}
    \end{table}

    \begin{table}[!ht]
    \centering
    \caption{Observed best fit differential cross section for the \ptjz (\GeVns{}) observable. For its calculation, the last bin is assumed to have a bin width equal to that of the last but one bin.}
    \begin{tabular}{lc}
    \ptjz (\GeVns{})&Best fit (fb/\GeVns{})\\
    \hline
    $\sigma_{0 - 30}$&$1.09^{+0.09}_{-0.07}\syst^{+0.09}_{-0.09}\stat \times 10^{3}$\\
    $\sigma_{30 - 40}$&$4.81^{+8.97}_{-8.46}\syst^{+28.07}_{-25.92}\stat \times 10^{1}$\\
    $\sigma_{40 - 55}$&$7.37^{+0.69}_{-0.74}\syst^{+2.13}_{-2.52}\stat \times 10^{2}$\\
    $\sigma_{55 - 75}$&$1.33^{+0.34}_{-0.32}\syst^{+1.06}_{-1.01}\stat \times 10^{2}$\\
    $\sigma_{75 - 95}$&$9.60^{+3.00}_{-2.91}\syst^{+8.96}_{-9.60}\stat \times 10^{1}$\\
    $\sigma_{95 - 120}$&$6.13^{+1.25}_{-1.03}\syst^{+4.63}_{-4.47}\stat \times 10^{1}$\\
    $\sigma_{120 - 150}$&$8.20^{+1.23}_{-0.83}\syst^{+3.58}_{-3.50}\stat \times 10^{1}$\\
    $\sigma_{150 - 200}$&$7.25^{+0.00}_{-37.54}\syst^{+118.71}_{-117.44}\stat \times 10^{-1}$\\
    $\sigma_{> 200}$&$2.41^{+0.23}_{-0.17}\syst^{+0.55}_{-0.53}\stat \times 10^{1}$\\
    \hline
    \end{tabular}
    \label{tab:observed_smH_PTJ0_xsvalues}
    \end{table}

    \begin{table}[!ht]
    \centering
    \topcaption{Observed best fit differential cross section for the \yh observable.}
    \begin{tabular}{lc}
    \yh &Best fit (fb)\\
    \hline
    $\sigma_{0 - 0.15}$&$2.82^{+0.09}_{-0.07}\syst^{+0.38}_{-0.37}\stat \times 10^{4}$\\
    $\sigma_{0.15 - 0.3}$&$2.81^{+0.09}_{-0.08}\syst^{+0.39}_{-0.38}\stat \times 10^{4}$\\
    $\sigma_{0.3 - 0.45}$&$2.84^{+0.08}_{-0.07}\syst^{+0.41}_{-0.40}\stat \times 10^{4}$\\
    $\sigma_{0.45 - 0.6}$&$2.57^{+0.10}_{-0.05}\syst^{+0.42}_{-0.40}\stat \times 10^{4}$\\
    $\sigma_{0.6 - 0.75}$&$1.95^{+0.06}_{-0.07}\syst^{+0.43}_{-0.39}\stat \times 10^{4}$\\
    $\sigma_{0.75 - 0.9}$&$2.53^{+0.09}_{-0.05}\syst^{+0.45}_{-0.43}\stat \times 10^{4}$\\
    $\sigma_{0.9 - 1.2}$&$1.84^{+0.09}_{-0.04}\syst^{+0.33}_{-0.32}\stat \times 10^{4}$\\
    $\sigma_{1.2 - 1.6}$&$1.61^{+0.07}_{-0.05}\syst^{+0.32}_{-0.35}\stat \times 10^{4}$\\
    $\sigma_{1.6 - 2.0}$&$1.51^{+0.09}_{-0.03}\syst^{+0.76}_{-0.50}\stat \times 10^{4}$\\
    $\sigma_{2.0 - 2.5}$&$2.89^{+0.93}_{-0.50}\syst^{+5.53}_{-5.46}\stat \times 10^{3}$\\
    \end{tabular}
    \label{tab:observed_yH_xsvalues}
    \end{table}

    \begin{table}[!ht]
    \centering
    \topcaption{Observed best fit differential cross section for the \deta observable. For its calculation, the last bin is assumed to have a bin width equal to that of the last but one bin.}
    \begin{tabular}{lc}
    \deta &Best fit (fb)\\
    \hline
    $\sigma_{0 - 0.7}$&$4.52^{+0.22}_{-0.16}\syst^{+1.23}_{-1.20}\stat \times 10^{3}$\\
    $\sigma_{0.7 - 1.6}$&$2.87^{+1.59}_{-0.89}\syst^{+9.95}_{-9.96}\stat \times 10^{2}$\\
    $\sigma_{1.6 - 3.0}$&$2.42^{+0.17}_{-0.09}\syst^{+0.61}_{-0.64}\stat \times 10^{3}$\\
    $\sigma_{3.0 - 5.0}$&$1.03^{+0.09}_{-0.08}\syst^{+0.32}_{-0.27}\stat \times 10^{3}$\\
    $\sigma_{> 5.0}$&$-1.61^{+0.43}_{-0.48}\syst^{+1.74}_{-1.62}\stat \times 10^{2}$\\
    \end{tabular}
    \label{tab:observed_DEtajj_xsvalues}
    \end{table}

    \begin{table}[!ht]
    \centering
    \topcaption{Observed best fit differential cross section for the \mjj (\GeVns{}) observable. For its calculation, the last bin is assumed to have a bin width equal to that of the last but one bin.}
    \begin{tabular}{lc}
    \mjj (\GeVns{})&Best fit (fb/\GeVns{})\\
    \hline
    $\sigma_{0 - 75}$&$4.92^{+1.52}_{-1.40}\syst^{+12.36}_{-12.41}\stat \times 10^{0}$\\
    $\sigma_{75 - 120}$&$5.65^{+0.63}_{-0.31}\syst^{+2.24}_{-2.18}\stat \times 10^{1}$\\
    $\sigma_{120 - 180}$&$2.33^{+0.71}_{-0.30}\syst^{+1.45}_{-1.39}\stat \times 10^{1}$\\
    $\sigma_{180 - 300}$&$1.36^{+0.40}_{-0.16}\syst^{+0.87}_{-0.85}\stat \times 10^{1}$\\
    $\sigma_{300 - 500}$&$6.56^{+0.60}_{-0.28}\syst^{+2.79}_{-2.59}\stat \times 10^{0}$\\
    $\sigma_{500 - 1000}$&$1.75^{+0.24}_{-0.23}\syst^{+1.09}_{-1.18}\stat \times 10^{0}$\\
    $\sigma_{> 1000}$&$9.04^{+1.25}_{-1.03}\syst^{+5.22}_{-4.86}\stat \times 10^{-1}$\\
    \end{tabular}
    \label{tab:observed_mjj_xsvalues}
    \end{table}

    \begin{table}[!ht]
    \centering
    \topcaption{Observed best fit differential cross section for the \taujc (\GeVns{}) observable. For its calculation, the last bin is assumed to have a bin width equal to that of the last but one bin.}
    \begin{tabular}{lc}
    \taujc (\GeVns{})&Best fit (fb/\GeVns{})\\
    \hline
    $\sigma_{15 - 20}$&$5.95^{+0.34}_{-0.27}\syst^{+3.51}_{-3.30}\stat \times 10^{2}$\\
    $\sigma_{20 - 30}$&$4.02^{+0.26}_{-0.15}\syst^{+1.57}_{-1.51}\stat \times 10^{2}$\\
    $\sigma_{30 - 50}$&$1.91^{+0.09}_{-0.05}\syst^{+0.54}_{-0.47}\stat \times 10^{2}$\\
    $\sigma_{50 - 80}$&$6.07^{+0.29}_{-0.20}\syst^{+1.76}_{-1.67}\stat \times 10^{1}$\\
    $\sigma_{> 80}$&$2.22^{+0.21}_{-0.12}\syst^{+0.90}_{-0.85}\stat \times 10^{1}$\\
    \end{tabular}
    \label{tab:observed_TauCJ_xsvalues}
    \end{table}

\clearpage
\section{\mbox{Correlation matrices for the combinations of differential observables}}
\label{app:correlation_matrices}

\begin{figure}[!ht]
    \centering
    \includegraphics[width=0.49\textwidth]{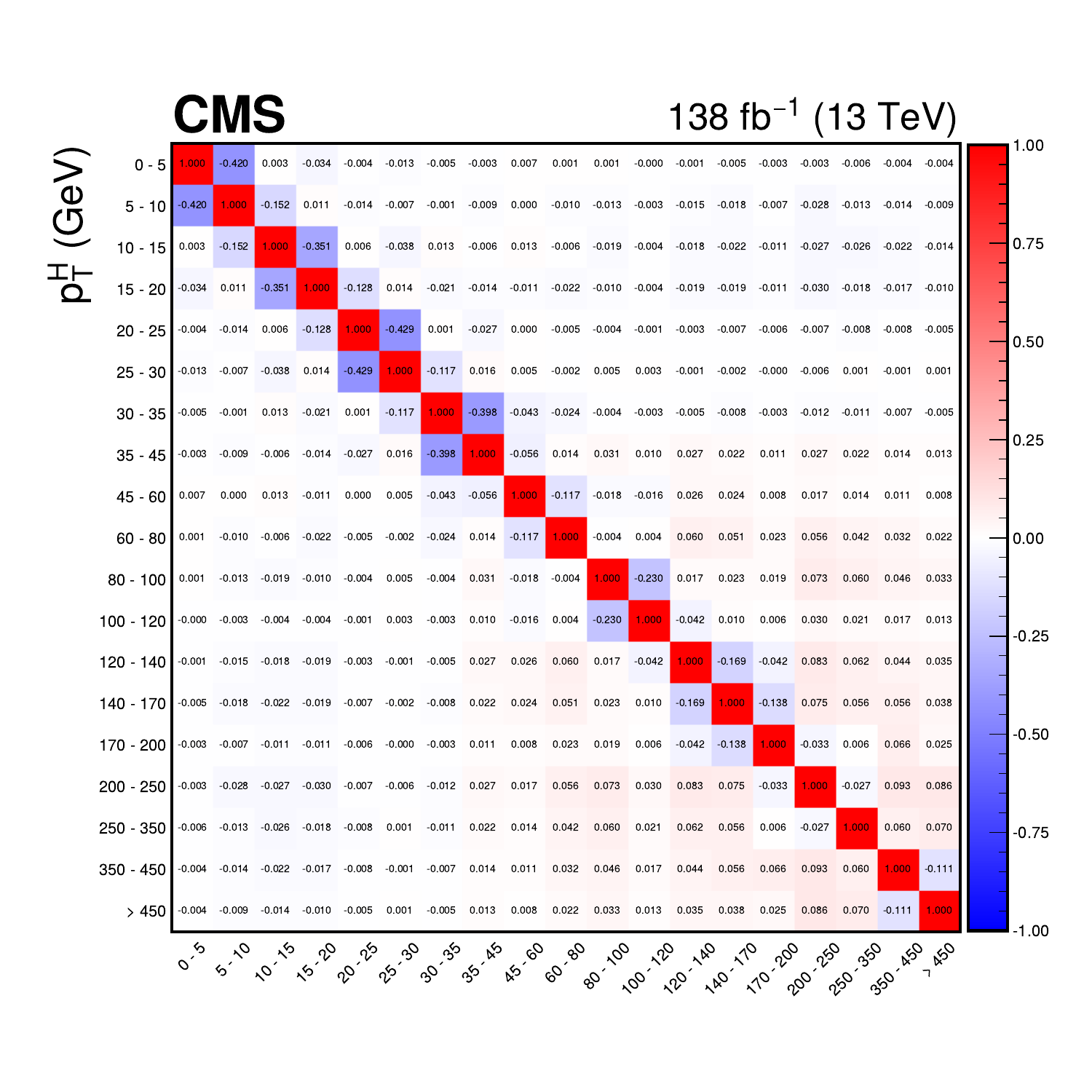}
    \includegraphics[width=0.49\textwidth]{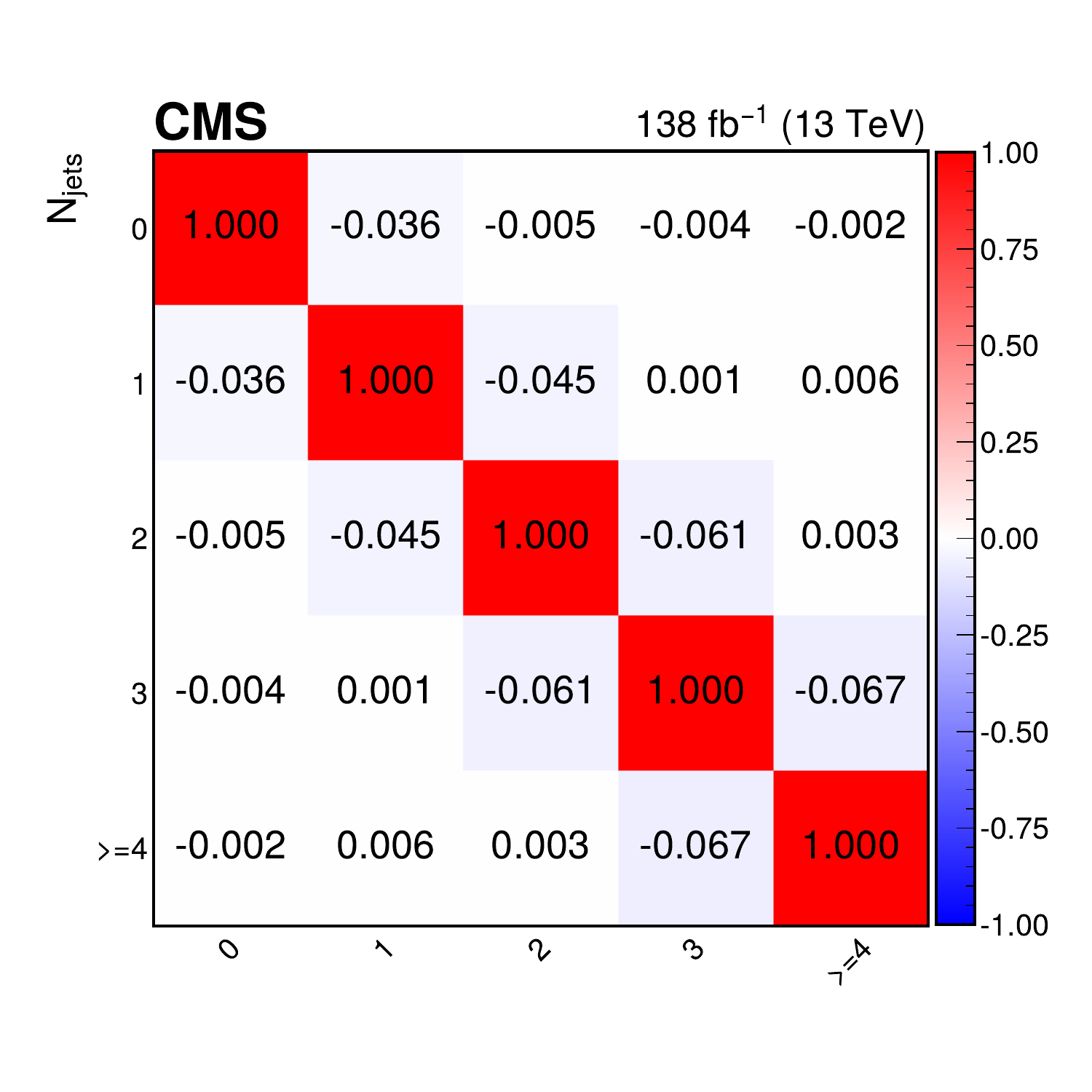} \\
    \includegraphics[width=0.49\textwidth]{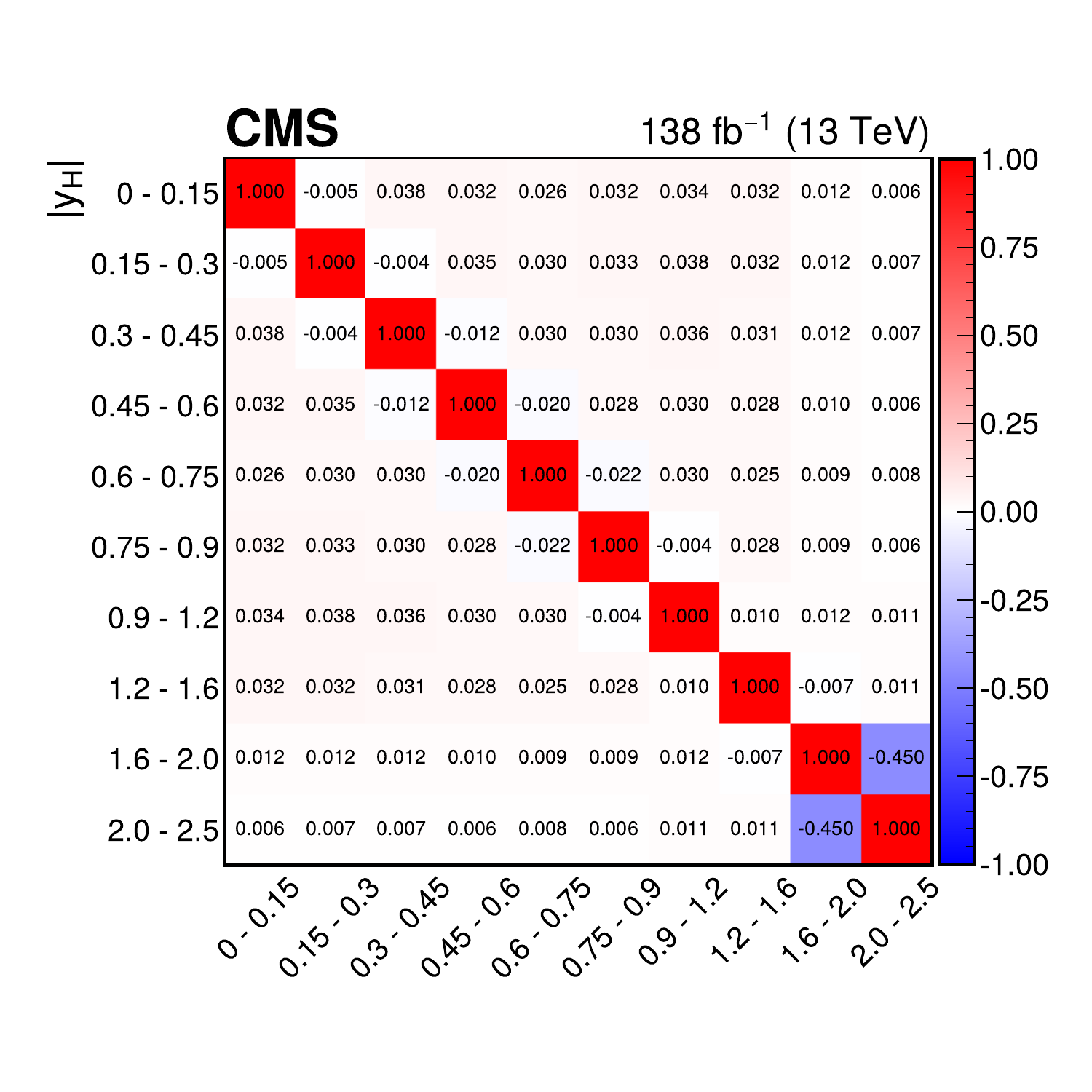}
    \includegraphics[width=0.49\textwidth]{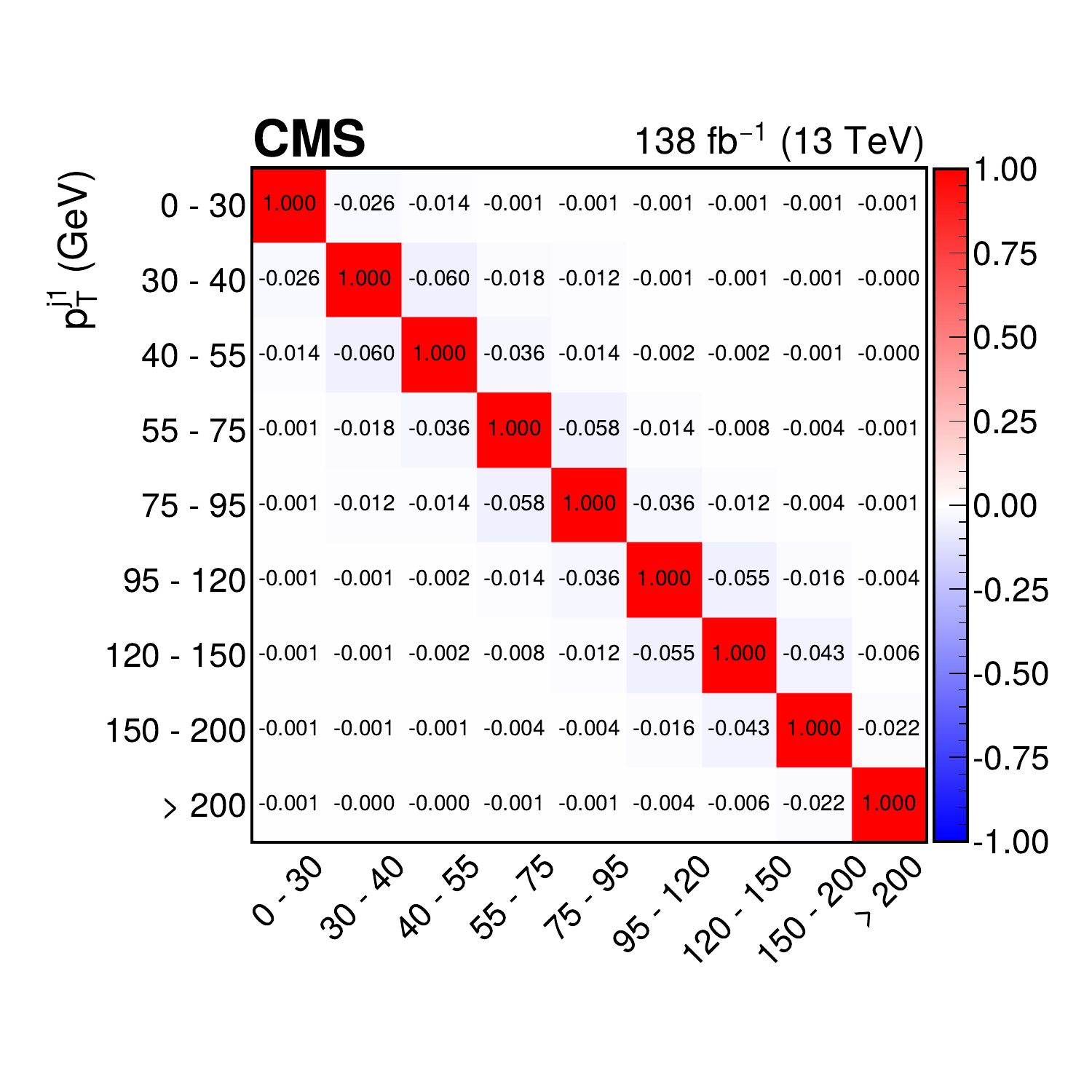}
    \caption{Bin-to-bin correlation matrices for the \pth (upper left), \njets\ (upper right), \yh\ (lower left), and \ptjz\ (lower right) spectra.}
    \label{fig:corr_matrices_1}
\end{figure}

\begin{figure}[!ht]
    \centering
    \includegraphics[width=0.49\textwidth]{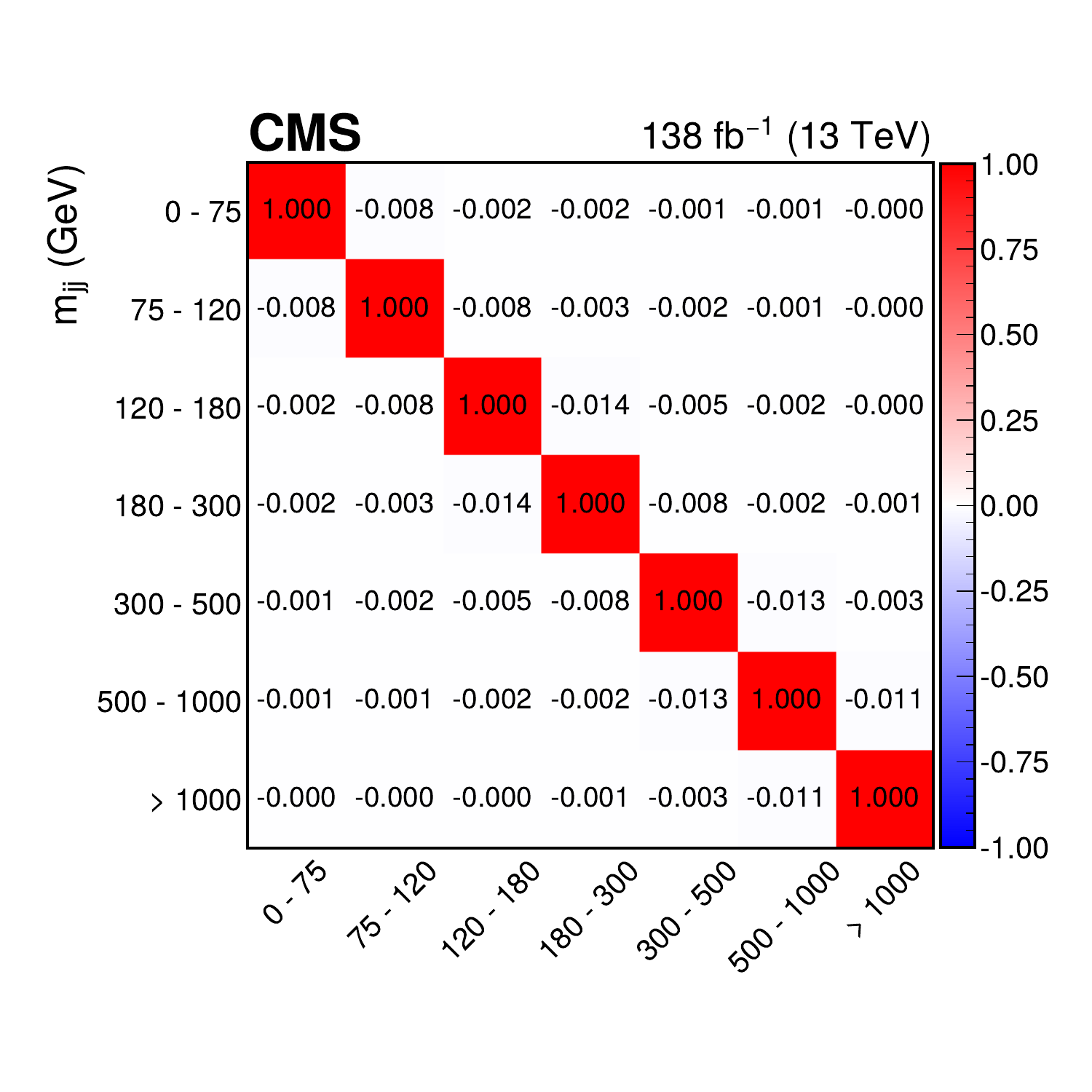}
    \includegraphics[width=0.49\textwidth]{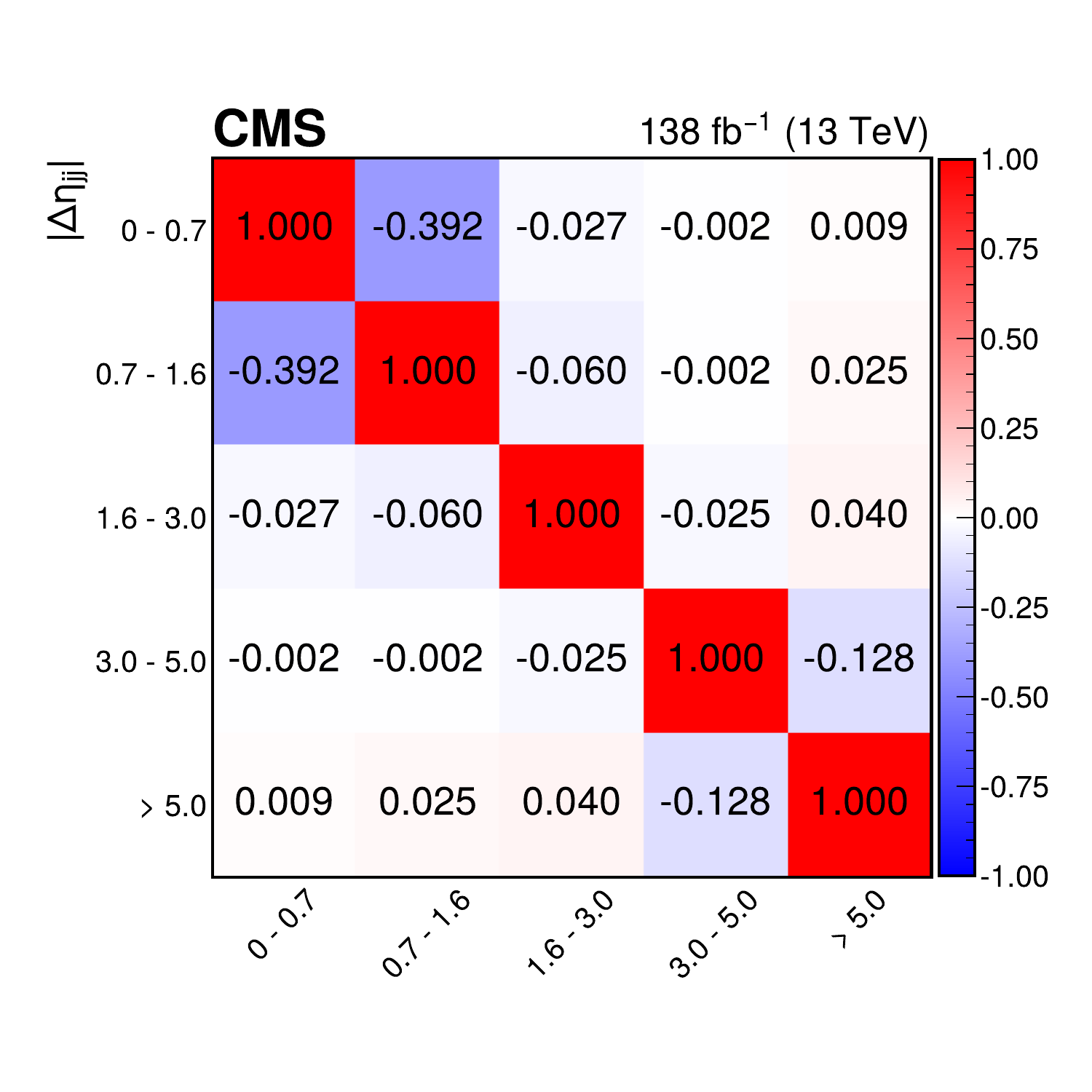} \\
    \includegraphics[width=0.49\textwidth]{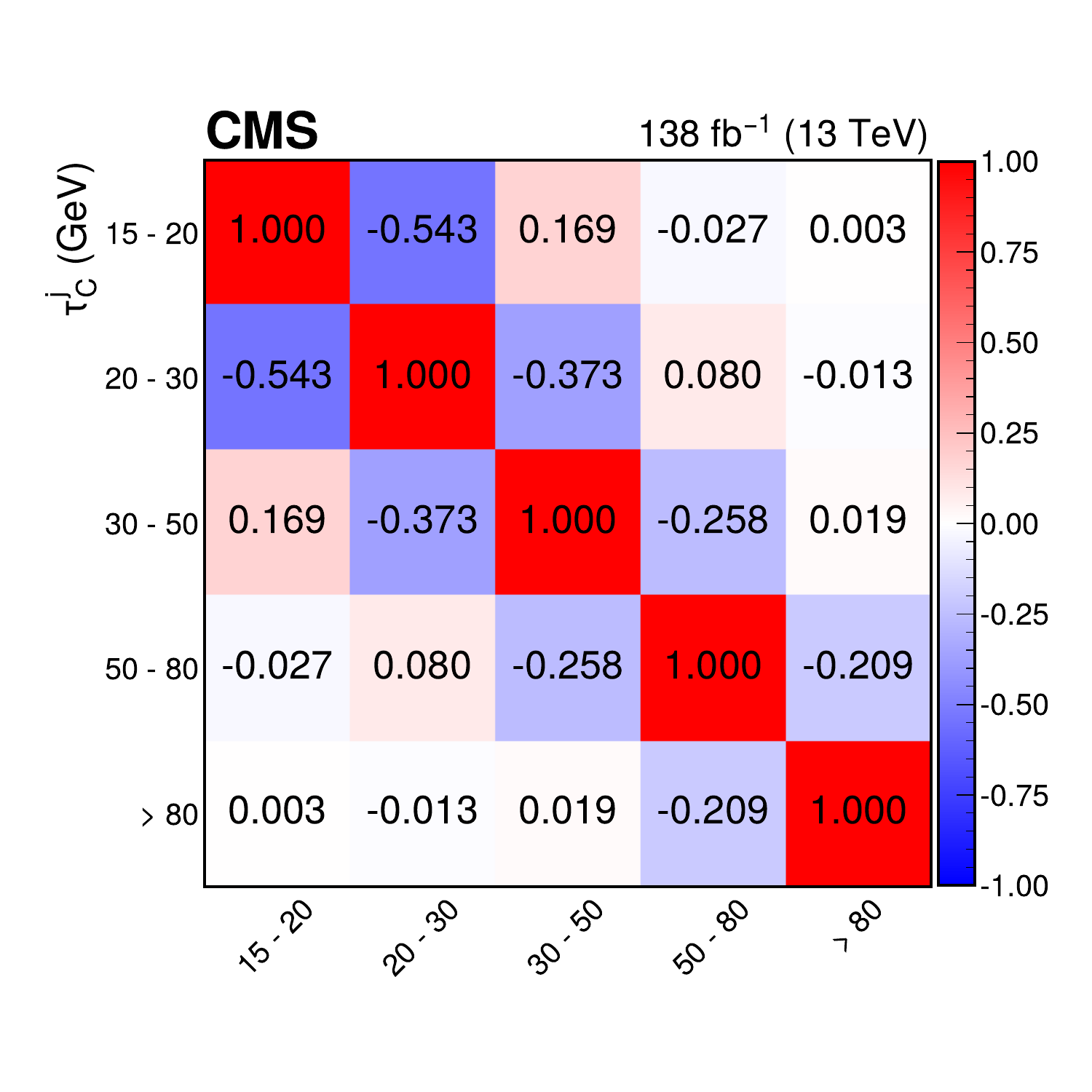}
    \caption{Bin-to-bin correlation matrices for the \mjj (upper left), \deta (upper right) and \taujc (lower) spectra.}
    \label{fig:corr_matrices_2}
\end{figure}

Figures~\ref{fig:corr_matrices_1} and \ref{fig:corr_matrices_2} show the bin-to-bin correlation matrices for considered observables.

\clearpage
\section{\texorpdfstring{\dpjj}{dpjj}\ SMEFT scans}
\label{app:smeft_deltaphijj_scans}

\begin{figure}[!ht]
    \centering
    \includegraphics[width=0.49\textwidth]{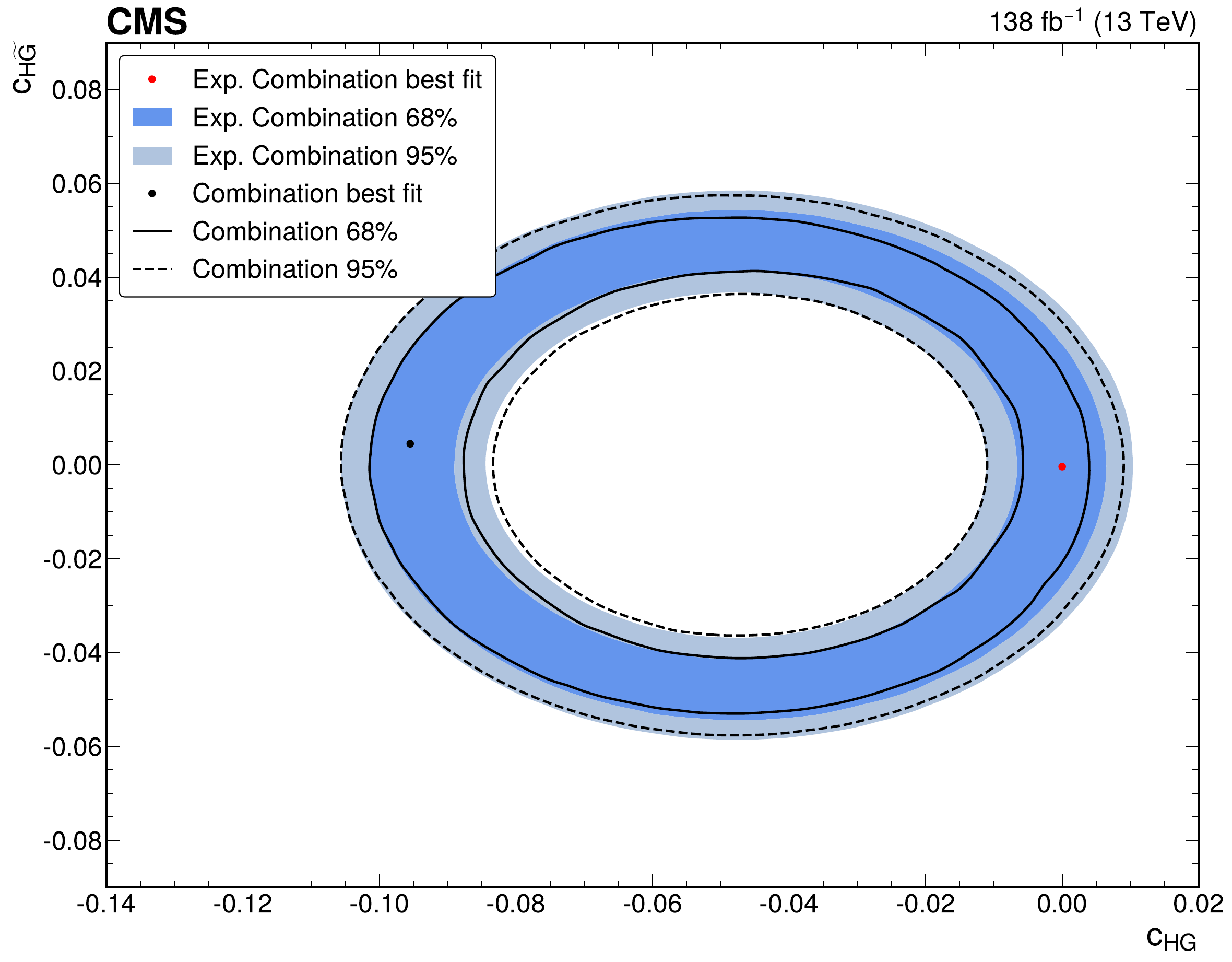}
    \includegraphics[width=0.49\textwidth]{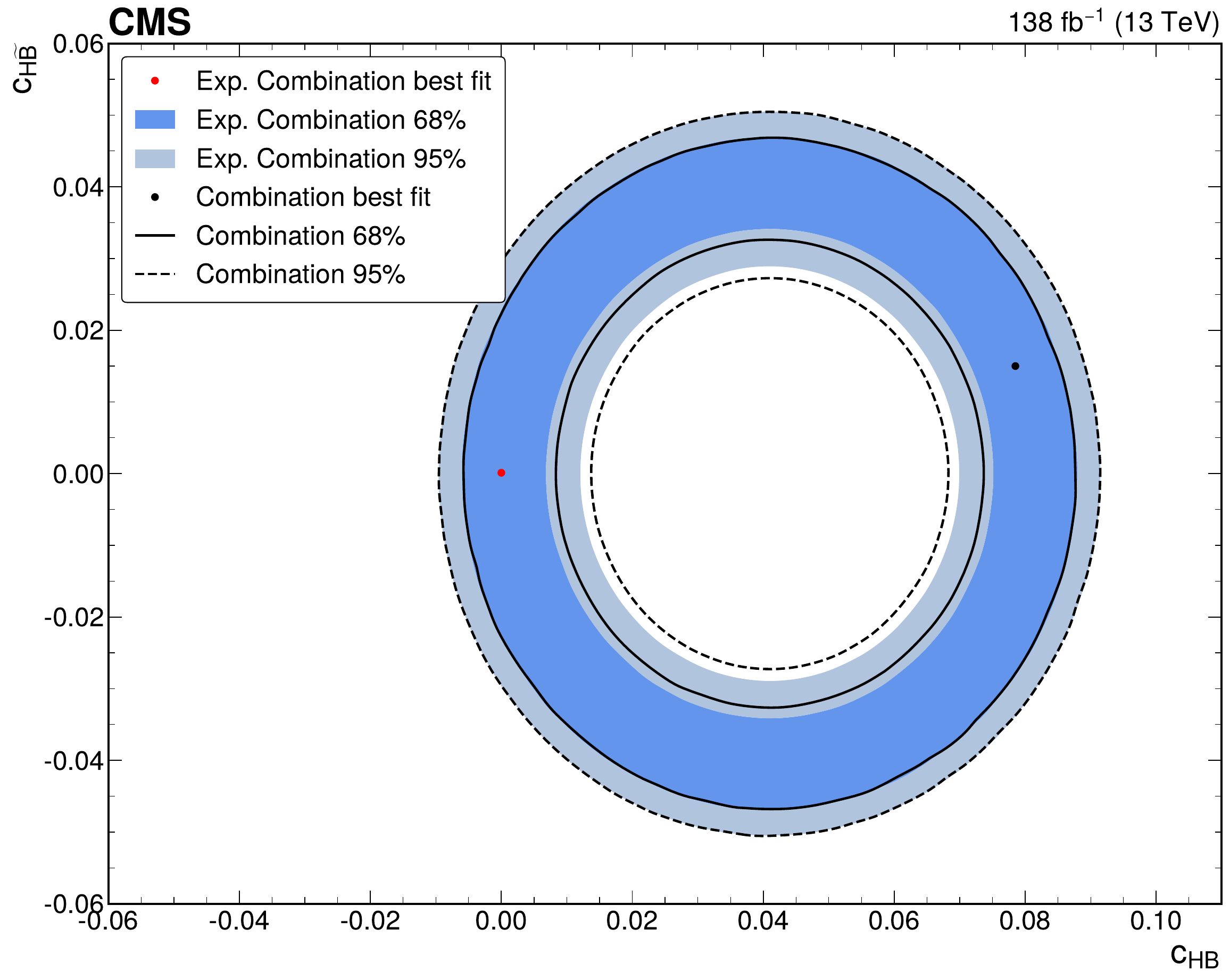} \\
    \includegraphics[width=0.49\textwidth]{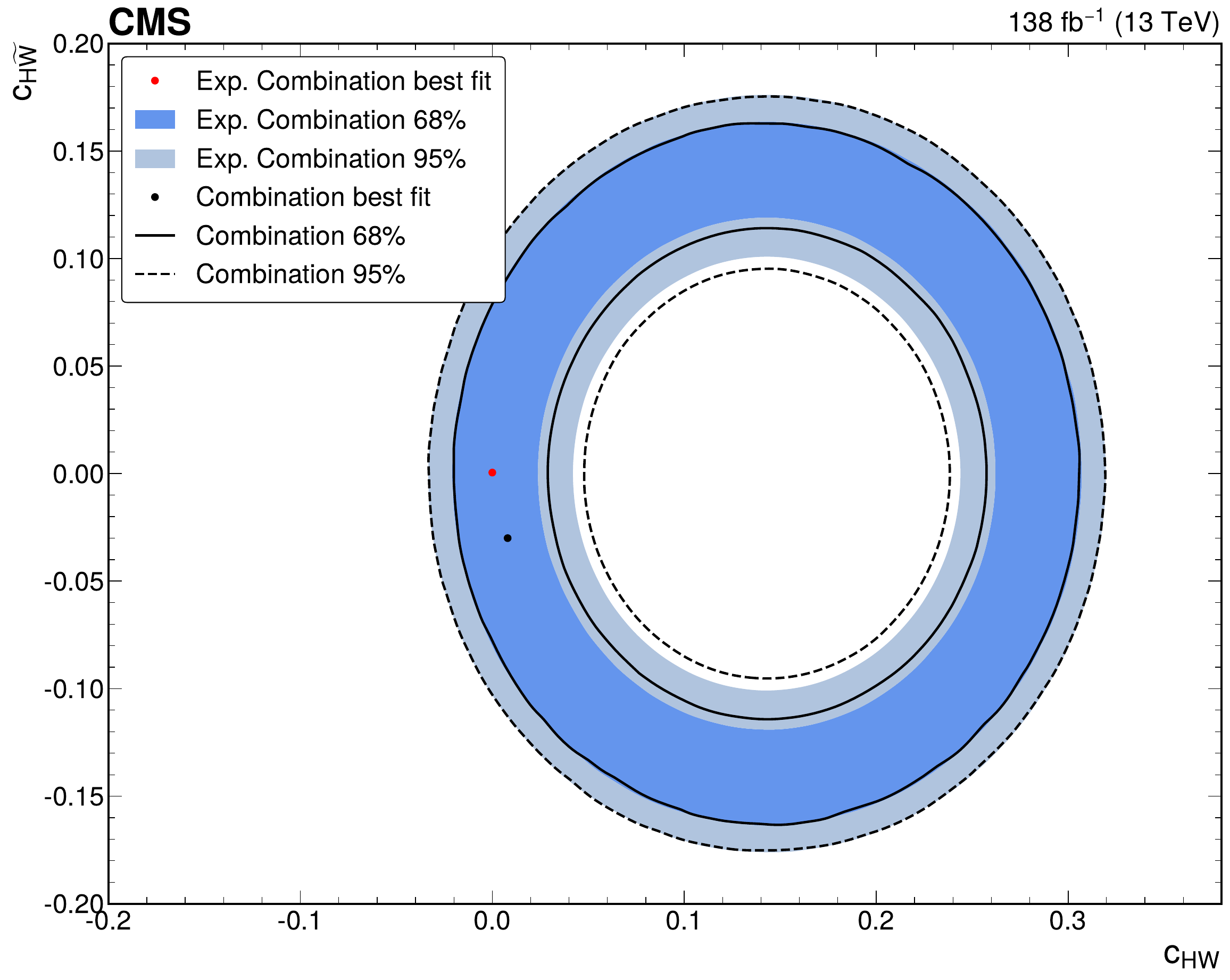}
    \includegraphics[width=0.49\textwidth]{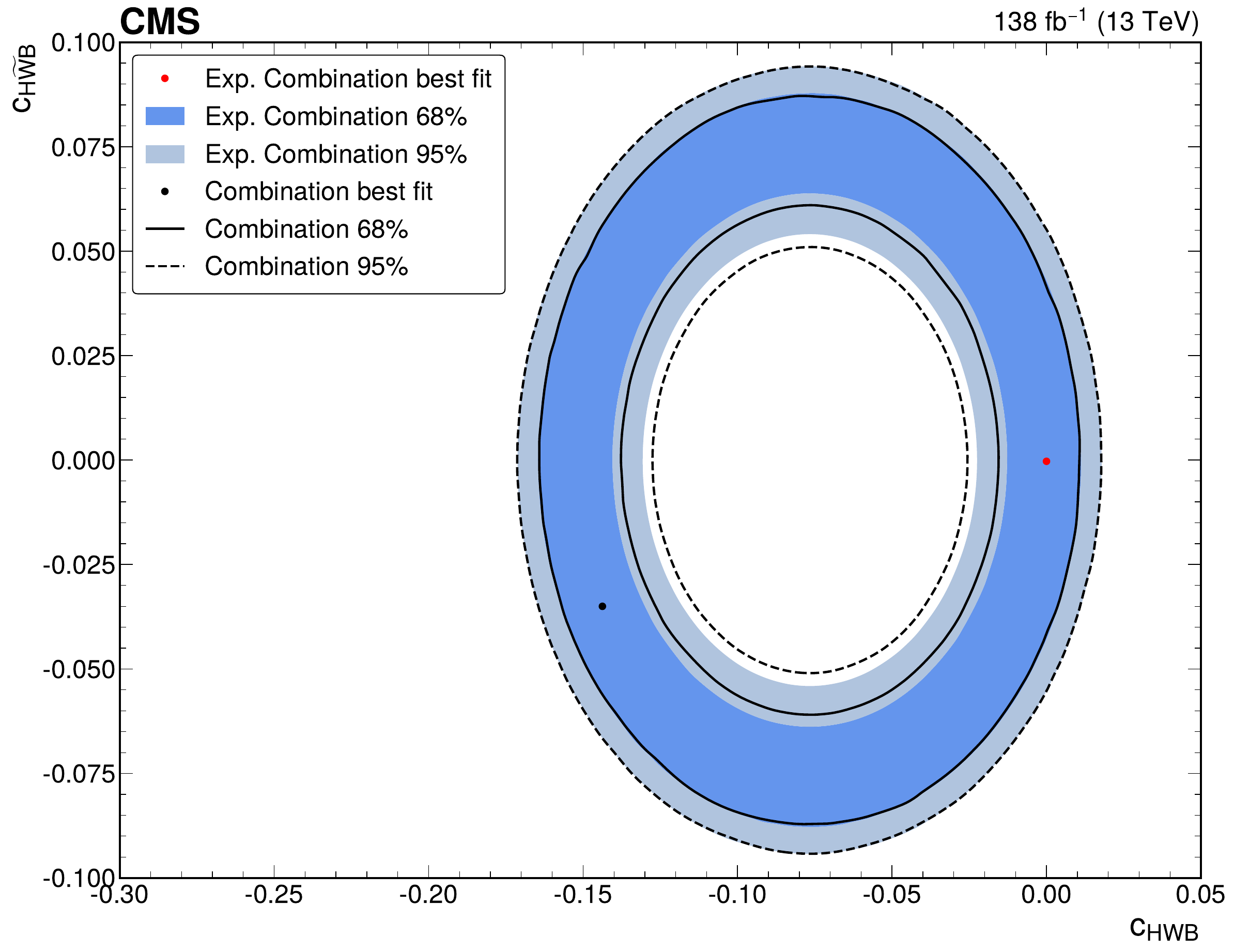}
    \caption{Two-dimensional scans for the \chg-\chgt\ (upper left), \chb-\chbt\ (upper right), \chw-\chwt\ (lower left), and \chwb-\chwtb\ (lower right) pairs with \dpjj spectra in \hgamgam\ and \hzz.}
    \label{fig:deltaphijj_scans_2d_expbkg}
\end{figure}

Figure~\ref{fig:deltaphijj_scans_2d_expbkg} shows the two-dimensional scans for the \chg-\chgt, \chb-\chbt, \chw-\chwt, and \chwb-\chwtb\ pairs with \dpjj spectra in \hgamgam\ and \hzz.

\clearpage
\section{SMEFT scans to eigenvectors}
\label{app:smeft_eigenvectors}

\begin{figure}[!htb]
    \includegraphics[width=0.45\textwidth]{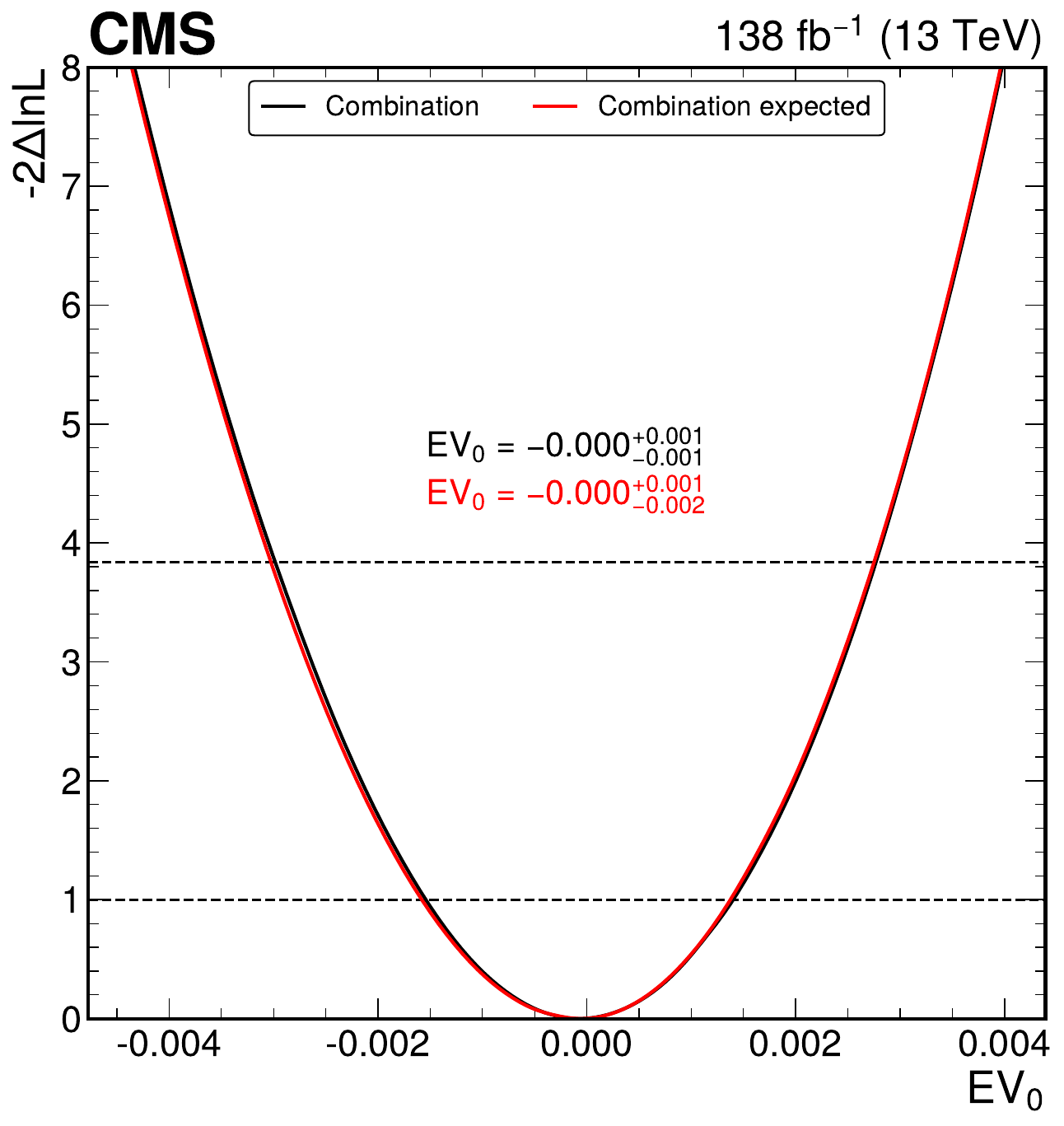}
    \includegraphics[width=0.45\textwidth]{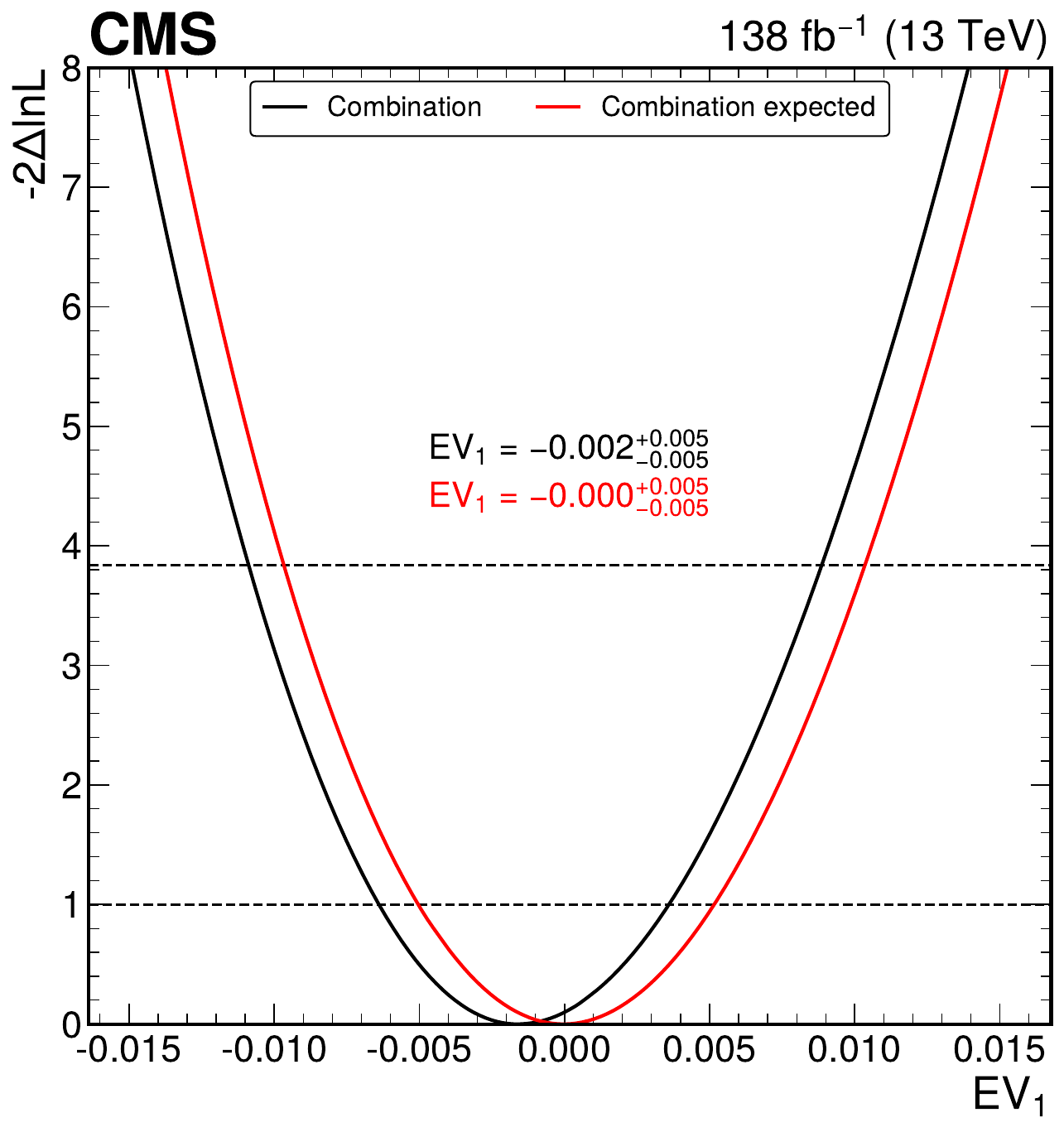} \\
    \includegraphics[width=0.45\textwidth]{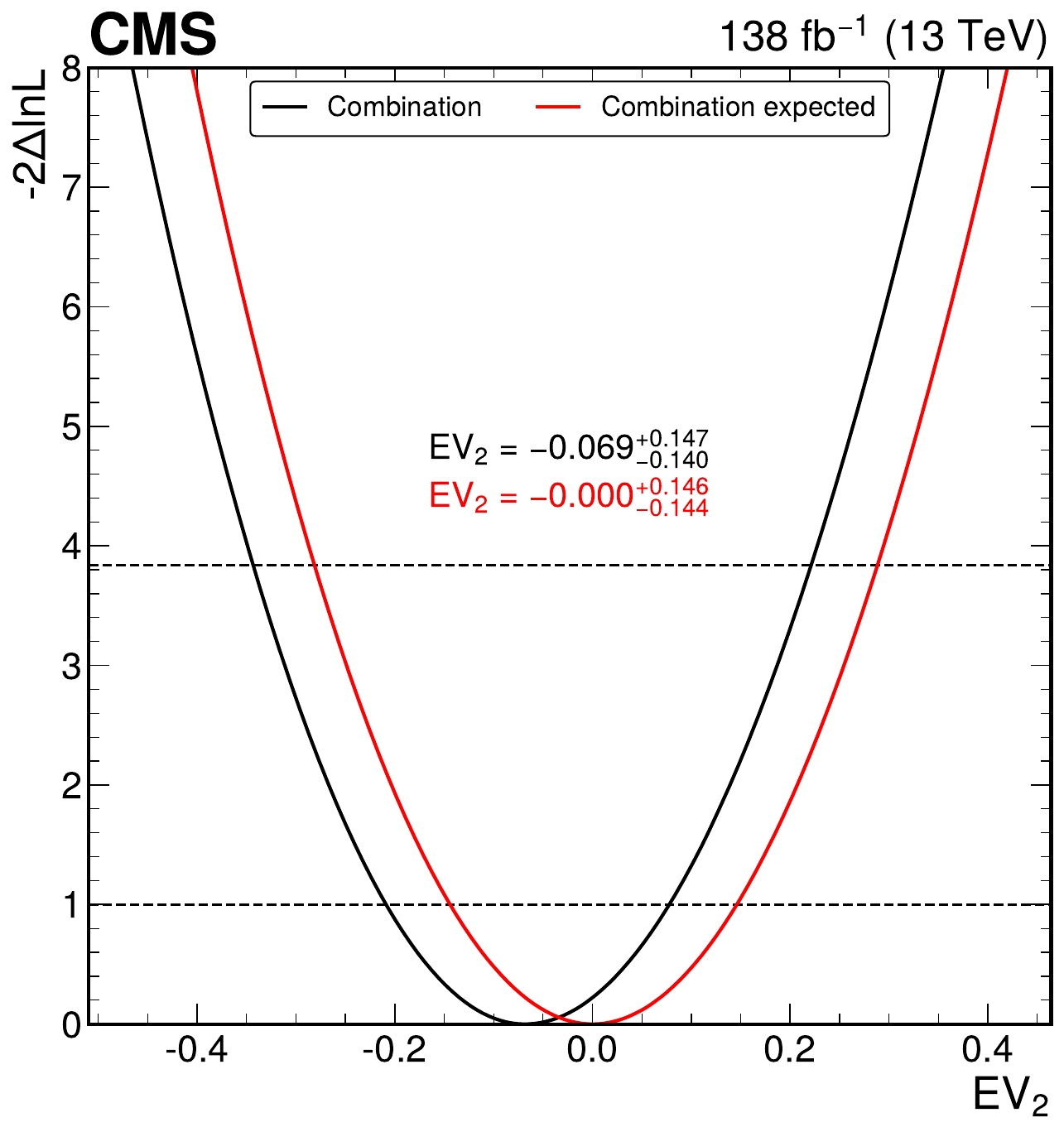}
    \includegraphics[width=0.45\textwidth]{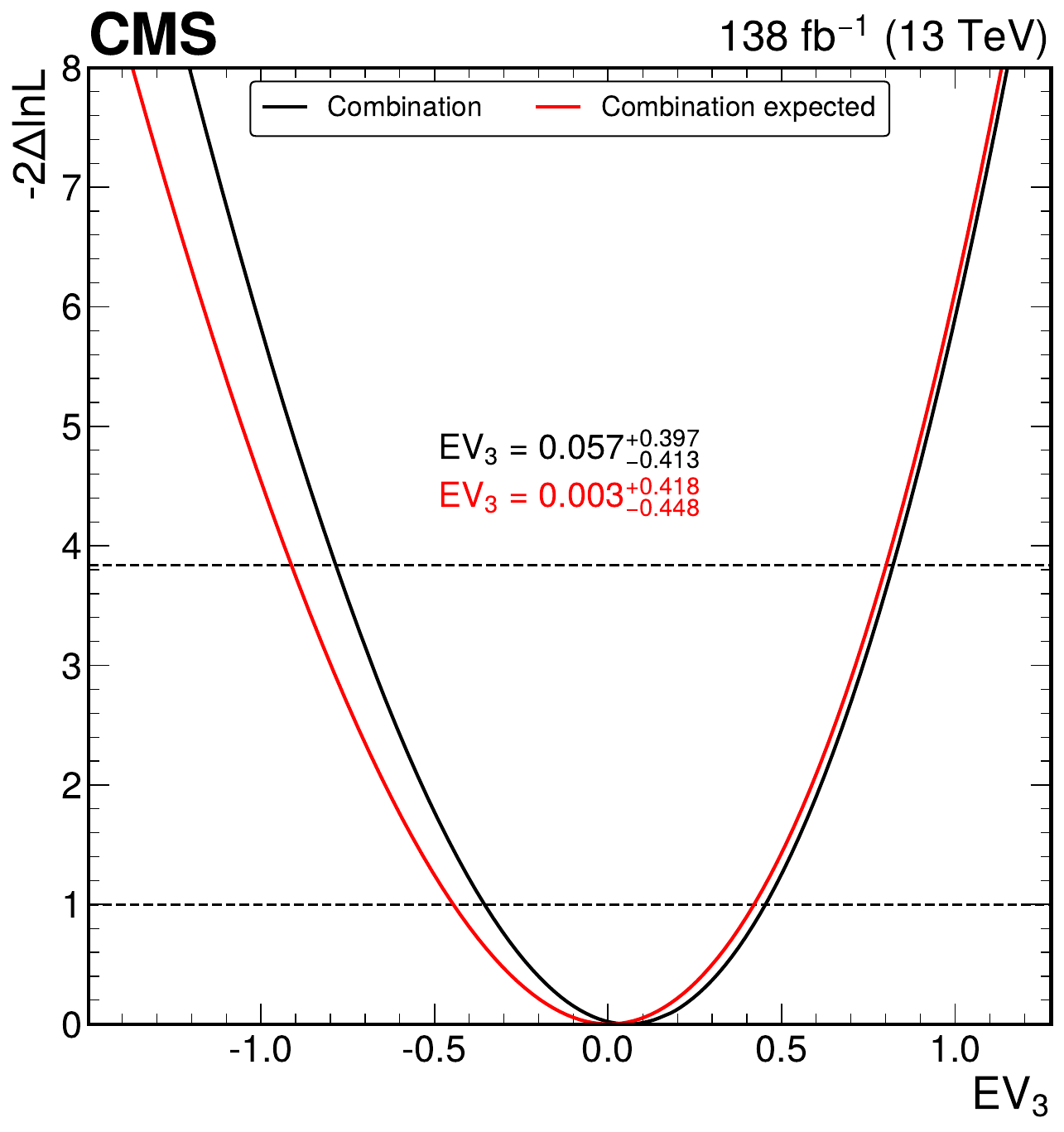} \\
    \caption{Observed and expected profile likelihood scans for eigenvectors 0 to 3.}
    \label{fig:fits_linear_1}
\end{figure}

\begin{figure}[!htb]
    \includegraphics[width=0.45\textwidth]{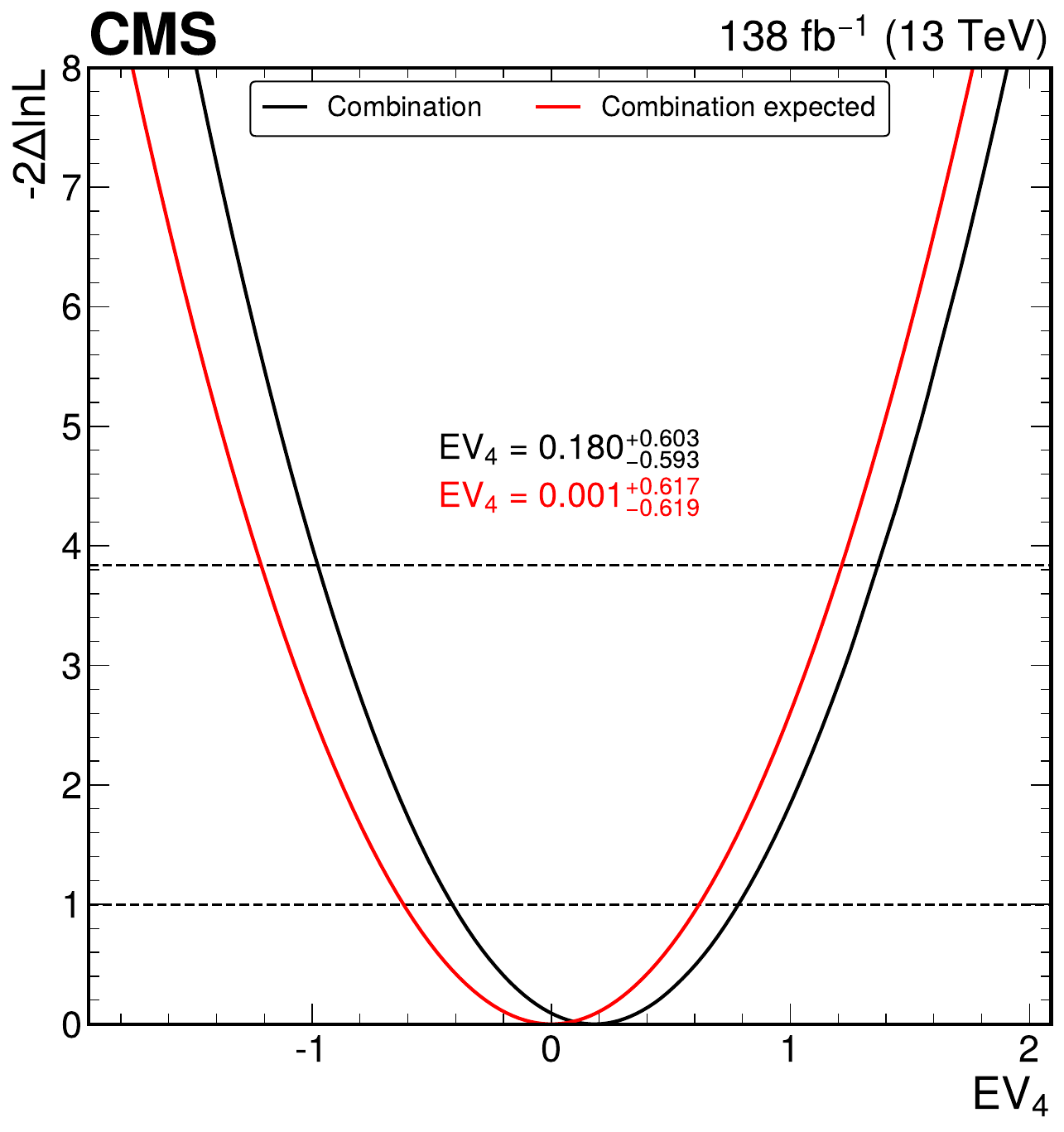}
    \includegraphics[width=0.45\textwidth]{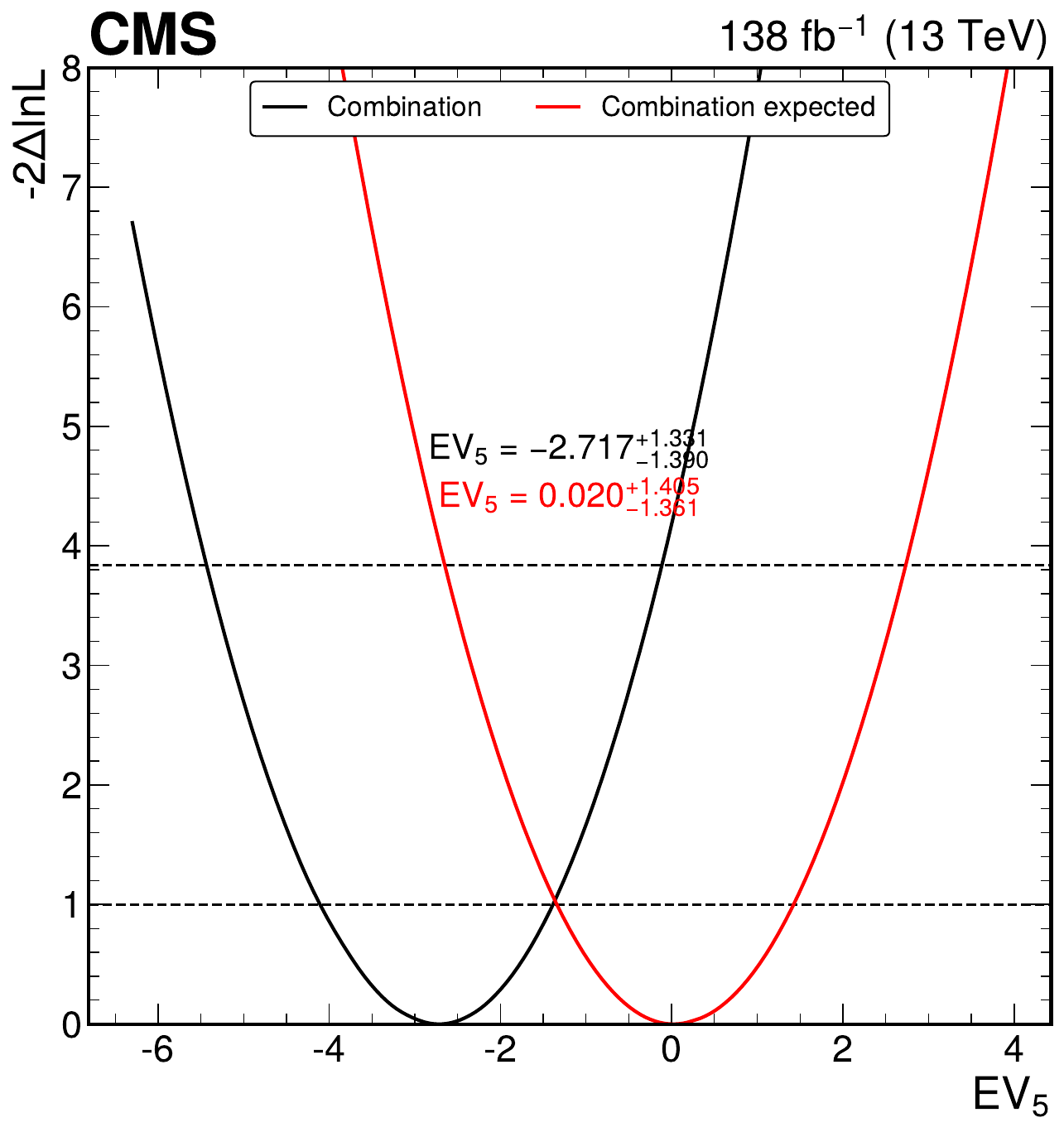} \\
    \includegraphics[width=0.45\textwidth]{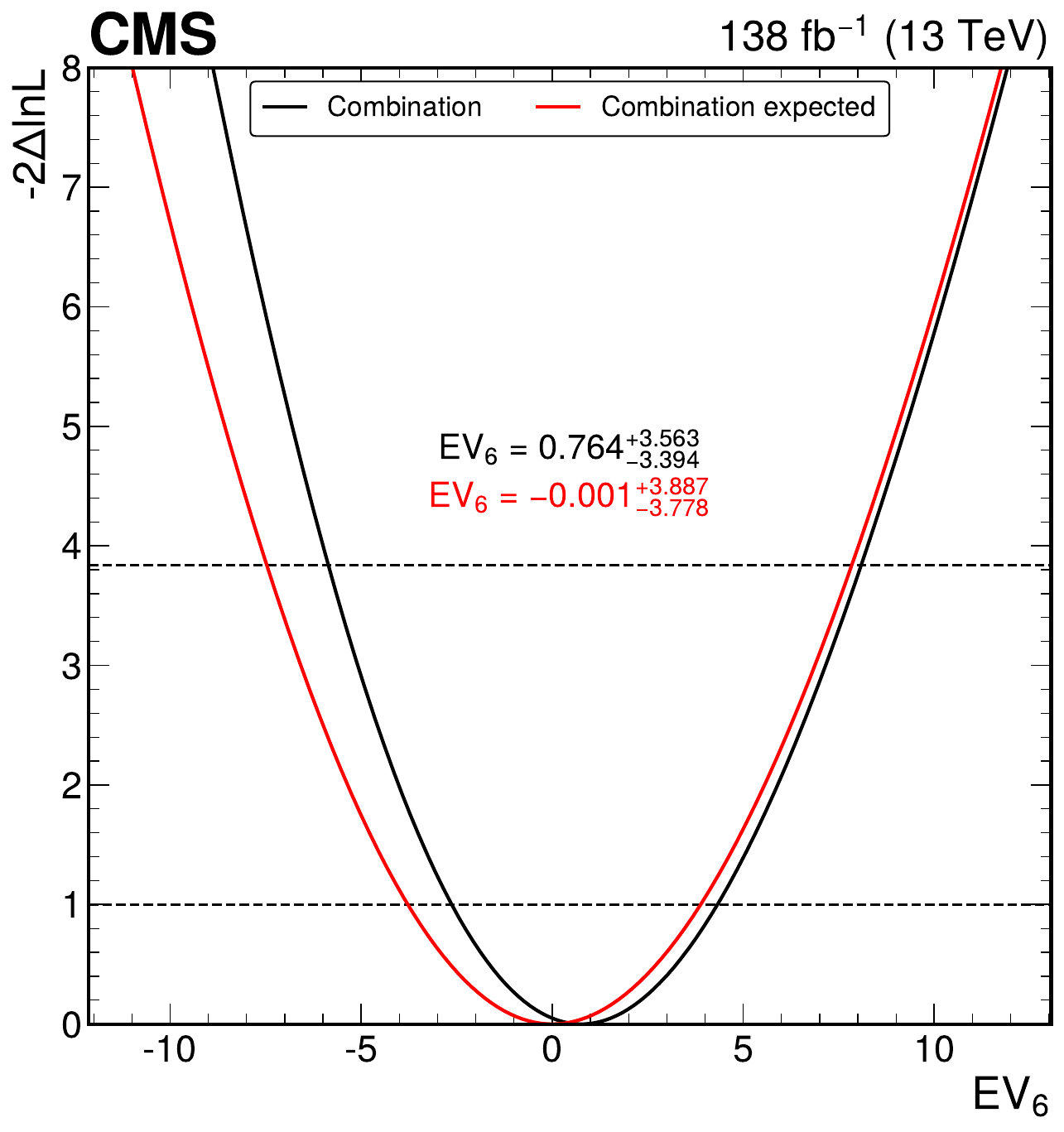}
    \includegraphics[width=0.45\textwidth]{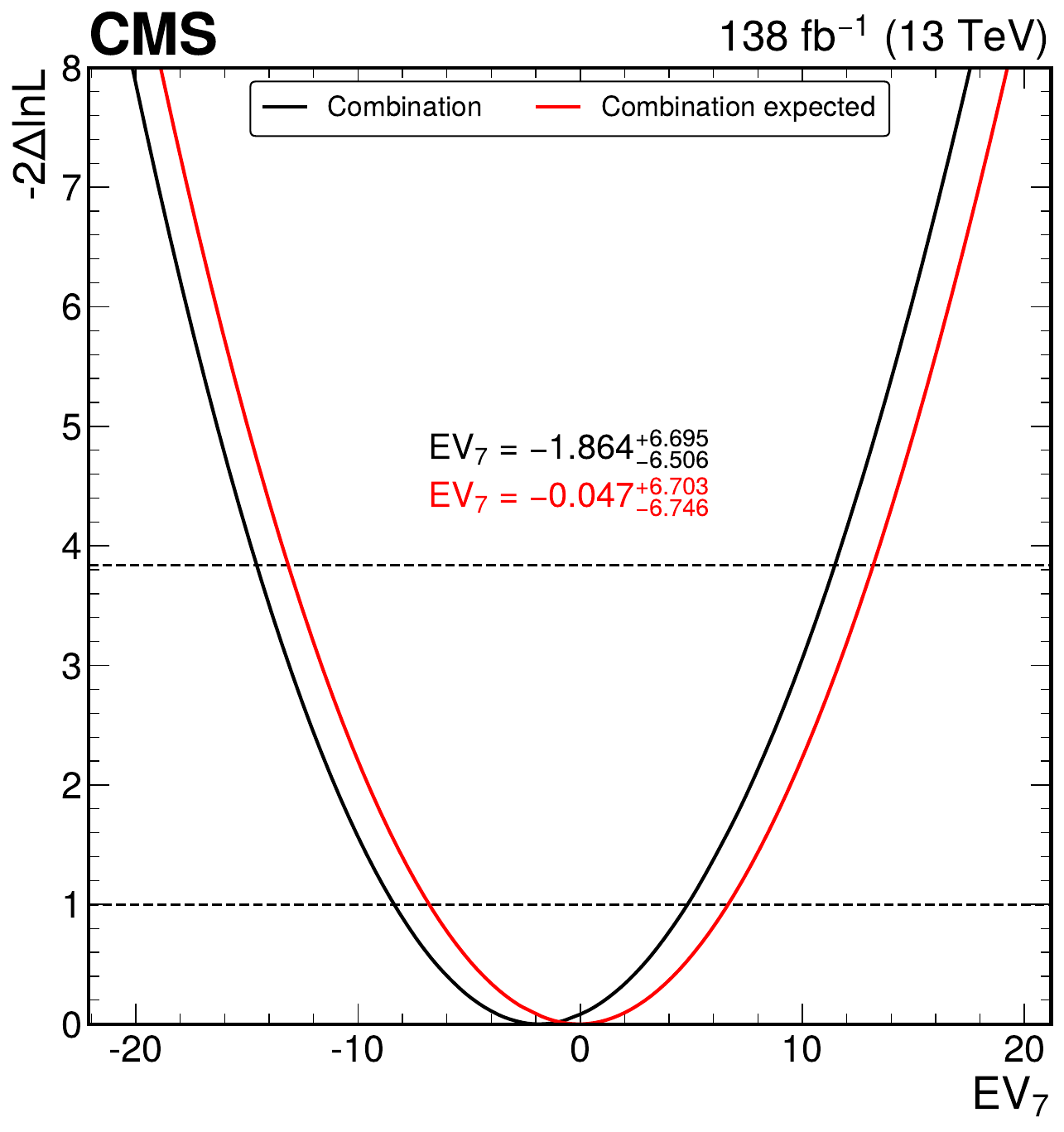} \\
    \caption{Observed and expected profile likelihood scans for eigenvectors 4 to 7.}
    \label{fig:fits_linear_2}
\end{figure}

\begin{figure}[!htb]
    \includegraphics[width=0.45\textwidth]{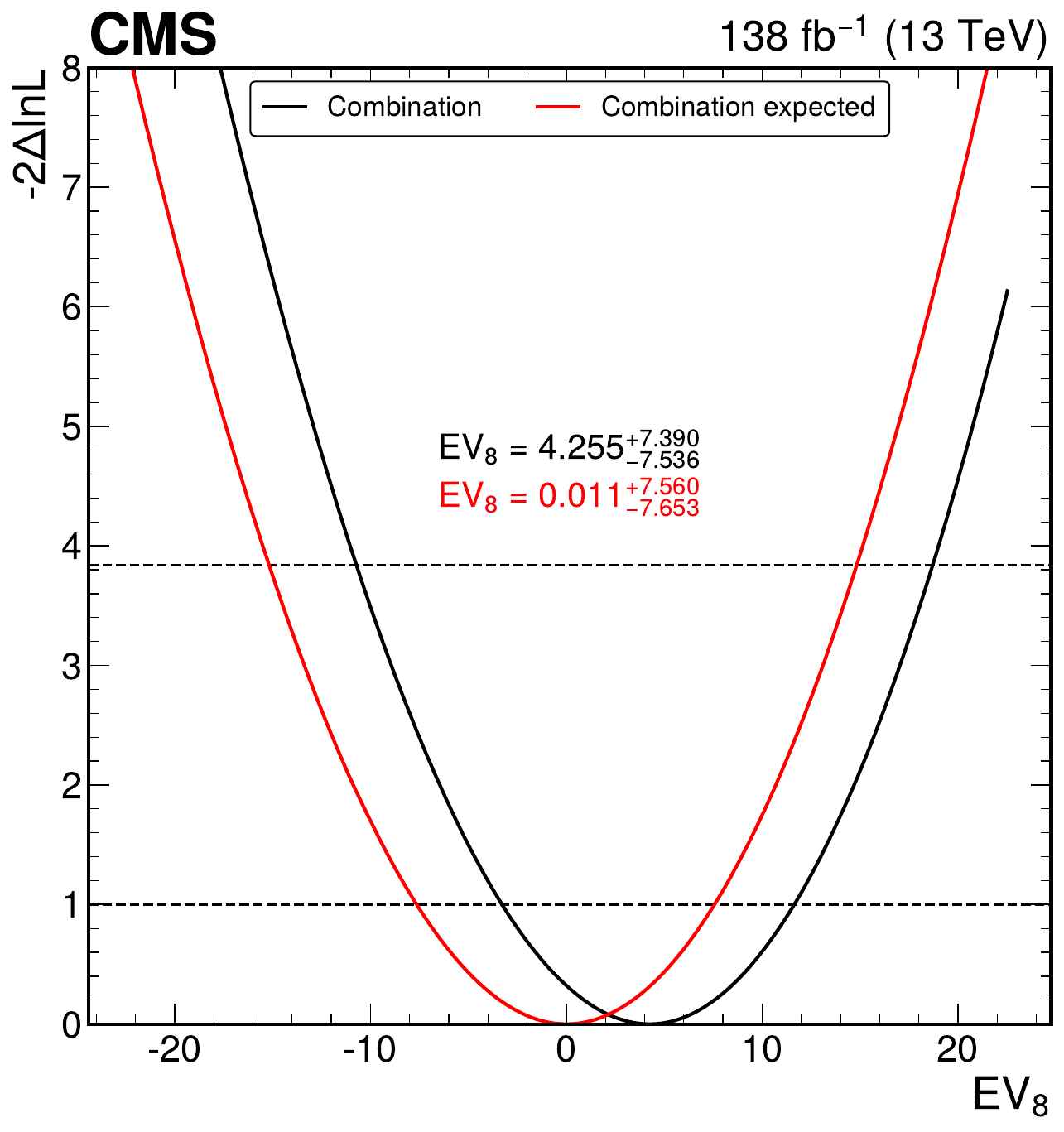}
    \includegraphics[width=0.45\textwidth]{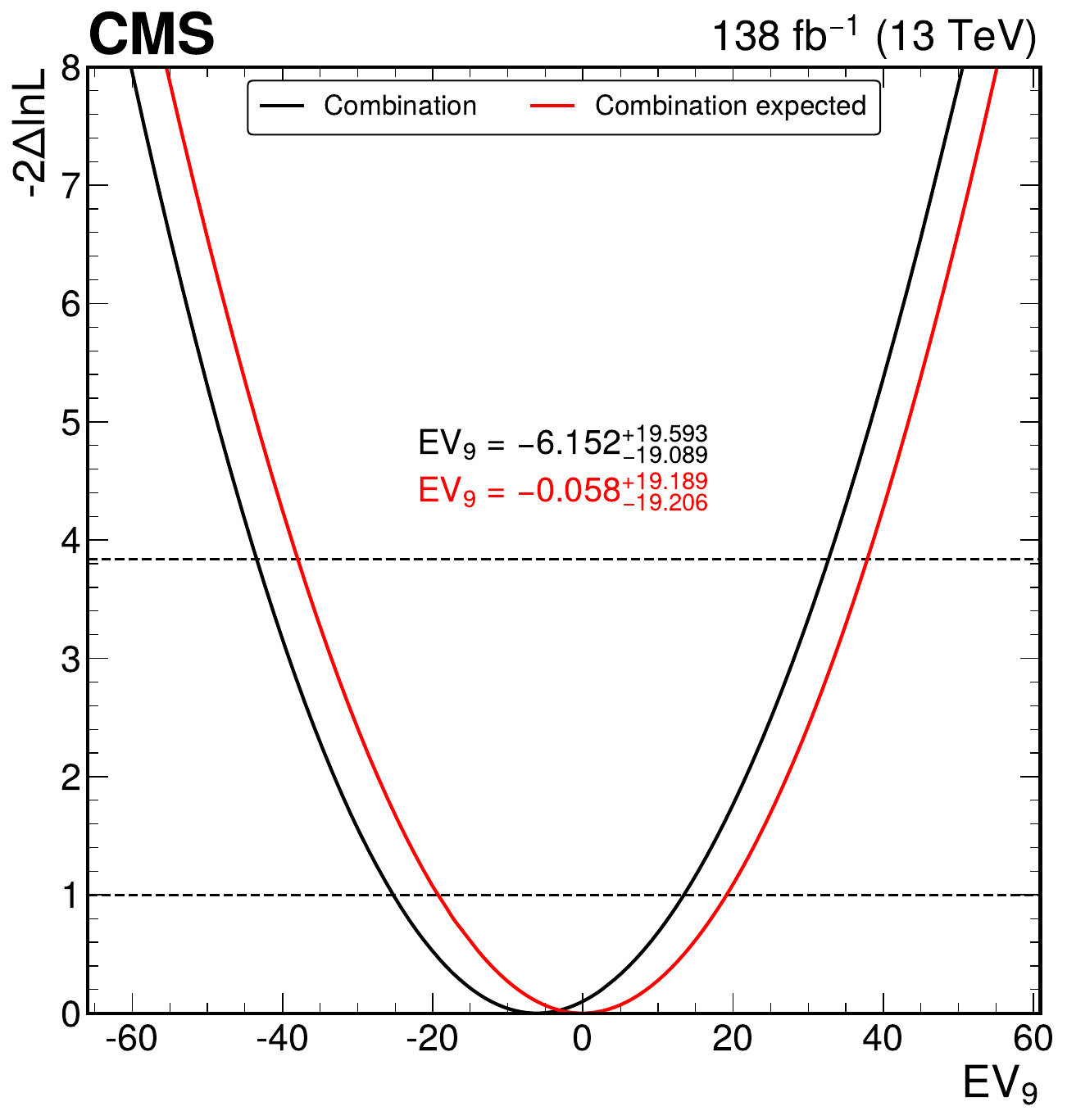} \\
    \caption{Observed and expected profile likelihood scans for eigenvectors 8 and 9.}
    \label{fig:fits_linear_3}
\end{figure}

Figures~\ref{fig:fits_linear_1}-\ref{fig:fits_linear_3} show the observed and expected profile likelihood scans for the first ten eigenvectors obtained from the PCA.

\cleardoublepage \section{The CMS Collaboration \label{app:collab}}\begin{sloppypar}\hyphenpenalty=5000\widowpenalty=500\clubpenalty=5000\input{HIG-23-013-public-authorlist.tex}\end{sloppypar}
\end{document}

%% file: HIG-23-013-public-authorlist.tex
\cmsinstitute{Yerevan Physics Institute, Yerevan, Armenia}
{\tolerance=6000
V.~Chekhovsky, A.~Hayrapetyan, V.~Makarenko\cmsorcid{0000-0002-8406-8605}, A.~Tumasyan\cmsAuthorMark{1}\cmsorcid{0009-0000-0684-6742}
\par}
\cmsinstitute{Institut f\"{u}r Hochenergiephysik, Vienna, Austria}
{\tolerance=6000
W.~Adam\cmsorcid{0000-0001-9099-4341}, J.W.~Andrejkovic, L.~Benato\cmsorcid{0000-0001-5135-7489}, T.~Bergauer\cmsorcid{0000-0002-5786-0293}, S.~Chatterjee\cmsorcid{0000-0003-2660-0349}, K.~Damanakis\cmsorcid{0000-0001-5389-2872}, M.~Dragicevic\cmsorcid{0000-0003-1967-6783}, P.S.~Hussain\cmsorcid{0000-0002-4825-5278}, M.~Jeitler\cmsAuthorMark{2}\cmsorcid{0000-0002-5141-9560}, N.~Krammer\cmsorcid{0000-0002-0548-0985}, A.~Li\cmsorcid{0000-0002-4547-116X}, D.~Liko\cmsorcid{0000-0002-3380-473X}, I.~Mikulec\cmsorcid{0000-0003-0385-2746}, J.~Schieck\cmsAuthorMark{2}\cmsorcid{0000-0002-1058-8093}, R.~Sch\"{o}fbeck\cmsAuthorMark{2}\cmsorcid{0000-0002-2332-8784}, D.~Schwarz\cmsorcid{0000-0002-3821-7331}, M.~Sonawane\cmsorcid{0000-0003-0510-7010}, W.~Waltenberger\cmsorcid{0000-0002-6215-7228}, C.-E.~Wulz\cmsAuthorMark{2}\cmsorcid{0000-0001-9226-5812}
% ...
\par}
\vskip\cmsinstskip
\dag:~Deceased\\
$^{1}$Also at Yerevan State University, Yerevan, Armenia\\
$^{2}$Also at TU Wien, Vienna, Austria\\
% ...
